\documentclass{jfm}

\usepackage{amssymb}
\usepackage{fancyhdr}
\usepackage{amsmath}
\usepackage{latexsym}
\usepackage{amstext}
\usepackage{stackrel}
\usepackage[usenames,dvipsnames]{color}
\usepackage[figuresright]{rotating}
\usepackage{graphicx}
\usepackage{natbib}
\usepackage{subfig}
\usepackage{color}
\usepackage{longtable}
\usepackage{multirow}		
\usepackage{array}	
\usepackage{esint}
\usepackage[mathscr]{eucal}
\newcommand\ste {\sqrt{\textrm{E}}}

\ifCUPmtlplainloaded \else
  \checkfont{eurm10}
  \iffontfound
    \IfFileExists{upmath.sty}
      {\typeout{^^JFound AMS Euler Roman fonts on the system,
                   using the 'upmath' package.^^J}%
       \usepackage{upmath}}
      {\typeout{^^JFound AMS Euler Roman fonts on the system, but you
                   dont seem to have the}%
       \typeout{'upmath' package installed. JFM.cls can take advantage
                 of these fonts,^^Jif you use 'upmath' package.^^J}%
      }
  \else
  \fi
\fi

\ifCUPmtlplainloaded \else
  \checkfont{msam10}
  \iffontfound
    \IfFileExists{amssymb.sty}
      {\typeout{^^JFound AMS Symbol fonts on the system, using the
                'amssymb' package.^^J}%
       \usepackage{amssymb}%
         \let\leq=\leqslant
         \let\geq=\geqslant
      }{}
  \fi
\fi

\ifCUPmtlplainloaded \else
  \IfFileExists{amsbsy.sty}
    {\typeout{^^JFound the 'amsbsy' package on the system, using it.^^J}%
     \usepackage{amsbsy}}
    {\providecommand\boldsymbol[1]{\mbox{\boldmath $##1$}}}
\fi

\providecommand\bcdot{\boldsymbol{\cdot}}

\newcommand\Rey{\mbox{\textit{Re}}}  
\newcommand\Deb{\mbox{\textit{De}}}  
\newcommand\El{\mbox{\textit{El}}}

\newsavebox{\astrutbox}
\sbox{\astrutbox}{\rule[-5pt]{0pt}{20pt}}

\title{Inertio-elastic instability of a vortex column}
\author[Anubhab Roy, Piyush Garg, Jhumpal Shashikiran Reddy and Ganesh Subramanian]%
{Anubhab Roy$^1$
\thanks{Present address: Department of Applied Mechanics, Indian Institute of Technology Madras, Chennai,
600 036, India}
and
Piyush Garg$^1$
and
Jhumpal Shashikiran Reddy$^1$
and
Ganesh Subramanian$^1$
\thanks{Email address for correspondence: sganesh@jncasr.ac.in}}
\affiliation{$^1$ Engineering Mechanics Unit, Jawaharlal Nehru Centre for Advanced Scientific Research, Jakkur, Bangalore, 560 064, India}

\pubyear{2018}
\volume{650}
\pagerange{119--126}
\date{?; revised ?; accepted ?. - To be entered by editorial office}

\begin{document}

\maketitle

\begin{abstract}
We analyze the instability of a vortex column in a dilute polymer solution at large $\Rey$ and $\Deb$ with $\El = \Deb/\Rey$, the elasticity number, being finite. Here, $\Rey = \Omega_0 a^2/\nu_s$ and $\Deb = \Omega_0 \tau$ are, respectively, the Reynolds and Deborah numbers based on the core angular velocity\,($\Omega_0$), the radius of the column\,($a$), the solvent-based kinematic viscosity\,($\nu_s = \mu_s/\rho$), and the polymeric relaxation time\,($\tau$). The stability of small-amplitude perturbations in this distinguished limit is governed by the elastic Rayleigh equation whose spectrum is parameterized by $\textrm{E} = \El(1-\beta)$, $\beta$ being the ratio of the solvent to the solution viscosity. The neglect of the relaxation terms, in the said limit, implies that the polymer solution supports undamped elastic shear waves propagating relative to the base-state flow. The existence of these shear waves leads to multiple\,(three) continuous spectra associated with the elastic Rayleigh equation in contrast to just one for the original Rayleigh equation. Further, unlike the neutrally stable inviscid case, an instability of the vortex column arises for finite $\textrm{E}$ due to a pair of elastic shear waves being driven into a resonant interaction under the differential convection by the irrotational shearing flow outside the core. An asymptotic analysis for the Rankine profile shows the absence of an elastic threshold; although, for small E, the growth rate of the unstable discrete mode is transcendentally small, being O$(\textrm{E}^2e^{-1/\textrm{E}^{\frac{1}{2}}})$. An accompanying numerical investigation shows that the instability persists for smooth vorticity profiles, provided the radial extent of the transition region (from the rotational core to the irrotational exterior) is less than a certain $\textrm{E}$-dependent threshold.
\end{abstract}


\section{Introduction}
The distinguishing trait of viscoelastic fluids is an underlying microstructure that often relaxes on macroscopic time scales. For dilute polymer solutions this microstructure consists of non-interacting macromolecules, the elasticity being endowed by the prolonged relaxation of these molecules on times scales which may range from milliseconds to seconds, and even longer, depending primarily on the polymer molecular weight and solvent viscosity. This elasticity has several striking consequences including rod climbing, die swell and the tubeless siphon effect which are well documented in textbooks\,(\cite{BIRD87}). \\

The role of elasticity in either suppressing or triggering instabilities in various flows has also been of immense interest due to implications for the polymer and food processing industries. About three decades back, the study of instabilities in polymer solutions received an impetus with the discovery of purely elastic instabilities in flows with curved streamlines, a scenario that includes the canonical viscometric flow geometries\,(\cite{LARSHAQ90, LARSON92, SHAQ96}). Such instabilities are driven by base-state hoop stresses, and the instability sets in above a threshold Deborah number\,($\Deb$), a dimensionless parameter denoting the ratio of the elastic to the flow time scales, even in the absence of inertia.  For large curvature ratios, a statistically stationary disorderly flow results for $\Deb$ values above the threshold, termed elastic turbulence\,(\cite{STEINBERG00}). \\

Elasticity also plays an important role in flows of polymer solutions where inertia is significant; a canonical and important example being the widely-studied phenomenon of turbulent drag reduction\,(\cite{VIRK75, LUMLEY69, MUNG08}). Simulations of polymers in turbulent channel flow show suppression of the coherent structures characteristic of the buffer layer, such as counter-rotating vortices and streaks aligned with the streamwise direction, in turn leading to a reduced wall shear stress \citep{MUNG08, SHAQFEH2004, BERIS1997}. Turbulent drag reduction corresponds mostly to the high $\Rey$ and moderate $\Deb$ regime. More recently, a novel spatio-temporally chaotic state, dubbed elasto-inertial turbulence, has been shown to exist for moderate $\Rey$ and high $\Deb$ \citep{MORO2013, ganesh18, hof2017, shankar2018, graham2019, chaudhary2021}. These observations call for a deeper mechanistic understanding of the inertial flows of polymeric solutions. The general question of hydrodynamic stability in the $(\Rey,\Deb)$ plane is thus an important one, and novel behavior is expected at large $\Rey$ and $\Deb$ \citep{graham2014, shankar2019}. Towards this goal, in this paper, we study the linear stability of a vortex column in an elastic liquid (a dilute polymer solution) at large $\Deb$ and $\Rey$ with the ratio $\Deb/\Rey$, known as the elasticity number\,($\El$), being finite. We demonstrate that the vortex column is susceptible to a novel inertio-elastic instability in this limit. \\

The early numerical investigations of curvilinear flows in the large $\Rey$, $\Deb$ regime concern the effects of elasticity on flow stability in the Taylor-Couette\,(\cite{WALTERS164}; \cite{WALTERS264}; \cite{DAVIES66}) and Dean geometries\,(\cite{WALTERS65}) for the inertially dominant case. Apart from analyzing the destabilizing effect of elasticity on the centrifugal mode, these early studies led to the identification, for finite $\El$, of the so-called inertio-elastic mode in Taylor-Couette flow\,(\cite{BERIS93}) that, unlike the centrifugal mode, had an oscillatory character at onset. More recently, it has been shown that the inertio-elastic mode exists for all Taylor-Couette configurations with a monotonically decreasing angular velocity profile (\cite{OGIPOT08}). There have been additional investigations of the stability of elastic free shear flows in the inertio-elastic regime. \cite{YARIN97} studied the effect of polymer additives on the dynamics of a vortex filament in an ambient shear flow, and showed that vortex stretching associated with the bending mode gets arrested by high longitudinal elastic stresses generated due to stretching of the deformed filament. \cite{AZHOM94} analyzed the instability of a shear layer, characterized by a hyperbolic tangent velocity profile, in a dilute polymer solution. The fluid rheology was modeled using both quasi-linear\,(Oldroyd-B, co-rotational Jeffreys) and non-linear\,(Giesekus) constitutive relations. The authors were the first to consider the aforementioned distinguished limit, $\Deb, \Rey \rightarrow \infty$ with $\El$ finite for the Oldroyd-B case, and derived the elastic Rayleigh equation that governs the stability of small-amplitude perturbations to a general parallel shearing flow in this limit. A numerical solution of this equation showed that, for finite $\El$, elasticity stabilized the original inertial\,(Kelvin-Helmholtz) instability of the shear layer by truncating the unstable range of wavenumbers. As argued by Hinch\,(appendix of \cite{AZHOM94}), the stabilizing action of elasticity for long wavelength perturbations is akin to the action of surface tension in damping out transverse modulations of a stretched membrane, the tension being on account of the stretched polymers in the region of high shear\,(the `membrane'). Later, motivated by the stability of vortical structures that result from the roll up of an elastic shear layer\,(\cite{HOMSY94}), \cite{HAJHOM97} studied the role of elasticity on the stability of an unbounded linear flow with elliptical streamlines. Earlier investigations of high-$\Rey$ flows of Newtonian fluids with elliptical streamlines\,(\cite{KER02}) show that the instability of such flows is related to the so-called elliptic instability where a vortex column is destabilized due to a resonant interaction between pairs of Kelvin modes\,(\cite{SAFFMAN75}). These resonances correspond to points of intersection of the Kelvin dispersion curves, and can only occur in three dimensions\,(a finite axial wavenumber). In the elastic case however, \cite{HAJHOM97} found the instability to exist even in two dimensions.\\

The study most closely related to the current one is that of \cite{RALLHIN95} who examined the stability of submerged planar and axisymmetric elastic jets. In addition to the expected elasticity-induced modification of the inertial instabilities known for these flow configurations, the authors identified an instability that owes its origin to an interplay of elasticity and inertia. Elastic stresses allow for the propagation of transverse shear waves along the otherwise unperturbed streamlines, and the differential convection by the base-state shear leads to the resonant interaction of a shear-wave-pair, in turn leading to exponential growth. The underlying physical mechanism for the instability of the Rankine vortex studied here has a similar character. An important advantage of the vortex column configuration is that it is inertially stable; as shown recently, the Rankine vortex only supports neutrally stable discrete and continuous spectrum modes, a combination of which may at best lead to algebraic growth for short times\,(\cite{GS2014, roy2014thesis}). Hence, unlike all earlier efforts \citep{AZHOM94, HAJHOM97, RALLHIN95} where the flow configurations are also susceptible to purely inertial instabilities, the novel inertio-elastic instability in the vortex case exists in isolation. The inertio-elastic instability also appears to require a base-state with a sharp spatial transition from a region, that strongly stretches the polymer molecules, to a relatively quiescent region where the polymer molecules are close to their equilibrium coil configurations. \\

In this paper, we examine the aforementioned inertio-elastic instability, analytically for a Rankine vortex, and numerically for other smooth vorticity profiles, for two-dimensional disturbances with zero axial wavenumber. The examination of smooth vorticity profiles allows one to verify the requirement of a sharp transition region which, for a vortex column, would be from a central region of solid-body rotation to an irrotational point-vortex exterior. The paper is organized as follows. In section \ref{sec:elas_form}, starting from the equations of motion and the Oldroyd-B constitutive relation, we derive the governing linearized equations for small-amplitude perturbations at finite $\Deb$ and $\Rey$. As a prelude to the material in section \ref{sec:CSrayleigh}, we discuss the continuous spectra of this system of equations, whose origin may be traced to the the spatially local character of the polymeric stress evolution. Next, we derive the elastic Rayleigh equation in plane-polar coordinates that governs the stability of a base-state vortical flow to infinitesimal perturbations in the limit $\Rey, \Deb \rightarrow \infty$ with $\El$ finite. The dimensionless parameter governing stability in this distinguished limit is $\textrm{E}=\El(1-\beta)$, and we examine perturbations with zero axial wavenumber. In section \ref{sec:CSrayleigh}, we examine the elastic Rayleigh equation equation in more detail. It is shown that in addition to the continuous spectrum of the familiar inviscid Rayleigh equation, one that spans the base-state range of velocities\, \citep{CASE60,DRAZINREID81,GSLiftup,GS2014}, the elastic Rayleigh equation exhibits a pair of continuous spectra that may be associated with neutrally stable `slow' and `fast' elastic shear waves propagating with speeds proportional to $\pm \sqrt{\textrm{E}}$ relative to the local flow. These additional travelling-wave spectra are in contrast to the finite-$\Deb$ continuous spectra, arising from a balance of inertia and elasticity. The travelling-wave spectra include singular modes propagating with speeds outside the base-state interval in apparent violation of the known generalization of Howard semi-circle theorem for finite elasticity that predicts a reduced radius of the semi-circle for finite $\textrm{E}$\,(\cite{RALLHIN95}). Numerical calculations using a spectral code reinforce the analytical predictions  for the elastic continuous spectra. In section \ref{discrete:instability}, we examine the exponential instability that arises for an elastic vortex column via a shear wave resonance; this involves numerically investigating smooth vorticity profiles for finite $\textrm{E}$ (section \ref{finiteE:numerics}), and an analytical investigation of the Rankine profile for small $\textrm{E}$ (section \ref{smallE:analysis}). The discrete unstable mode for the Rankine profile is a regularized version of the singular travelling-wave modes analyzed in section \ref{sec:CSrayleigh}, that propagates with an angular speed slower than that of the core by an amount of $O(\textrm{E}^{1/2})$ and with a transcendentally small growth rate of $O(\textrm{E}^2 e^{-\textrm{E}^{-1/2}})$ determined using a matched asymptotics expansions approach. The latter transcendental scaling leads to a precipitous drop in the growth rate for small $\textrm{E}$ in the numerics, similar to that observed earlier for the submerged elastic jet (\cite{RALLHIN95}). Numerical results for more general vorticity profiles, where the discontinuity in the base state vorticity of the Rankine vortex is smoothened into a transition layer of width $d$, show that the instability persists for non-Rankine profiles. Finally, in section \ref{sec:conclu}, we summarize the main results with a discussion of future lines of research that emerge from this effort. The vortex stability problem analyzed here also has astrophysical ramifications owing to the direct analogy between the governing equations of polymer dynamics for large $\Deb$ and those of magnetohydrodynamics at large magnetic Reynolds numbers\,($\Rey_m$), and these are also discussed in section \ref{sec:conclu}. Note that the general reader, not interested in details of the continuous spectrum, may directly move to the stability analysis in section \ref{discrete:instability} after section \ref{sec:elas_form}.

\section{Problem formulation: the elastic Rayleigh equation} \label{sec:elas_form}
The equations of motion and continuity for a polymer solution of density $\rho$ are given by:
\begin{eqnarray}
\rho\frac{D\mathbf{v}}{Dt}&=&-\nabla p^{*} +\nabla \bcdot \mbox{\boldmath$\sigma$}_d, \label{eq:old1} \\
\nabla \bcdot \mathbf{v}&=&0, \label{eq:old2}
\end{eqnarray}
where ${\boldsymbol \sigma}_d$, the deviatoric stress, is assumed to satisfy the Oldroyd-B constitutive equation and the rescaled pressure, $p^{*}$ accounts for the additional (osmotic) pressure induced by polymer molecules. The Oldroyd-B relation is one of the simplest constitutive relations that offers a semi-quantitative description of nearly constant viscosity dilute polymer solutions known as Boger fluids\,(\cite{BOGER2009}), and corresponds to a microscopic description where the polymer molecules are modeled as non-interacting Hookean dumbbells\,(\cite{BIRD87};\cite{LARSON88}). For simple shear flow, the Oldroyd-B relation predicts a constant shear viscosity and first normal stress difference, and a zero second normal stress difference. It is convenient to write ${\boldsymbol \sigma}_d$ in (\ref{eq:old1}) as the sum of solvent and polymer contributions, $\mbox{\boldmath$\sigma$}_d=2\mu_s\mathbf{E}+G\mathbf{A}$. Here, ${\boldsymbol E} = \nabla {\boldsymbol v} + (\nabla {\boldsymbol v})^\dag$ is the rate of strain tensor with $\mu_s$ being the solvent viscosity. The polymer stress contribution is proportional to the shear modulus $G$ with the conformation tensor  $\mathbf{A}\propto \langle \mathbf{R R} \rangle$, ${\boldsymbol R}$ being the dumbbell end-to-end vector. Since the dumbbells respond affinely to an imposed velocity field in the absence of relaxation, ${\boldsymbol A}$ is governed by: 
\begin{equation}
\overset{\mathcal{5}}{\mathbf{A}} =-\frac{1}{\tau}(\mathbf{A}-\mathbf{I}). \label{eq:oldA_2}
\end{equation}
where $\tau$ is the relaxation time and `$\overset{\mathcal{5}}{}$' denotes the upper-convected derivative defined by:
\begin{equation}
\overset{\mathcal{5}}{\mathbf{X}}\equiv\frac{D\mathbf{X}}{Dt}-(\nabla\mathbf{v})^{\dag} \bcdot \mathbf{X}-\mathbf{X} \bcdot (\nabla\mathbf{v}),
\end{equation}
with $\frac{D}{Dt}$ being the material derivative. One may now rewrite the governing set of equations as:
\begin{eqnarray}
\rho\frac{D\mathbf{v}}{Dt}&=&-\nabla p^{*}+\mu_s\nabla^2\mathbf{v}+G\nabla \bcdot \mathbf{A}, \label{eq:oldA_1}\\
\nabla \bcdot \mathbf{v}&=&0, \label{eq:oldA_3} \\
\frac{D\mathbf{A}}{Dt}-(\nabla\mathbf{v})^{\dag} \bcdot \mathbf{A}-\mathbf{A} \bcdot (\nabla\mathbf{v}) &=& -\frac{1}{\tau}(\mathbf{A}-\mathbf{I}). \label{eq:oldA_22}
\end{eqnarray}  

To examine the linearized evolution of two-dimensional disturbances in the swirling flow of an Oldroyd-B fluid, we write $\mathbf{v}=\overline{\mathbf{u}}+\mathbf{u}, \mathbf{A}=\overline{\mathbf{A}}+\mathbf{a}$ for the velocity and polymer stress fields with the overbar quantities denoting the unperturbed base state.  In the regime of interest in this paper ($\Rey, \Deb \rightarrow \infty$), any general axisymmetric swirling flow, $\overline{\mathbf{u}}=(0,\Omega(r) r, 0)$ with $\Omega(r)$ the angular velocity, is an exact solution of the equations of motion. The associated base-state stresses are given by:
\begin{eqnarray}
\overline{\mathbf{A}}=\left[\begin{array}{cc} 1 & r\Omega'\tau \\ r\Omega'\tau & 1+2(r\Omega'\tau)^2 \end{array}\right],
\label{eq:basestress}
\end{eqnarray}
in a cylindrical coordinate system where $'$ denotes a radial derivative. The base-state hoop stress component\,($\overline{\boldsymbol A}_{\theta\theta}$), on account of the quadratic scaling with the shear rate, becomes dominant for large shear rates\,($r\Omega'\tau \gg 1$) except when $\Omega' = 0$ which corresponds to the trivial case of solid-body rotation. In what follows, we analyze the linear stability of the Rankine vortex profile for which $\Omega(r) = \Omega_0$ for $r<a$\,(the rigidly rotating core) and $\Omega(r) = \Omega_0(a/r)^2$ for $r \geq a$\,(the irrotational exterior). We also examine the linear stability of more general vorticity profiles numerically. \\

The governing equation for the perturbation velocity field is:
\begin{eqnarray}
\frac{\partial \mathbf{u}}{\partial t}\!+\!\Omega\frac{\partial \mathbf{u}}{\partial\theta}+\mathbf{u} \bcdot \nabla\overline{\mathbf{u}}=-\frac{1}{\rho}\nabla p+\nu_s\nabla^2\mathbf{u}+\frac{G}{\rho}\nabla \bcdot \mathbf{a}.
\end{eqnarray} 
For the two-dimensional perturbations under consideration, $\mathbf{u}\equiv(u_r,u_{\theta})$, and the formulation is more convenient in terms of the axial vorticity, $w_z$, which satisfies the following equation:
\begin{eqnarray}
\hspace*{-0.5in}\left(\frac{\partial }{\partial t}\!+\!\Omega\frac{\partial }{\partial\theta}\!\right)\!\!w_z\!+\!u_r DZ\!\!&=&\nu_s \nabla^2w_z+\frac{G}{\rho}\left\{\nabla\wedge(\nabla \bcdot \mathbf{a})\right\}_z\nonumber \\
&=&\!\nu_s\nabla^2w_z\!+\!\frac{G}{\rho}\!\left[\!\frac{1}{r^2}\frac{\partial^2}{\partial r\partial\theta}(rN_1)\!+\!\frac{1}{r}\frac{\partial}{\partial r}\!\!\left(\frac{1}{r}\frac{\partial}{\partial r}(r^2a_{r\theta})\!\!\right)\!\!-\!\frac{1}{r^2}\frac{\partial^2}{\partial\theta^2}a_{r\theta}\!\right]\!. \label{ax:vorticity}
\end{eqnarray}
Here, $D\!Z=r\Omega''+3\Omega'$ is the base-state vorticity gradient and $N_1=a_{\theta\theta}-a_{rr}$ is the perturbation to the first normal stress difference. For the Rankine vortex, the base-state vorticity\,($Z$) and vorticity gradient\,($D\!Z$) are $Z(r) = 2\Omega_0{\mathcal H}(a-r)$ and $D\!Z(r) = - 2\Omega_0 \delta(r-a)$, ${\mathcal H}(z)$ and $\delta(z)$ being the Heaviside and Dirac delta functions, respectively. 

The perturbation elastic stress components that appear in (\ref{ax:vorticity}) obey the following equations:
\begin{eqnarray}
&&\left(\frac{\partial }{\partial t}+\Omega\frac{\partial }{\partial\theta}+\frac{1}{\tau}\right)a_{rr}-2\left\{\overline{A}_{rr}\frac{\partial u_r}{\partial r}+\frac{\overline{A}_{r\theta}}{r}\frac{\partial u_r}{\partial\theta}\right\}=0, \label{perta:1}\\
&&\left(\frac{\partial }{\partial t}+\Omega\frac{\partial }{\partial\theta}+\frac{1}{\tau}\right)a_{r\theta}+\left\{\overline{A}'_{r\theta}u_r-\overline{A}_{r\theta}\left(\frac{\partial u_r}{\partial r}+\frac{u_r}{r}\right)-\frac{\overline{A}_{\theta\theta}}{r}\frac{\partial u_r}{\partial\theta}\right\}+\nonumber\\
&&\left\{\overline{A}_{rr}\left(\frac{u_{\theta}}{r}-\frac{\partial u_{\theta}}{\partial r}\right)-\frac{\overline{A}_{r\theta}}{r}\frac{\partial u_{\theta}}{\partial\theta}\right\}-r\Omega'a_{rr}=0, \label{perta:2}\\
&&\left(\frac{\partial }{\partial t}+\Omega\frac{\partial }{\partial\theta}+\frac{1}{\tau}\right)a_{\theta\theta}-2\left\{\overline{A}_{r\theta}\left(\frac{\partial u_{\theta}}{\partial r}-\frac{u_{\theta}}{r}\right)+\frac{\overline{A}_{\theta\theta}}{r}\frac{\partial u_{\theta}}{\partial\theta}\right\}+\nonumber\\
&&\left(\overline{A}'_{\theta\theta}-\frac{2\overline{A}_{\theta\theta}}{r}\right)u_r-2r\Omega'a_{r\theta}=0. \label{perta:3}
\end{eqnarray}
We use the vortex core radius $a$ as the length scale and the turnover time based on the core angular frequency $\Omega_0^{-1}$ as a time scale. Next, assume a normal mode form, $h(r,\theta)=\hat{h}(r)e^{i(m\theta-\omega t)}$, for the various perturbation fields, where $m$ is the azimuthal wavenumber and $Im(\omega)>0$ corresponds to an exponentially growing perturbation. Thus, from (\ref{ax:vorticity}) and (\ref{perta:1})-(\ref{perta:3}), we obtain the following equations governing the $r$-dependent perturbation amplitudes:
\begin{eqnarray}
&&\Sigma r\mathscr{L}(r\hat{u}_r)\!+\!mrDZ\hat{u}_r\!=\!\frac{i}{\textrm{Re}}r\mathscr{L}^2\!(r\hat{u}_r)\!-\!\!\frac{im}{\textrm{Ma}^2_e}\!\!\left[\!-mD^*\hat{N}_1\!+\!iDD^{*}(r\hat{a}_{r\theta})\!+\!\!\frac{im^2}{r}\,\hat{a}_{r\theta}\!\!\right]\!\!, \label{eq_ur}\\
&&\Sigma_2\hat{a}_{rr}=2i\left\{\overline{A}_{rr}D+\frac{im\overline{A}_{r\theta}}{r}\right\}\hat{u}_r,\label{eq_arr}\\
&&\Sigma_2\hat{a}_{r\theta}=-\frac{r\overline{A}_{rr}}{m}DD^*\hat{u}_r-\left\{\frac{m}{r}\overline{A}_{\theta\theta}+i\overline{A}'_{r\theta}\right\}\hat{u}_r+i\hat{a}_{rr}r\Omega', \label{eq_art}\\
&&\Sigma_2\hat{a}_{\theta\theta}=-\frac{2r\overline{A}_{r\theta}}{m}DD^*\hat{u}_r-i\left\{\overline{A}'_{\theta\theta}+2\overline{A}_{\theta\theta}D\right\}\hat{u}_r+2i\hat{a}_{r\theta}r\Omega', \label{eq_att}
\end{eqnarray}  
where $D=\displaystyle\frac{d}{dr},\, D^*=\displaystyle\frac{d}{dr}+\displaystyle\frac{1}{r},\,\Sigma(r)=\omega-m\Omega(r)$ and $\Sigma_2(r)=\omega-m\Omega(r)+\displaystyle\frac{i}{\Deb}$ and and the non-dimensional, base-state, polymeric stresses are given by \eqref{eq:basestress} with $\tau$ replaced by $\Deb$. Here, we have used the relation $\hat{w}_z=(i/m)\mathscr{L}(r\hat{u}_r)$ between the axial vorticity and radial velocity perturbations for zero axial wavenumber with $\mathscr{L}=DD^*-(m^2-1)/r^2$\,(\cite{GS2014}). The non-dimensional parameters in (\ref{eq_ur})-(\ref{eq_att}) are the Deborah number, $\Deb=\Omega_0\tau$ which is the ratio of the relaxation to the flow time scale, the Reynolds number, $\Rey=\Omega_0a^2/\nu$ based on the total viscosity $\mu = \mu_s + \mu_p = \mu_s + G\tau$, and the elastic Mach number, $Ma_e=\Omega_0a/c_{\textrm{elas}}$ where $c_{\textrm{elas}}=\sqrt{G/\rho}$ is the shear wave speed in a quiescent elastic medium. (Note that $\Deb$ is used here instead of the Weissenberg number which is sometimes used in both the drag reduction and elastic instability literature; the distinction between the two parameters is not relevant in the present context). Similar to its counterpart in compressible flows, $Ma_e$ may be interpreted as the ratio of a characteristic flow velocity scale to the the speed of propagation of infinitesimal amplitude shear stress\,(or vorticity) fluctuations in a quiescent incompressible elastic medium. The elastic Mach number may be written in terms of $\Rey$ and $\Deb$ as $Ma_e^2 = \frac{1}{1-\beta}\Deb \Rey$, $\beta$ being the ratio of the solvent to the total viscosity, and for a fixed $\beta$, the evolution of the perturbations, as governed by (\ref{eq_ur})-(\ref{eq_att}), depends therefore on $\Rey$ and $\Deb$. $\beta = 0$ corresponds to a UCM fluid while $\beta = 1$ corresponds to a Newtonian fluid; in the latter case, $Ma_e \rightarrow \infty$, and (\ref{eq_ur}) with only the first term on the RHS is the Orr-Sommerfeld equation in cylindrical coordinates.\\

The above system of equations may be combined into a single fourth-order differential equation governing $\hat{u}_r$, the cylindrical analog of the viscoelastic Orr-Sommerfeld equation that has been examined earlier in the context of plane parallel flows\,(\cite{REN1986}). Apart from discrete modes, for any finite $\Deb$ and $\Rey$, the system (\ref{eq_ur})-(\ref{eq_att}), similar to the case of parallel flow shear flows, possesses a pair of continuous spectra. The latter are given by $r\,\epsilon\,r_c$ with $r_c$ defined by:
\begin{eqnarray}
\Sigma_2(r_c) &=& 0, \label{GL_csspectrum}\\
\Sigma(r_c)+\frac{\mathrm{i}}{\beta \Deb} &=& 0, \label{Viscous_csspectrum}
\end{eqnarray}
where $\Sigma_2(r)$ and $\Sigma(r)$ are as defined above\,(\cite{REN1986}; \cite{WIL99}; \cite{KUP05}). For $\Rey = \infty$, there also arises the well-known inviscid continuous spectrum\,(CS), given by $\Sigma(r_c) = 0$, and spanning the base-state range of angular velocities\,(\cite{GSLiftup, GS2014}); see section \ref{sec:CSrayleigh}). However, with $\Rey$ finite and for a bounded domain, the continuous spectra arise solely due to the additional viscoelastic terms \citep{shankar2019}. \\

The relation (\ref{GL_csspectrum}) defines the Gorodtsov-Leonov\,(GL) continuous spectrum\,(\cite{GL67}), named after the authors who originally discovered it for inertialess plane Couette flow of a UCM fluid. Although usually studied in the aforementioned specific context owing to its analytical tractability\,(\cite{GRAHAM98}), the GL spectrum continues to exist for finite $\Rey$, and for both parallel shear flows and the azimuthal shearing flows considered here. Its origin is the assumed local nature of the polymeric stress field in almost all constitutive equations in polymer rheology\,(\cite{BIRD87}). The evolution of the polymeric stress field in the absence of center-of-mass diffusion is, in fact, similar to that of the vorticity field in the inviscid limit, and both cases, in principle, allow for arbitrarily large gradients across streamlines\,(\cite{GS2014}; \cite{GSLiftup}). Based on this analogy, one expects CS-modes with singularities in the polymeric stress fields. The singular GL-eigenfunctions, in addition to being convected with the flow velocity at $r= r_c$, decay at a rate $\Deb^{-1}$ due to relaxation, asymptoting to neutral stability for $\Deb \rightarrow \infty$. Further, similar to the plane parallel case, the GL spectrum in cylindrical coordinates is characterized by the Frobenius exponents $0, 1, 3$ and $4$, the streamline curvature being negligible on the length scales defining the validity of the local Frobenius analysis. An additional continuous spectrum arises due to a finite solvent viscosity\,(\cite{WIL99}), being pushed off to infinity in the UCM limit\,($\beta \rightarrow 0$). Again, similar to the parallel flow case, this viscous singular continuous spectrum has Frobenius exponents - 0, 1, 2 and $3-2/\beta$. The final fractional Frobenius exponent indicates the existence of an algebraic branch point and an associated branch cut \citep{KUP05}. The CS-eigenfunctions in this case require therefore a principal-finite-part interpretation\,(\cite{ENG71}; \cite{GS2014}).\\

We now proceed to the regime of interest, $\Rey \rightarrow \infty$ with $\El = \Deb/\Rey = \nu \tau/a^2$ fixed, in which case (\ref{eq_ur})-(\ref{eq_att}), with the neglect of the $O(\Deb^{-1})$ terms denoting microstructural relaxation, yields the following equation governing $u_r$:
\begin{eqnarray}
&&\Sigma^3\left[\Sigma r\left(r^2D^2\hat{u}_r+3rD\hat{u}_r-(m^2-1)\hat{u}_r\right)+mrDZ\hat{u}_r\right]=2m^2E\Omega'\Bigl[\Sigma^2\Bigl\{r^2\Omega'D^2\hat{u}_r+r(r\Omega''+DZ)D\hat{u}_r\Bigr.\nonumber\\
&&\Bigl.-(m^2-1)\Omega'\hat{u}_r\Bigr\}+mr\Omega'\Sigma\Bigl\{2r\Omega'D\hat{u}_r+3(DZ-2\Omega')\hat{u}_r\Bigr\}+2m^2r^2\Omega'^3\hat{u}_r\Bigr],  \label{eq:rayleigh2}
\end{eqnarray}
which is the elastic Rayleigh equation with $\textrm{E} = \El(1-\beta)$; the terms proportional to $\textrm{E}$ in (\ref{eq:rayleigh2}) denote the contributions due to elasticity. Thus, the original fourth order ODE resulting from (\ref{eq_ur})-(\ref{eq_att}) reduces to a second-order ODE in the limit $\Deb \rightarrow \infty$, $\El,\beta$ fixed, implying that the neglect of relaxation is a singular limit. From what is known for the eigenfunctions of the Rayleigh and Orr-Sommerfeld equations\,(see section $5$ in \cite{GSLiftup}), one expects a non-trivial relationship between the spectrum of the elastic Rayleigh equation, and that for large but finite $\Deb$. The viscosity ratio $\beta$ no longer plays a fundamental role as for finite $\Deb$ where a non-zero $\beta$ leads to an additional continuous spectrum; in the above limit, one may interpret a change in $\beta$ in terms of a re-scaled $\textrm{E}$.\\

For typical inertio-elastic flows, $\El$ is the governing parameter only for small but finite $\Deb$ when relaxation effects are dominant and the fluid rheology is describable in terms of a retarded motion expansion\,(see, for instance, \cite{VIVEK2015}). For large $\Deb$, relaxation is unimportant with the dynamics in the polymeric fluid at finite $\Rey$ being governed by elastic shear waves damped due to the solvent viscosity alone. In this limit, $Ma_e$ becomes the governing parameter\,(\cite{JOSEPHBook}). It is thus a little surprising that in the limit of large $\Deb$ considered here, $\El$ rather than $Ma_e$ turns out to be relevant for (\ref{eq:rayleigh2}), suggesting the continued importance of relaxation\,($\El \propto \tau$). However, $\El$ in (\ref{eq:rayleigh2}) is more appropriately interpreted in terms of a viscoelastic Mach number where the sonic speed corresponds to shear waves propagating in a pre-stressed elastic medium. The base-state stress level is $\bar{A}_{\theta\theta}\sim O(\Deb^2)$ as given earlier, and the shear wave speed relevant to the perturbation dynamics is given by $\sqrt{G\bar{A}_{\theta\theta}/\rho} \sim \Deb\sqrt{G/\rho}$. The relevant Mach number is $O(\Deb).O(\sqrt{G/\rho}/\Omega_0a) \sim (\Deb/\Rey)^{\frac{1}{2}} \sim \El^{\frac{1}{2}}$. Thus, the rather paradoxical dependence on $\El$, and thence on the relaxation time, occurs via the shear wave speed being dependent on the base-state hoop stress. Note that the assumption of the base-state stresses being dependent on the relaxation time, but the perturbations not being sensitive to it, is analogous to the usual assumption for Newtonian fluids where the base-state is assumed to be determined by slow viscous diffusion at large $Re$, but the latter is assumed to play a negligible role in the dynamics of perturbations.\\

To analyze the properties of the elastic Rayleigh equation, with an emphasis on the underlying CS-spectra, we rewrite (\ref{eq:rayleigh2}) in terms of the radial displacement, $\xi\equiv i\hat{u}_r/\Sigma$, in the following much more compact form, first identified by \cite{RALLHIN95} in the context of parallel shearing flows:
\begin{eqnarray}
&&D\left[r^3PD\xi\right]=r(m^2-1)P\xi, \label{eq:xir}
\end{eqnarray}
where $P=\Sigma^2-2m^2\textrm{E}\Omega'^2$. From (\ref{eq:xir}), one may easily construct the following modified version of Howard's semi-circle theorem\,(\cite{HOWARD61}) for swirling flows:
\begin{eqnarray}
\left(\omega_r-\frac{m(\Omega_{\textrm{max}}+\Omega_{\textrm{min}})}{2}\right)^2+\omega_i^2 \leq m^2\left(\frac{\Omega_{\textrm{max}}-\Omega_{\textrm{min}}}{2}\right)^2-2m^2\textrm{E}\,{\Omega'_{\textrm{min}}}^2. \label{How:circle}
\end{eqnarray}
The role of elasticity is to shrink the inviscid semi-circle of instability, implying a relative stabilization. The analog of (\ref{eq:xir}) has been used by \cite{RALLHIN95} in their study of elastic instabilities in jets. Stability to exponentially growing perturbations results when the semi-circle radius decreases to zero, and this happens at a finite $E$ provided $\Omega'_{min}$ is non-zero. A novel mechanism of instability, associated with the elastic Rayleigh equation, and that may arise even for non-inflectional profiles, is that resulting from the resonant interaction of elastic shear waves. The shear waves propagate more rapidly\,(relative to the flow) with increasing $E$, and the onset of stability above coincides with the inability of the base-state shear, beyond a threshold $E$, to bring a pair of such waves into resonance by causing them to propagate at the same speed\,(\cite{REN2008}).\\

\section{Elastic Rayleigh Equation: the continuous spectra}  \label{sec:CSrayleigh}
In contrast to the inviscid Rayleigh operator which supports a single continuous spectrum ranging over the base-state interval of angular velocities\,(\cite{CASE60};\cite{GS2014}), the elastic Rayleigh operator supports three distinct continuous spectra. The first of these is the original inviscid continuous spectrum modified by elasticity. There exist in addition a pair of continuous spectra associated with fore- and aft-travelling elastic shear waves in the azimuthal direction. On writing $\Sigma^2-2m^2\textrm{E}\Omega'^2$ in (\ref{eq:xir}) as $(\Sigma + m\Omega'\sqrt{2\textrm{E}})(\Sigma - m\Omega'\sqrt{2\textrm{E}})$, it is seen that these shear waves propagate at angular frequencies of $\pm (m\Omega')\sqrt{2\textrm{E}}$ relative to the base-state angular frequency, $m\Omega(r)$, at $r$. Unlike the finite-$\Deb$ continuous spectra discussed in section \ref{sec:elas_form}, these elastic-Rayleigh travelling-wave spectra arise due to a balance of the inertial and elastic terms, and must therefore disappear for any finite $\Deb$. Thus, in the presence of any amount of relaxation, the original travelling-wave CS-modes are no longer true eigenfunctions, and must instead be expressible in terms of a superposition of finite $\Deb$ discrete modes. Aside from the obvious reduction in the order of the equation, this again highlights the singular relation between the spectrum of the elastic Rayleigh equation discussed below and the finite $\Deb$ spectrum, spectrum associated with the viscoelastic Orr-Sommerfeld equation.\\

The elastic Rayleigh equation for arbitrary $E$ belongs to the confluent Heun class with a pair of regular singular points at the travelling wave locations given by $\Sigma \pm m\Omega'\sqrt{2\textrm{E}} = 0$, and an irregular one at infinity (\cite{SLAVNOV}). The associated insolubility implies that we only analyze the continuous spectra for small but finite $\textrm{E}$. The inertial terms become arbitrarily small sufficiently close to the critical radius, defined by $\Sigma(r_c) = 0$, and for any $\textrm{E}$ however small, the effects of elasticity become comparable to those of inertia in a boundary layer around $r_c$. Thus, the elastic continuous spectra are analyzed below via a matched asymptotic expansions approach which involves matching the leading-order inviscid solutions, in regions away from $r_c$, to those in an $O(\textrm{E}^{\frac{1}{2}})$ interior elastic boundary layer around the critical radius. The analysis of the continuous spectrum lends additional insight into the structure of the unstable (discrete) mode described in detail in the next section. The discrete mode mirrors the structure of the elastic CS-modes in the limit of a vanishingly small growth rate.\\

To begin with, we summarize briefly the inviscid 2D CS-spectrum of the Rankine vortex for $\textrm{E} = 0$ as found by Roy \& Subramanian\,(2014), and referred to as the $\Lambda_1$-family therein. In light of the scalings used in section \ref{sec:elas_form}, the non-dimensional angular velocity profile for the Rankine vortex is given by $\Omega(r)=\mathcal{H}(1-r)+\frac{1}{r^2}\mathcal{H}(r-1)$, ${\mathcal H}(z)$ being the Heaviside function. For an azimuthal wavenumber $m$, the 2D CS-modes span the angular frequency range $(0,m\Omega_0)$, and have a twin-vortex-sheet structure. The vortex sheets are cylindrical, being threaded by axial lines, with one sheet located at the edge of the core and the other at the critical radius in the irrotational exterior. A given CS-mode rotates with the base-angular velocity corresponding to the critical radius. Therefore, the radial velocity and axial vorticity eigenfunctions are of the form $[u_r(r;r_c),w_z(r;r_c)] = [\hat{u}_r(r;r_c);\hat{w}_z(r;r_c)]e^{\mathrm{i}(m\theta-\omega t)}$ with $\omega = \Omega(r_c)$, $r_c = (m/\omega)^{\frac{1}{2}}$, and
\begin{align}
\hat{u}_r(r;r_c)=&\, d r^{m-1}\hspace*{0.2in} r < 1, \label{urE0:1} \\
=&\, c_1 r^{m-1} + c_2 \frac{1}{r^{m+1}} \hspace*{0.2in}1< r < r_c, \label{urE0:2} \\
=&\, \frac{1}{r_c}\!\frac{1}{r^{m+1}}  \hspace*{0.2in}r > r_c, \label{urE0:3} \\
\hat{w}_z(r;r_c) =&\, \bigl[ \frac{2\mathrm{i}d}{(\omega- m)}\,\delta(r-1) - A(r_c) \delta(r-r_c) \bigr], \label{wzE0}
\end{align}
where 
\begin{align}
d =&\, \frac{1}{r_c} + \frac{\mathrm{i}A(r_c)}{2}\left[ r_c^{m+1}
- \frac{1}{r_c^{m-1}} \right], \\
c_1 =&\, -\frac{\mathrm{i}A(r_c)}{2}\!\frac{1}{r_c^{m-1}}, \label{expr:c1} \\
c_2 =&\, \frac{1}{r_c}+\frac{\mathrm{i}A(r_c)}{2} r_c^{m+1},  \label{expr:c2} \\
A(r_c) =&\, \frac{(2\mathrm{i}/r_c)[(m-1)- \omega]}{(1/r_c^{m-1})+ [(m-1) - \omega]r_c^{m+1}}.
\end{align}
The radial velocity eigenfunction itself is continuous at $r = r_c$, but the discontinuities in slope at both $r =1$ and $r=r_c$ correspond to delta-function\,(vortex-sheet) contributions in the perturbation axial vorticity field. The first delta function is an artifact of the kink in the base-state\,(Rankine) profile, and it is the amplitude of the second vortex sheet, $A(r_c)$, that is of interest and that characterizes the 2D CS-spectrum. Note that $A(r_c)$ equals zero when $\omega = (m-1)$, which corresponds to the regular Kelvin mode with a critical radius given by $r_{ck} = (\frac{m}{m-1})^{\frac{1}{2}}$. There is a direct analogy between the 2D CS-spectra of plane Couette flow as found by Case\,(1960) and that of the Rankine vortex summarized above. In both cases, the vorticity eigenfunctions are vortex-sheets convected with the base-state flow - a single plane vortex sheet for Couette flow and a pair of cylindrical sheets for the Rankine vortex. The crucial difference is the additional presence of the Kelvin mode above in the latter case. While the inviscid spectrum for plane Couette flow\,(and other non-inflectional profiles) is purely continuous with the amplitude of the vortex-sheet remaining non-zero over the entire range of wave speeds, there exists a unique angular frequency, the Kelvin mode for a given $m$, for which the vortex-sheet amplitude is zero for the case of the Rankine vortex. The Kelvin mode, together with the 2D CS-modes provide a complete basis for an arbitrary axial vorticity field in two dimensions\,(\cite{GS2014}).\\

For small but finite $\textrm{E}$, we examine the solutions of (\ref{eq:xir}) separately in the outer region where $r - r_c \sim O(1)$, and in the inner region (the elastic boundary layer) where $r-r_c \sim O(\textrm{E}^{\frac{1}{2}})$, before matching them to determine the unknown coefficients in the respective domains. We assume the critical radius, $r_c$, to be such that the elastic boundary layer lies an $O(1)$ distance away from the edge of the core\,($r=1$). The solutions in the outer regions, at leading order, are thus identical to those given in (\ref{urE0:1})-(\ref{urE0:3}) except that (\ref{urE0:2}) and (\ref{urE0:3}) are not valid right until $r_c$. Likewise, apart from the core contribution\,(the term proportional to $\delta(r-1)$ in (\ref{wzE0})), the vorticity field is localized in the elastic boundary layer around $r_c$, although no longer a delta function at this location. Thus, (\ref{urE0:2}) is now valid in the range $r > 1, (r_c - r) \gg O(\textrm{E}^{\frac{1}{2}})$ while (\ref{urE0:3}) is now valid in the range $(r -r_c) \gg O(\textrm{E}^{\frac{1}{2}})$. Note that since $P = \Sigma^2-2m^2\textrm{E}\Omega'^2$ in (\ref{eq:xir}), the direct effect of elasticity enters the outer regions only at $O(\textrm{E})$. At $O(\textrm{E}^{\frac{1}{2}})$, the outer solutions still satisfy the inviscid Rayleigh equation, and the effects of elasticity only enter via matching to the far-field forms of the elastic boundary layer solution. For purposes of matching below, it is convenient to normalize the perturbation in the core region, so that $\hat{u}_r(r;r_c) = r^{m-1}$ for $r < 1$ instead of (\ref{urE0:1})\,(this being valid to all orders in $\textrm{E}$). As a result, instead of (\ref{urE0:2}) and (\ref{urE0:3}), we have:
\begin{align}
\hat{u}_{r1}(r;r_c)=&\, c^{(0)}_1 r^{m-1} \!\!+\! \frac{ (1\!-\!c^{(0)}_1)}{r^{m+1}} \!+\! \textrm{\textrm{E}}^{\frac{1}{2}}c^{(1)}_1 ( r^{m-1}\!\!-\!\! \frac{1}{r^{m+1}})\! +\! O(\textrm{E}) \hspace*{0.15in} r>1, r_c - r \gg O(\textrm{E}^{\frac{1}{2}}), \label{urE0:2m} \\
\hat{u}_{r2}(r;r_c)=&\, \frac{c_2^{(0)}+ \textrm{E}^{\frac{1}{2}} c^{(1)}_2}{r^{m+1}}+  O(\textrm{E})\hspace*{0.2in}r -r_c \gg O(\textrm{E}^{\frac{1}{2}}), \label{urE0:3m} 
\end{align}
for the irrotational radial velocity perturbation outside the core and outside the elastic boundary layer. The choice of constant in (\ref{urE0:2m}) reflects consistency with the normalized core perturbation at $r=1$. Since one is looking for elastic generalizations of the $\Lambda_1$ CS-modes, we also impose continuity of the radial velocity for $r \rightarrow r_c$, as seen from the outer region. This implies $c^{(0)}_1 r_c^{m-1} \!\!+\! \frac{ (1\!-\!c^{(0)}_1)}{r_c^{m+1}} = \frac{c_2^{(0)}}{r_c^{m+1}} =\hat{u}_{r}(r_c;r_c)$, which also leads to the relation between the two constants involved: $c_2^{(0)} = c_1^{(0)}\!(r_c^{2m}\!-\!1) +1$.\\

It may be shown that a balance between the inertial and elastic terms occurs at leading order when $r - r_c \sim O(\textrm{E}^{\frac{1}{2}})$, and one therefore defines the boundary layer variable $\eta = \frac{r-r_c}{(2\textrm{E})^{\frac{1}{2}}}$. On using the expansion $\Sigma(r) \approx -m\Omega'_c(r-r_c) - m\Omega''_c\frac{(r-r_c)^2}{2} \approx -m\Omega_c'(2\textrm{E})^{\frac{1}{2}}\eta[1 + \frac{\Omega_c''}{\Omega_c'}(\frac{\textrm{E}}{2})^{\frac{1}{2}}\eta]$, the original relationship between the radial velocity and displacement, $\xi(r) = (\mathrm{i}\hat{u}_r)/\Sigma(r)$, takes the form $\tilde{u}_r(\eta) = - \mathrm{i}m\Omega_c'\eta [ 1+ \frac{\Omega_c''}{\Omega_c'}(\frac{\textrm{E}}{2})^{\frac{1}{2}}\eta]\tilde{\xi}(\eta)$ in the boundary layer, to $O(\textrm{E})$, with $\xi(r) = (2\textrm{E})^{\frac{1}{2}}\tilde{\xi}(\eta)$. Further, $P = (\omega - m\Omega(r))^2 - 2m^2 \textrm{E}[\Omega'(r)]^2 \approx 2m^2\textrm{E}{\Omega_c'}^2[ \eta^2-1 + \frac{\Omega_c''}{\Omega_c'}(2\textrm{E})^{\frac{1}{2}}\eta^3]$. To $O(\textrm{E}^{\frac{1}{2}})$, the governing equation, (\ref{eq:xir}), then takes the form:
\begin{equation}
\frac{d}{d\eta}\left[ \{r_c + (2\textrm{E})^{\frac{1}{2}}\eta\}^3\left[(\eta^2-1)- \frac{3\sqrt{2}}{r_c}\textrm{E}^{\frac{1}{2}}\right] \frac{d\tilde{\xi}}{d\eta}\right] = O(\textrm{E}), \label{EBL:goveq}
\end{equation}
in terms of the re-scaled radial displacement, $\tilde{\xi}({\eta})$, in the boundary layer. Here, we have used that $\Omega'_c = -\frac{2}{r_c^3}, \Omega_c'' = \frac{6}{r_c^4}$ for a Rankine vortex. The form of (\ref{EBL:goveq}) evidently suggests a series expansion of the form $\tilde{\xi}({\eta}) = \tilde{\xi}^{(0)}({\eta}) + \textrm{E}^{\frac{1}{2}}\tilde{\xi}^{(1)}({\eta}) + O(\textrm{E})$, and one obtains the following governing equations at $O(1)$ and $O(\textrm{E}^{\frac{1}{2}})$, respectively:
\begin{align}
\frac{d}{d\eta}\left[(\eta^2-1) \frac{d\tilde{\xi}^{(0)}}{d\eta} \right] =&\,0, \label{eqn:tildexi0} \\
\frac{d}{d\eta}\left[(\eta^2-1) \frac{d\tilde{\xi}^{(1)}}{d\eta} \right] =&\, \frac{3\sqrt{2}}{r_c}\! \frac{d}{d\eta}\!\!\left[ \eta\frac{d\tilde{\xi}^{(0)}}{d\eta} \right]. \label{eqn:tildexi1}
\end{align}

The solution of (\ref{eqn:tildexi0}) is given by:
\begin{equation}
\tilde{\xi}^{(0)}(\eta) = A^{(0)}_1 + A^{(0)}_2 \ln \lvert \frac{\eta-1}{\eta +1} \lvert, \label{disp:EBL}
\end{equation}
implying a radial velocity in the boundary layer of the form:
\begin{equation}
\tilde{u}_r^{(0)}(\eta) = -\mathrm{i}m\Omega_c' \eta\! \left[ A^{(0)}_1 + A^{(0)}_2 \ln \lvert \frac{\eta -1}{\eta +1} \lvert \right],  \label{vel:EBL}
\end{equation}
at leading order. The values $\eta = \pm 1$ denote the locations of the travelling wave singularities where the radial velocity is logarithmically divergent. This divergence is expected since the Frobenius exponents associated with each of the regular singularities are both 0. The singularities divide the boundary layer into three regions, with the solution forms in the individual regions may be written explicitly as:
\begin{align}
\tilde{u}_{r-}^{(0)}(\eta) =&\,  \eta \left[ \hat{A}^{(0)}_{1-} + \hat{A}^{(0)}_2 \ln \frac{\eta-1}{\eta +1} \right] \hspace*{0.2in} \eta < -1, \label{eq:BLlhs} \\
\tilde{u}_r^{(0)}(\eta) =&\, \eta \left[ \hat{A}^{(0)}_{1} + \hat{A}^{(0)}_2 \ln \frac{1-\eta}{1+\eta} \right] \hspace*{0.2in} -1 < \eta < 1, \label{eq:BLcent}\\
\tilde{u}_{r+}^{(0)}(\eta) =&\,  \eta \left[ \hat{A}^{(0)}_{1+} + \hat{A}^{(0)}_2 \ln \frac{\eta-1}{\eta+1} \right] \hspace*{0.2in} \eta > 1, \label{eq:BLrhs}
\end{align}
where $\hat{A}^{(0)}_{i}= -\mathrm{i}m\Omega_c'A^{(0)}_{i}$. Note that we have chosen the same constant for the singular logarithmic solution in all three parts of the boundary layer with the logarithm being real valued in each region\,(this being possible by suitable choice of the regular constants). Such a choice is consistent with the purely inviscid case\,($\textrm{E} =0$) where, for a general non-linear shear flow, the constant multiplying the logarithmically singular Tollmein solution is the same across the critical level, and it is the jump in the constant multiplying the regular solution that generates the inviscid CS-spectrum of the Rayleigh equation\,(\cite{BALMMORRIS95}). It will be seen below that the constants in the two peripheral regions\,($\hat{A}^{(0)}_{1\pm}$,$\hat{A}^{(0)}_2$) are constrained by matching, and an appropriate choice can accommodate the differing slopes of the outer solutions on either side. The regular constant in the central part of the boundary layer, at leading\,($\hat{A}^{(0)}_1$) and higher orders, can be chosen independently, however, and this additional degree of freedom is crucial to the existence of additional continuous spectra for any finite $\textrm{E}$. \\

Using (\ref{disp:EBL}) in (\ref{eqn:tildexi1}), and solving, gives:
\begin{equation}
\tilde{\xi}^{(1)}(\eta) = A^{(1)}_1 + \left[ A^{(1)}_2 + \frac{3A^{(0)}_2}{\sqrt{2}r_c}\eta \right]\ln \lvert \frac{\eta-1}{\eta +1} \lvert  - \frac{3\sqrt{2}A^{(0)}_2}{r_c}\frac{1}{\eta^2\!-\!1}, \label{disp:EBL1}
\end{equation}
where the terms proportional to $A^{(1)}_1$ and $A^{(1)}_2$ denote the homogeneous solution. Although not made explicit in (\ref{disp:EBL1}), a distinction will again being made between solution forms in the the three parts of the elastic boundary layer similar to that done at leading order. The radial velocity in the elastic boundary layer, at $O(\textrm{E}^{\frac{1}{2}})$, is given by:
\begin{align}
\tilde{u}_r^{(1)}(\eta) =&\, \eta \left[\hat{A}^{(1)}_1 + \left( \hat{A}^{(1)}_2 + \frac{3\hat{A}^{(0)}_2}{\sqrt{2}r_c}\eta\right)\ln \lvert \frac{\eta-1}{\eta +1} \lvert  - \frac{3\sqrt{2}\hat{A}^{(0)}_2}{r_c}\frac{1}{\eta^2\!-\!1} \right],
\label{vel:EBL1}
\end{align}
where $\hat{A}_i^{(1)} = -\mathrm{i}m\Omega'_c A_i^{(1)}$. \\

The matching requirement between the outer regions and the elastic boundary layer may be stated as: $\lim_{r \rightarrow r_c} \hat{u}_{r1}(r;r_c) = \lim_{\eta \rightarrow - \infty} \tilde{u}_{r-}(\eta)$ and $\lim_{r \rightarrow r_c} \hat{u}_{r2}(r;r_c) = \lim_{\eta \rightarrow \infty} \tilde{u}_{r+}(\eta)$ to $O(\textrm{E}^{\frac{1}{2}})$. As pointed out earlier, the only difference between the 2D inviscid spectra of plane Couette flow and the Rankine vortex is the existence of a lone discrete mode - the Kelvin mode - in the latter case. In order to discriminate between the generalizations of the inviscid CS-modes and the Kelvin mode for non-zero $\textrm{E}$, one needs to carry out the matching to $O(\textrm{E}^{\frac{1}{2}})$. At $O(1)$, the matching process ensures the continuity of the radial velocity across the elastic boundary layer, and it is only at $O(\textrm{E}^{\frac{1}{2}})$ that the two ends of the elastic boundary layer\,($\eta \rightarrow \pm \infty$) sense the difference in the slopes of the outer eigenfunctions, $\hat{u}_{r1}(r;r_c)$ and $\hat{u}_{r2}(r;r_c)$, for $r$ approaching $r_c$. It is precisely this jump in slope that differentiates the CS-modes from the Kelvin mode. \\

The far-field forms of the solutions in the peripheral regions of the elastic boundary layer are given by:
\begin{align}
\lim_{\eta \rightarrow \pm \infty} \tilde{u}_{r-}(\eta) =&\,\hat{A}_{1\pm}^{(0)}\eta-2 \hat{A}_2^{(0)} + \textrm{E}^{\frac{1}{2}}\left[ \!-2\hat{A}_2^{(1)} \!+\!\left(\hat{A}_{1\pm}^{(1)} - \frac{3\sqrt{2}}{r_c}\hat{A}_2^{(0)}\right)\!\!\eta\!\right],  \label{eq:BLlimit}
\end{align}
where the neglected terms only affect the matching at $o(\textrm{E}^{\frac{1}{2}})$. The above expressions are to be matched to the limiting forms of the outer solutions obtained from (\ref{urE0:2m}) and (\ref{urE0:3m}) in the limit $r \rightarrow r_c + (2\textrm{E})^{\frac{1}{2}}\eta$, which are given by:
\begin{align}
\hat{u}_{r1}(r;r_c) =&\, \hat{u}_r(r_c;r_c)+\! \textrm{E}^{\frac{1}{2}}\!\!\!\left[\!\left(\!\!c_1^{(0)}\!(m\!\!-\!\!1)r_c^{m-2} \!-\! \frac{(m\!+\!1)(1\!-\!c_1^{(0)})}{r_c^{m+2}}\!\right)\!\!\sqrt{2}\eta\!+\! c_1^{(1)}( r_c^{m-1}\!\! -\!\! \frac{1}{r_c^{m+1}}) \!\!\right] \!\!+\! O(\textrm{E}), \label{out:lim1} \\
\hat{u}_{r2}(r;r_c) =&\, \hat{u}_r(r_c;r_c) + \textrm{E}^{\frac{1}{2}}\!\!\left[ - \frac{\sqrt{2}(m+1)c_2^{(0)}}{r_c^{m+2}}\,\eta + \frac{c_2^{(1)}}{r_c^{m+1}} \right] + O(\textrm{E}).
\label{out:lim2}
\end{align}
Matching (\ref{eq:BLlimit}) and (\ref{out:lim1})-(\ref{out:lim2}), at leading order, gives $\hat{A}_{1\pm}^{(0)} = 0$ and $\hat{A}_2^{(0)} = -\frac{\hat{u}_r(r_c;r_c)}{2}$. A consistent match of the constant term at $O(\textrm{E}^{\frac{1}{2}})$ gives $\hat{A}_2^{(1)} = c_1^{(1)} = c_2^{(1)} = 0$. Matching the term proportional to $\eta$ at $O(\textrm{E}^{\frac{1}{2}})$ gives:
\begin{align}
\hat{A}_{1-}^{(1)} =&\, \frac{3\sqrt{2}}{r_c}\hat{A}_2^{(0)} + \sqrt{2}\left(\!\!c_1^{(0)}\!(m\!\!-\!\!1)r_c^{m-2} \!-\! \frac{(m\!+\!1)(1\!-\!c_1^{(0)})}{r_c^{m+2}}\!\right), \\
=&\,-\!\!\frac{3}{\sqrt{2}r_c}\hat{u}_r(r_c;r_c) \!+ \!\sqrt{2}\!\left[\!\frac{(m\!-\!1)r_c^{2m} + (m\!+\!1)}{r_c^{m+2}}\frac{r_c^{m+1}\hat{u}_r(r_c;r_c)-1}{r_c^{2m}-1} - \frac{m\!+\!1}{r_c^{m+2}} \!\right], \label{BLconst:1} \\
\hat{A}_{1+}^{(1)} =&\,\frac{3\sqrt{2}}{r_c}\hat{A}_2^{(0)} - \frac{(m+1)\sqrt{2}}{r_c^{m+2}}\,c_2^{(0)}, \\
=&\,-\frac{(2m+5)}{\sqrt{2}r_c} \hat{u}_r(r_c;r_c), \label{BLconst:2}
\end{align} 
where the two boundary-layer constants are rewritten in terms of $\hat{u}_r(r_c;r_c)$, the amplitude of the radial velocity eigenfunction at the critical radius.\\

Having determined the constants above, the forms of the eigenfunction in the outer regions, to $O(\textrm{E}^{\frac{1}{2}})$, are given by:
\begin{align}
\hat{u}_{r1}(r;r_c)=&\, c^{(0)}_1 r^{m-1} \!\!+\! \frac{ (1\!-\!c^{(0)}_1)}{r^{m+1}}; \hspace{0.2in} \hat{u}_{r2}(r;r_c)=\, \frac{c_2^{(0)}}{r^{m+1}}, \label{outer:final}
\end{align}
where $c_1^{(0)}$ and $c_2^{(0)}$ have been defined in terms $\hat{u}_r(r_c;r_c)$ above. The form of the eigenfunction within the elastic boundary layer may be written down in the following piecewise form: 
\begin{align}
\tilde{u}_{r-}(\eta) =&\hat{A}_2^{(0)}\!\mbox{Pf}.\eta\!\ln \!\frac{\eta-1}{\eta +1} + \textrm{E}^{\frac{1}{2}}\eta\!\!\left[\!\hat{A}_{1-}^{(1)} \!+ \!\frac{3}{\sqrt{2}r_c}\hat{A}_2^{(0)}\!\mbox{Pf}.\!\left(\eta\ln \frac{\eta-1}{\eta +1} -  \frac{2}{\eta^2\!-\!1} \!\right)\!\!\right]\,\hspace*{0.2in} \eta < -1, \label{BL:eigen1} \\
\tilde{u}_r(\eta) =&\eta\!\!\left[\!\hat{A}_1 \!+\! \hat{A}_2^{(0)}\!\mbox{Pf}.\ln \frac{1-\eta}{1+\eta}\right] + \textrm{E}^{\frac{1}{2}}\eta\!\!\left[\frac{3}{\sqrt{2}r_c}\hat{A}_2^{(0)}\!\mbox{Pf}.\!\left(\eta\ln \frac{1-\eta}{1+\eta} -  \frac{2}{\eta^2\!-\!1} \!\right) \!\!\right]\,\hspace*{0.2in} -1 < \eta < -1, \label{BL:eigen2}\\
\tilde{u}_{r+}(\eta) =&\hat{A}_2^{(0)}\!\mbox{Pf}.\eta\!\ln \frac{\eta-1}{\eta+1} + \textrm{E}^{\frac{1}{2}}\eta\!\!\left[\!\hat{A}_{1+}^{(1)} \!+\! \frac{3}{\sqrt{2}r_c}\hat{A}_2^{(0)}\!\mbox{Pf}.\!\left(\eta\ln \frac{\eta-1}{\eta +1} -  \frac{2}{\eta^2\!-\!1}\!\right)\!\!\right]\,\hspace*{0.2in} \eta > 1,
\label{BL:eigen3}
\end{align}
again to $O(\textrm{E}^{\frac{1}{2}})$. Note that the terms linear in $\eta$, at leading order and at $O(\textrm{E}^{\frac{1}{2}})$, have been combined into a single term, $\hat{A}_1\eta$, in (\ref{BL:eigen2}). While $\hat{A}_2^{(0)} = -\frac{\hat{u}_r(r_c;r_c)}{2}$, and $\hat{A}_{1\pm}^{(1)}$ in the above expressions are given by (\ref{BLconst:1}) and (\ref{BLconst:2}), respectively, $\hat{A}_{1}$ in (\ref{BL:eigen2}) remains arbitrary. The prefix $\mbox{Pf}.$ in (\ref{BL:eigen1})-(\ref{BL:eigen3}) denotes a principal-finite-part interpretation which, as will be seen below, is required in interpreting the axial vorticity field within the elastic boundary layer. The expressions (\ref{BL:eigen1})-(\ref{BL:eigen3}), taken together with the expressions for the constants involved, show that the CS-modes for small but finite $\textrm{E}$ involve two parameters. These may be taken as $[\hat{u}_r(r_c;r_c) - r_c^{-(m+1)}]$ and $\hat{A}_1$, where $1/r_c^{m+1}$ is the normalized radial velocity associated with the Kelvin mode at $r=r_c$. The first parameter, of course, already exists for $\textrm{E} = 0$, and is proportional to the amplitude of the second vortex sheet, $A(r_c)$, used earlier for the description of the CS-modes for $\textrm{E}= 0$\,(see (\ref{urE0:1})-(\ref{urE0:3})). The second parameter arises only for non-zero $\textrm{E}$ and affects the detailed structure of the (elastic)\,boundary layer vorticity field.\\

The finite-$\textrm{E}$ generalization of the regular Kelvin mode may be obtained by taking $r_c = r_{ck}$, $c_1^{(0)} = 0$, $c_2^{(0)} = 1$. This implies $\hat{A}_2^{(0)} = -1/(2r_{ck}^{m+1})$, $\hat{A}_1^{(1-)} = \hat{A}_1^{(1+)} = -(2m+5)/(\sqrt{2}r_{ck}^{m+2})$. In the outer region, one now has $\hat{u}_{r1}(r;r_c)= \hat{u}_{r2}(r;r_c)= 1/r^{m+1}$, and within the boundary layer, (\ref{urE0:1})-(\ref{urE0:3}) take the form:
\begin{align}
\tilde{u}_{rk-}(\eta) \!=&\!-\!\frac{1}{2r_{ck}^{m+1}}\mbox{Pf}.\eta\!\ln \frac{\eta-1}{\eta +1}\! -\! \frac{\textrm{E}^{\frac{1}{2}}\eta}{\sqrt{2}r_{ck}^{m+2}}\!\!\!\left[\!(2m\!+\!5)\!+ \!\!\frac{3}{2}\mbox{Pf}.\!\!\left(\eta\ln\! \frac{\eta\!-\!1}{\eta \!+\!1} \!- \!\! \frac{2}{\eta^2\!-\!1} \!\right)\!\!\right]\,\hspace*{0.1in} \eta < -1, \label{Kel:BL1} \\
\tilde{u}_{rk}(\eta)\!=&\eta\left[\hat{A}_1\!-\!\frac{1}{2r_{ck}^{m+1}}\mbox{Pf}.\!\ln \frac{1-\eta}{1+\eta} \right] \!-\!\! \frac{3E^{\frac{1}{2}}\eta}{2\sqrt{2}r_{ck}^{m+2}}\mbox{Pf}.\left(\eta\!\ln \frac{1-\eta}{1+\eta} \!- \! \frac{2}{\eta^2\!-\!1} \!\right) \!\,\hspace*{0.1in} -1 < \eta < -1, \label{Kel:BL2}\\
\tilde{u}_{rk+}(\eta)\!=&\!-\!\frac{1}{2r_{ck}^{m+1}}\mbox{Pf}.\eta\!\ln \frac{\eta-1}{\eta+1}\! -\! \frac{\textrm{E}^{\frac{1}{2}}\eta}{\sqrt{2}r_{ck}^{m+2}}\!\!\!\left[\!(2m\!+\!5) \!+\!\! \frac{3}{2}\mbox{Pf}.\eta\!\left(\eta\ln\! \frac{\eta\!-\!1}{\eta +1} \!-\!\!  \frac{2}{\eta^2\!-\!1}\!\right)\!\!\right]\,\hspace*{0.1in} \eta > 1, \label{Kel:BL3}
\end{align}
The above expressions suggest that the finite-$\textrm{E}$ generalization of the regular Kelvin mode is a one-parameter family of singular eigenfunctions, $\hat{A}_{1}$ being the parameter, with singularities at $r = r_{ck} \pm \sqrt{2\textrm{E}}\,(\eta = \pm 1)$ for small $\textrm{E}$.\\

The above interpretation for finite $\textrm{E}$ also becomes clear on consideration of the axial vorticity field associated with (\ref{BL:eigen1})-(\ref{BL:eigen3})\,(recall that the velocity field in the outer regions is irrotational to $O(\textrm{E}^{\frac{1}{2}})$). For $\textrm{E} = 0$, the axial vorticity field is, of course, a delta function at $r =r_c$. For small but finite $\textrm{E}$, the vorticity field is still localized in the elastic boundary layer which may be regarded as a vortex sheet on the scale of the outer region. One may therefore discriminate between the 2D CS-modes and Kelvin mode based on the strength of this equivalent vortex sheet, defined as the total vorticity contained within the $O(\textrm{E}^{\frac{1}{2}})$ boundary layer over a single wavelength in the azimuthal direction. For $\textrm{E} \ll 1$, this integrated vorticity contribution is proportional to $\textstyle\int_{r_c - r \gg O(\textrm{E}^{\frac{1}{2}})}^{r - r_c \gg O(\textrm{E}^{\frac{1}{2}})}\hat{w}_z(r;r_c)r dr = (2\textrm{E})^{\frac{1}{2}}\textstyle\int_{-\infty}^{\infty}\tilde{w}_z(\eta)[r_c \!+ \!(2\textrm{E})^{\frac{1}{2}}\eta] d\eta$.\\

Using the relation $\hat{w}_z=(\mathrm{i}/m)\mathscr{L}(r\hat{u}_r)$, together with the small $\textrm{E}$ expansion for the radial velocity in the boundary layer, one obtains $\tilde{w}_z(\eta) = (2\textrm{E})^{-1}[\tilde{w}_z^{(0)}(\eta) + \textrm{E}^{\frac{1}{2}}\tilde{w}_z^{(1)}(\eta)]$ where:
\begin{align}
\tilde{w}_z^{(0)}(\eta) =&\, \frac{\mathrm{i}r_c}{m}\frac{d^2 \tilde{u}^{(0)}_r}{d\eta^2}, \\
=&\,\frac{\mathrm{i}r_c}{m}\left[\!\! -\mbox{Pf.} \frac{4\hat{A}_2^{(0)}}{(\eta^2-1)^2}\! +\! \hat{A}_1[\delta(\eta+1)- \delta(\eta-1) +\delta'(\eta+1) + \delta'(\eta-1)]\right], \label{wz0_EBL} \\
\tilde{w}_z^{(1)}(\eta) =&\,\frac{\mathrm{i}}{m}\!\! \left[ \sqrt{2}\!\!\left(\!\!\eta \frac{d^2 \tilde{u}^{(0)}_r}{d\eta^2} + 3\frac{d \tilde{u}^{(0)}_r}{d\eta} \!\!\right) + r_c \frac{d^2 \tilde{u}^{(1)}_r}{d\eta^2}\right], \\
=&\, {\mathcal P}\frac{\mathrm{i\hat{A}_2^{(0)}}}{m}\!\!\left[\sqrt{2}\!\!\left( \frac{2\eta(3\eta^2-5)}{(\eta^2-1)^2} +  3\ln \lvert \frac{\eta-1}{\eta+1} \rvert \right)+ 2r_c\!\!\left( \ln \lvert \frac{\eta-1}{\eta+1} \rvert + \frac{2\eta(\eta^4 \!-\!4\eta^2\!-\!1)}{(\eta^2-1)^3}\right)\right] \nonumber\\
&+\frac{\mathrm{i}r_c}{m}[\hat{A}_{1-}^{(1)}\{\delta'(\eta+1)-\delta(\eta+1)\} + \hat{A}_{1+}^{(1)}\{\delta'(\eta-1)+ \delta'(\eta-1)\}], \label{wz1_EBL}
\end{align}
where ${\mathcal P}$ denotes a Cauchy-principal-value interpretation. The integrated boundary-layer vorticity, for small $\textrm{E}$, may be written as:
\begin{align}
&(2\textrm{E})^{\frac{1}{2}}\displaystyle\int_{-\infty}^{\infty}\!\!\!\tilde{w}_z(\eta)[r_c \!+ \!(2\textrm{E})^{\frac{1}{2}}\eta] d\eta \\
=&\,\frac{1}{(2\textrm{E})^{\frac{1}{2}}}\left[r_c\!\!\!\displaystyle\int_{-\infty}^{\infty}\!\!\!\tilde{w}_z^{(0)}(\eta)d\eta + \textrm{E}^{\frac{1}{2}}\!\!\left(\!\sqrt{2}\!\!\displaystyle\int_{-\infty}^{\infty}\!\!\!\tilde{w}_z^{(0)}(\eta)\eta d\eta + r_c\!\!\!\displaystyle\int_{-\infty}^{\infty}\!\!\!\tilde{w}_z^{(1)}(\eta)d\eta \right)\right], \\
=&\, -\frac{4\mathrm{i}r_c^2}{m(2\textrm{E})^{\frac{1}{2}}}\mbox{Pf}.\displaystyle\int_{-\infty}^\infty \frac{d\eta}{(\eta^2-1)^2} + \frac{\mathrm{i}r_c}{m}(\hat{A}^{(1)}_{1+} - \hat{A}^{(1)}_{1-}) + O(\textrm{E}^{\frac{1}{2}}), \label{TotalBL_vort} \\
=&\,  \frac{\mathrm{i}r_c}{m}(\hat{A}^{(1)}_{1+} - \hat{A}^{(1)}_{1-}) + O(\textrm{E}^{\frac{1}{2}}), 
\end{align}
where the Cauchy-principal-value interpretation implies that only the terms proportional to $\hat{A}^{(1)}_{1\pm}$ in (\ref{wz1_EBL}) contribute. Physically, the contributions that are odd in $\eta$ denote jet-like structures, either localized at the travelling wave singularities\,(proportional to $\delta'(\eta \pm1)$) or non-local, within the elastic boundary layer, and their contribution to the outer velocity field is negligibly small for $\textrm{E} \ll 1$. The second term in (\ref{TotalBL_vort}) is expected, and denotes the strength of the elastic boundary layer, interpreted as an equivalent vortex sheet, on the outer scale. This vortex-sheet contribution is present only for the CS-modes, and vanishes for the Kelvin mode in which case $\hat{A}^{(1)}_{1+} = \hat{A}^{(1)}_{1-}$. The first term in (\ref{TotalBL_vort}) is larger, being $O(\textrm{E}^{-\frac{1}{2}})$, but does not contribute to the outer velocity field owing to the principal finite-part interpretation. Thus, the vortex-sheet contribution, proportional to $\hat{A}^{(1)}_{1+} - \hat{A}^{(1)}_{1-}$, is the only relevant one as far as the induced velocity field in the outer region is concerned, ensuring consistency with the inviscid scenario in the limit $\textrm{E} \rightarrow 0$. In terms of the vorticity field in the elastic boundary layer requiring a principal-finite-part interpretation, the relation between the finite $\textrm{E}$ CS-modes, and those for $\textrm{E}=0$, is similar to that between the CS-modes of a smooth vorticity profile and those of a Rankine vortex\,(\cite{GS2014}).\\

Having established in detail a connection between the solutions of the elastic Rayleigh equation for small but finite $\textrm{E}$, and the original inviscid spectrum of the Rankine vortex\,(including the Kelvin mode), we proceed towards an alternate intepretation of the CS-eigenfunctions for non-zero $\textrm{E}$. As will be seen, this interpretation is more general in that it is not reliant on $\textrm{E}$ being small. The presence of two parameters, $\hat{u}_r(r_c;r_c) - 1/r_c^{m+1}$ and $\hat{A}_1$ in (\ref{outer:final})-(\ref{BL:eigen3}), implies the existence of a pair of continuous spectra for finite $\textrm{E}$, in contrast to the single one for $\textrm{E}=0$. It is convenient to regard each of these as corresponding to a particular choice of $\hat{A}_1$ in (\ref{BL:eigen2}), and therefore, as being parameterized by $\hat{u}_r(r_c;r_c) - 1/r_c^{m+1}$ alone. The natural choices are $\hat{A}_1 = \textrm{E}^{\frac{1}{2}}\hat{A}_{1+}^{(1)}$ and $\hat{A}_1 =\textrm{E}^{\frac{1}{2}}\hat{A}_{1-}^{(1)}$ which ensure the smooth connection of the regular solution across $\eta = 1$ and $-1$, respectively. For either choice, the absence of a kink ensures the absence of delta-function-like contributions to the tangential velocity and vorticity fields at the relevant travelling wave singularity. Of course, the finite-$\textrm{E}$ eigenfunction is still singular owing to the logarithmic terms in (\ref{BL:eigen1})-(\ref{BL:eigen3}). From here onwards, the singularities at $\eta = 1$ and $-1$ will be associated with the fast\,(or forward) and slow\,(or backward) shear wave, respectively, for obvious reasons.  The choice $\hat{A}_1 = \hat{A}_{1+}^{(1)}$ implies that one only has a kink at the slow shear wave. Since it is this kink that allows for singular eigenfunctions even as $r_c$ spans a continuous interval, the  interval of existence for this slow-shear-wave-spectrum\,(SSWS) may be obtained by the requirement that the kink\,(the location of the slow shear wave singularity)  lie in the physical domain. The slow shear wave propagates with an angular frequency of $\omega = m[\Omega(r) + \Omega'(r)(2\textrm{E})^{\frac{1}{2}}] = m[1/r^2 - 2(2\textrm{E})^{\frac{1}{2}}/r^3]$, which must then lie in the base-state range of angular frequencies $(0,m)$. The shear wave frequency evidently approaches zero for $r \rightarrow \infty$, while it equals $m$ for $r = r_{cs}$ with $1/r_{cs}^2 - 2(2\textrm{E})^{\frac{1}{2}}/r_{cs}^3= 1$. For small $\textrm{E}$ this gives $r_{cs} = 1 - (2\textrm{E})^{\frac{1}{2}}$, corresponding to an angular frequency of $1+2(2\textrm{E})^{\frac{1}{2}}$. So, the SSWS frequency interval is $[0,m(1+2(2\textrm{E})^{\frac{1}{2}})]$. Analogous arguments for the fast shear wave yield the frequency interval for the fast-shear-wave-spectrum\,(FSWS) as $[0,m(1-2(2\textrm{E})^{\frac{1}{2}})]$ for small $\textrm{E}$. The SSWS and FSWS eigenfunctions differ in structure only within the $O(\textrm{E}^{\frac{1}{2}})$ boundary layer, where they are given by
\begin{align}
\tilde{u}_{r-}(\eta) =&\hat{A}_2^{(0)}\!\mbox{Pf}.\eta\!\ln \! \frac{\eta\!-\!1}{\eta \!+\!1}\! +\! \textrm{E}^{\frac{1}{2}}\eta\!\!\left[\!\hat{A}_{1-}^{(1)} \!+ \!\!\frac{3}{\sqrt{2}r_c}\hat{A}_2^{(0)}\!\mbox{Pf}.\!\!\left(\eta\ln \!\lvert \frac{\eta\!-\!1}{\eta\!+\!1} \rvert \!-\!\!  \frac{2}{\eta^2\!-\!1} \!\right)\!\!\right]\,\,\hspace*{0.05in} \eta < -1 , \label{SSWS:1}\\
\tilde{u}_{r+}(\eta) =&\hat{A}_2^{(0)}\!\mbox{Pf}.\eta\!\ln \lvert\!\frac{\eta\!-\!1}{\eta\!+\!1}\rvert \!+\! \textrm{E}^{\frac{1}{2}}\eta\!\!\left[\!\hat{A}_{1+}^{(1)} \!+\!\! \frac{3}{\sqrt{2}r_c}\hat{A}_2^{(0)}\!\mbox{Pf}.\!\!\left(\eta\ln \frac{\eta\!-\!1}{\eta\!+\!1} \!-\!  \frac{2}{\eta^2\!-\!1}\!\right)\!\!\right]\,\hspace*{0.05in} \eta > - 1,
\label{SSWS:2}
\end{align}
and
\begin{align}
\tilde{u}_{r-}(\eta) =&\hat{A}_2^{(0)}\!\mbox{Pf}.\eta\!\ln \!\lvert \frac{\eta\!-\!1}{\eta\!+\!1} \rvert\!\!+\! \textrm{E}^{\frac{1}{2}}\eta\!\!\left[\!\hat{A}_{1-}^{(1)} \!+\! \!\frac{3}{\sqrt{2}r_c}\hat{A}_2^{(0)}\!\mbox{Pf}.\!\left(\!\eta\ln \!\lvert \frac{\eta-1}{\eta +1} \rvert\! -\!  \frac{2}{\eta^2\!-\!1} \!\right)\!\!\right]\,\,\hspace*{0.05in} \eta < 1, \label{FSWS:1}\\
\tilde{u}_{r+}(\eta) =&\hat{A}_2^{(0)}\!\mbox{Pf}.\eta\!\ln \frac{\eta-1}{\eta+1} \!+\! \textrm{E}^{\frac{1}{2}}\eta\!\!\left[\!\hat{A}_{1+}^{(1)} \!+\!\! \frac{3}{\sqrt{2}r_c}\hat{A}_2^{(0)}\!\mbox{Pf}.\!\left(\!\!\eta\ln \!\frac{\eta\!-\!1}{\eta \!+\!1} \!- \! \frac{2}{\eta^2\!-\!1}\!\right)\!\!\right]\,\hspace*{0.05in} \eta > 1,
\label{FSWS:2}
\end{align}
respectively, to $O(\textrm{E}^{\frac{1}{2}})$. The SSWS interval above extends outside the base-state interval of angular frequencies, clearly implying that the modified semi-circle theorem stated earlier doesn't apply to the CS-modes. The equivalence of the CS-interval to the semi-circle radius, for the purely inviscid case, is thus a coincidence. It is worth noting that the expressions (\ref{SSWS:1})-(\ref{FSWS:2}) are valid only when the elastic boundary layer is an $O(1)$ distance away from the core. Strictly speaking, derivation of the finite-$\textrm{E}$ eigenfunctions for frequencies close to the upper end of the SSWS and FSWS intervals requires application of the core boundary condition to a uniformly valid representation constructed from both the outer and boundary layer solutions. This is a detail, however, and can be done\,(\cite{reddy15}). \\

\begin{figure}
     \centering
     \subfloat[]{
          \includegraphics[height=1.2in]{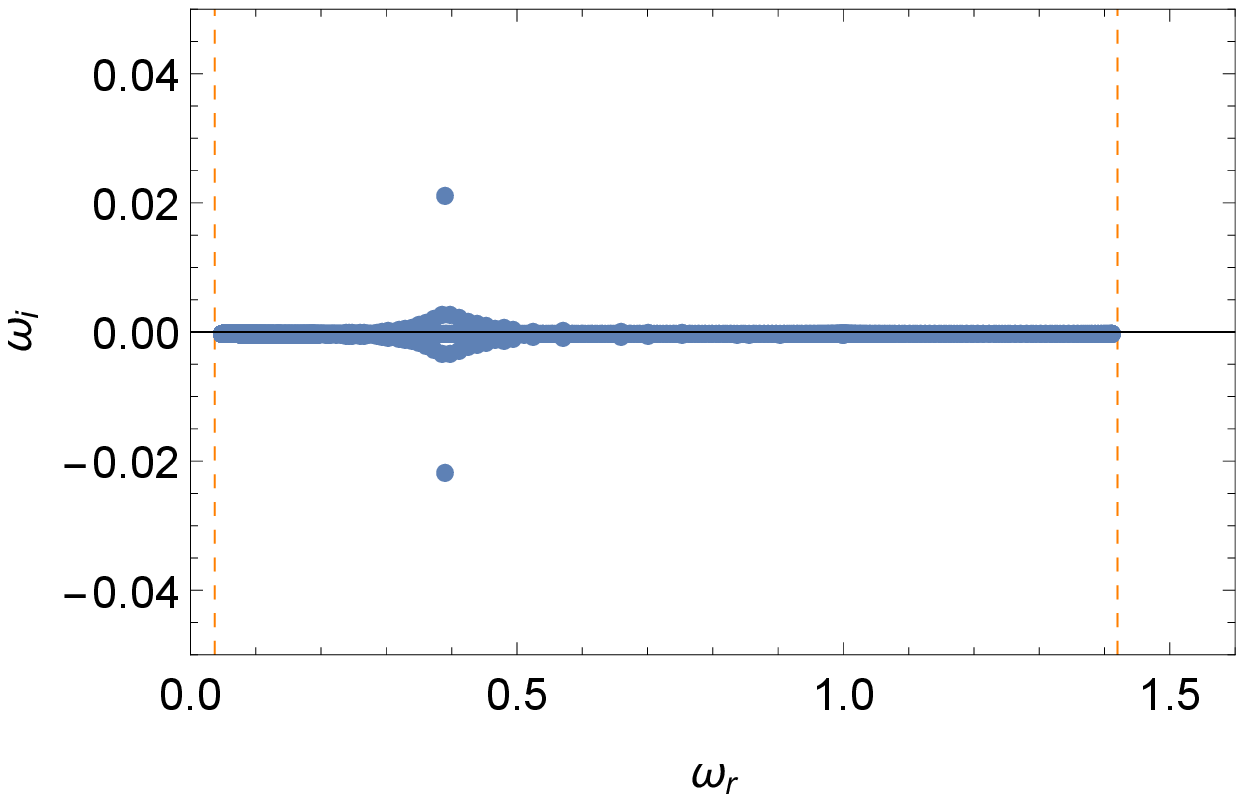}}
          \subfloat[]{
          \includegraphics[height=1.2in]{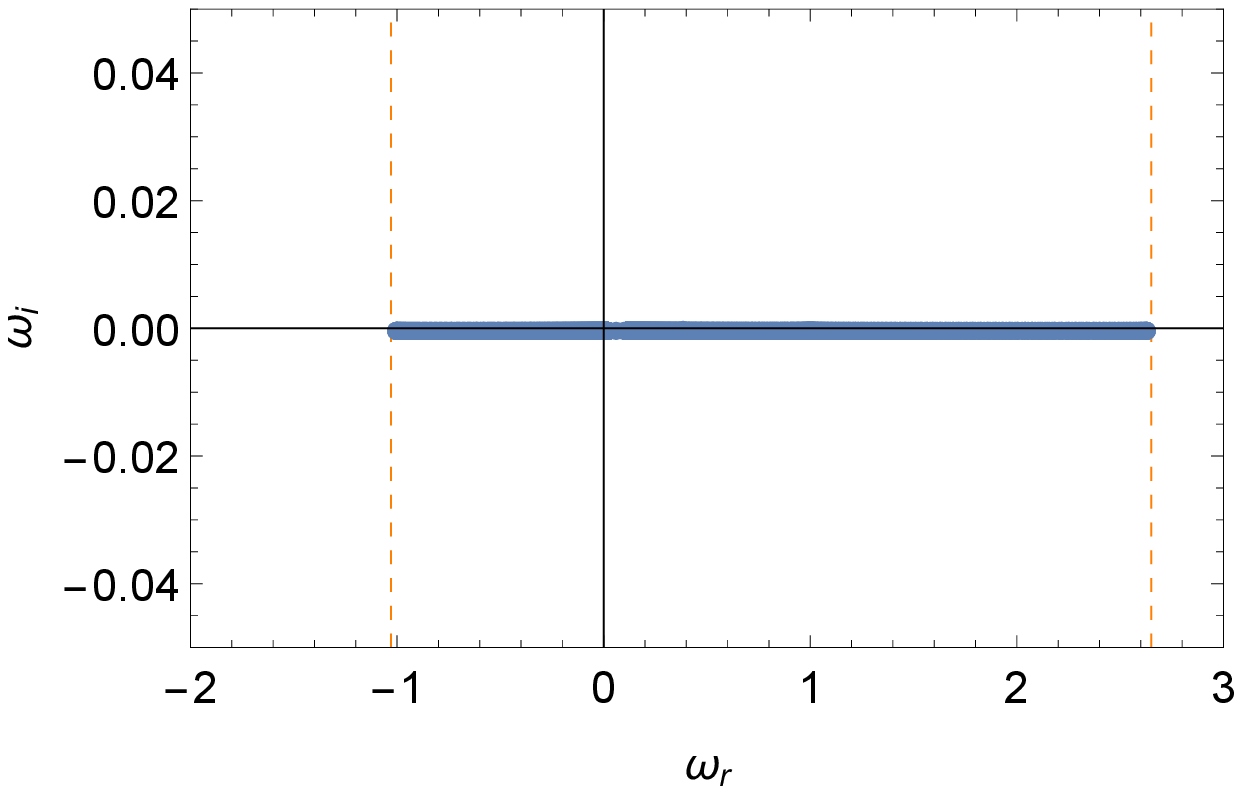}}
          
    \caption{The elastic Rayleigh spectrum for the hyperbolic tangent vorticity profile (defined in \eqref{eq:ZTanh}) with  $d=0.025$ for (a) $\textrm{E}=0.1$ and (b) $\textrm{E}=1$; $m=2$ and $N=1500$. In addition to the continuous spectrum, we see a pair of discrete modes for $\textrm{E}=0.1$. For the numerical calculation, the domain chosen is $r \in (0,r_\infty)$ with $r_\infty = 4a$. Hence the continuous spectrum eigenvalues lie between $\omega_{min}$ and $\omega_{max}$ which can be found analytically, using the expressions derived in the main text, and are marked on the plot.} \label{Fig:spectratanh}
\end{figure}

\begin{figure}
     \centering
     \subfloat[]{
          \includegraphics[height=1.2in]{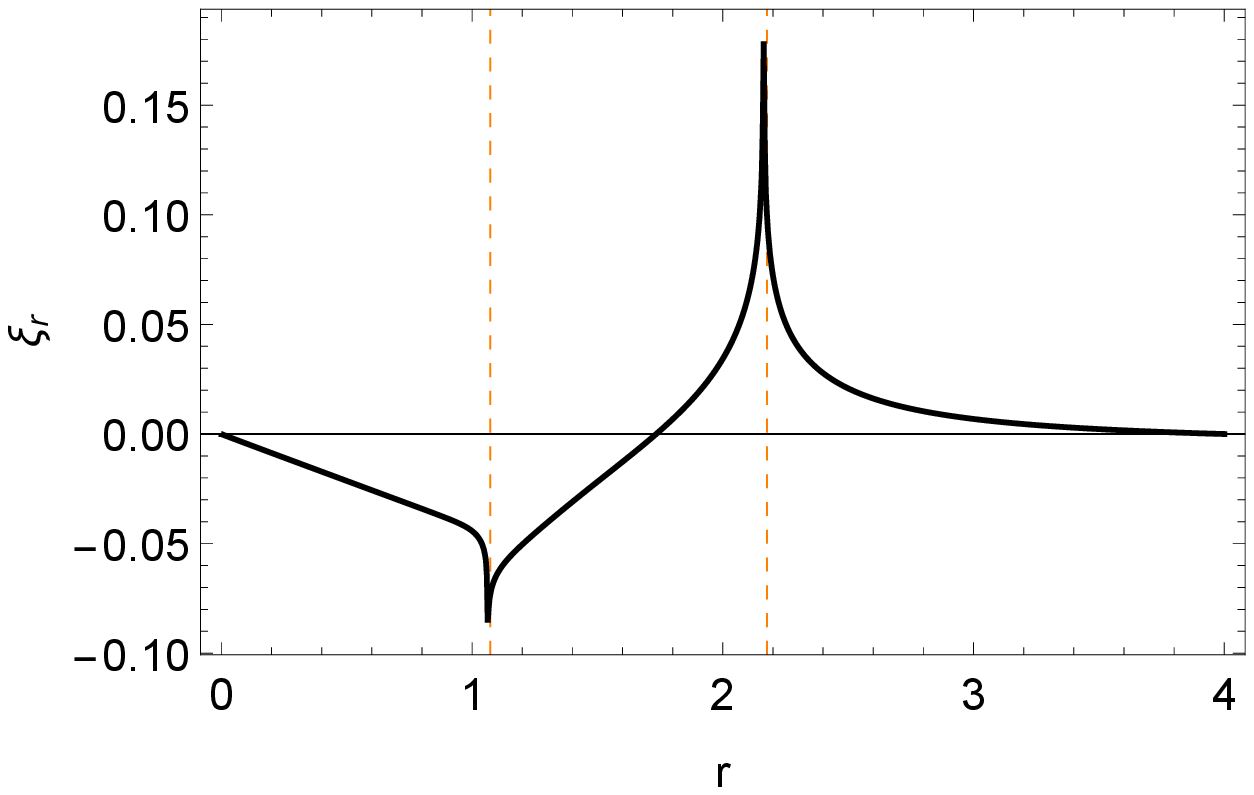}}
           \hspace{+.2in}
     \subfloat[]{
          \includegraphics[height=1.2in]{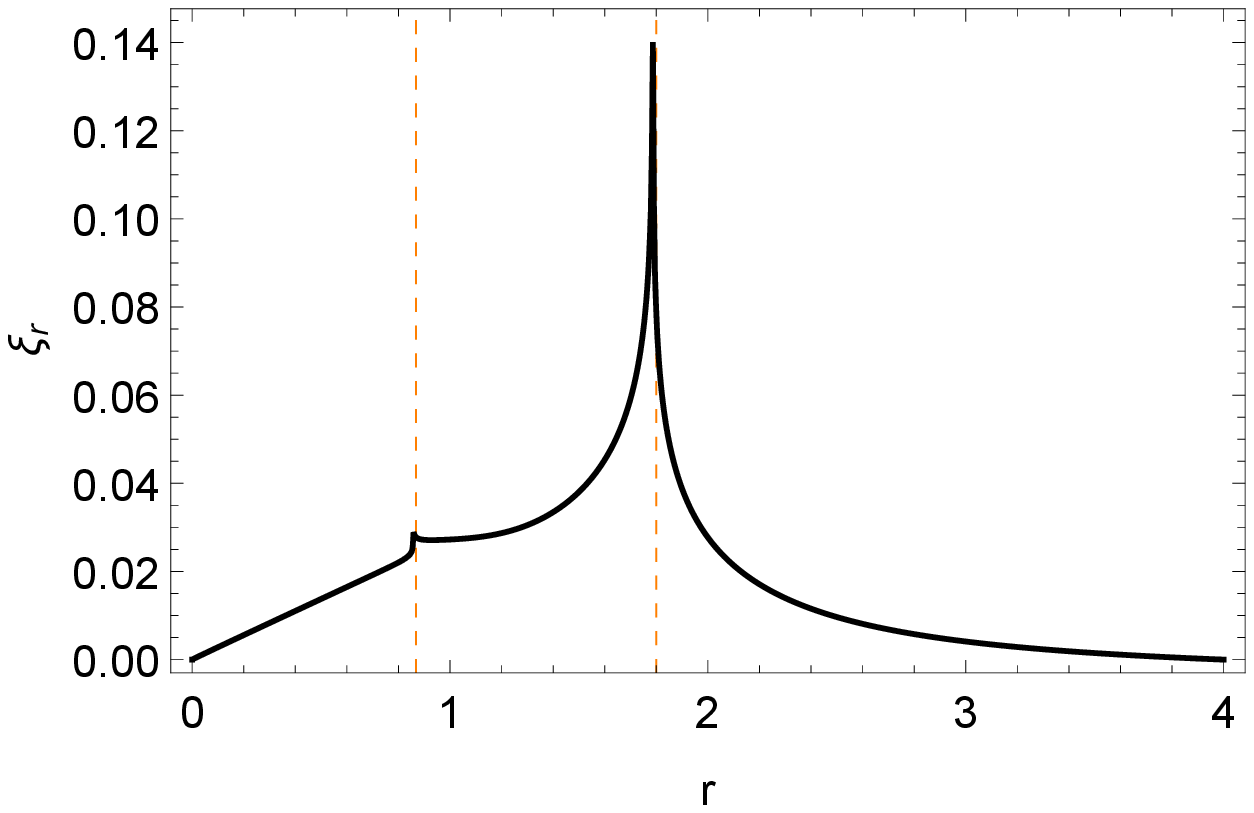}}
       \hspace{+.2in}
      \subfloat[]{
          \includegraphics[height=1.2in]{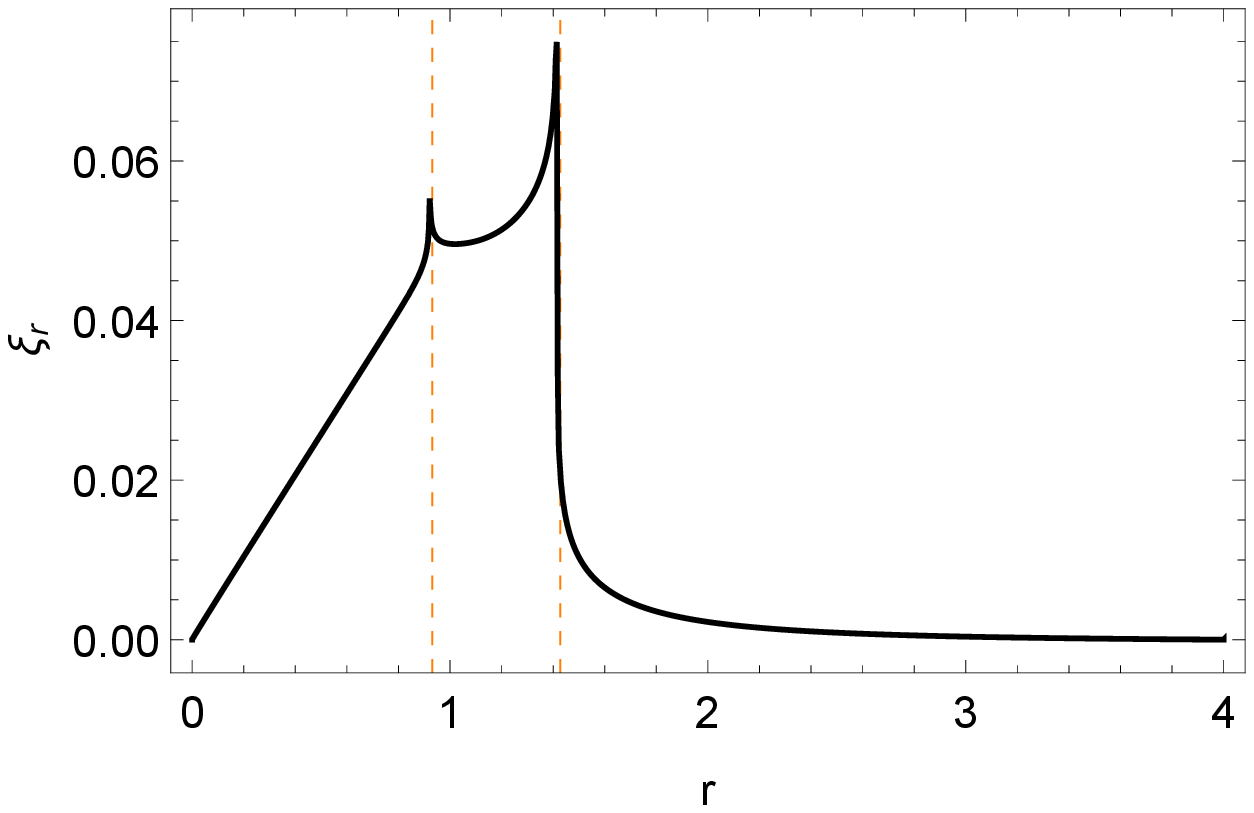}}  
 \hspace{+.2in}
           \subfloat[]{
          \includegraphics[height=1.2in]{./figures/cseigenfncrankine22.eps}}
           \hspace{+.2in}
     \subfloat[]{
          \includegraphics[height=1.2in]{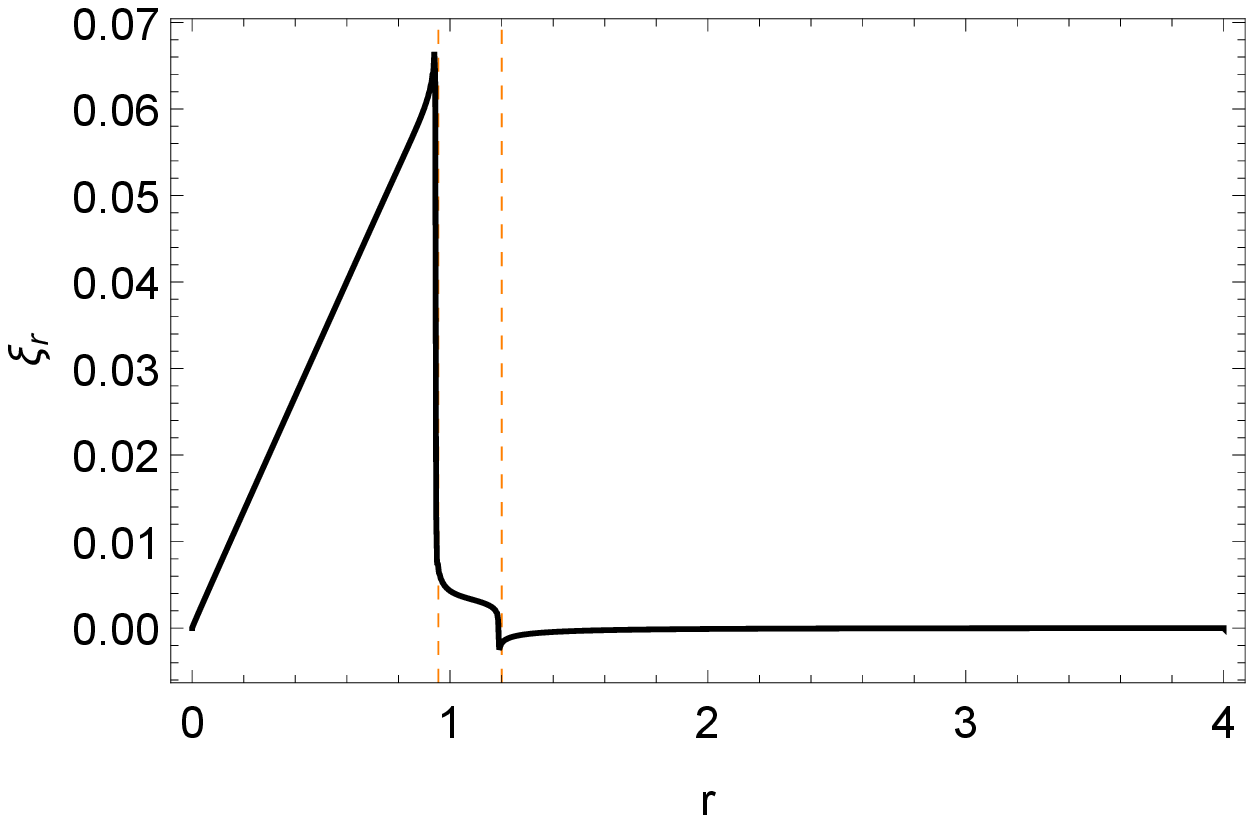}}
       \hspace{+.2in}
      \subfloat[]{
          \includegraphics[height=1.2in]{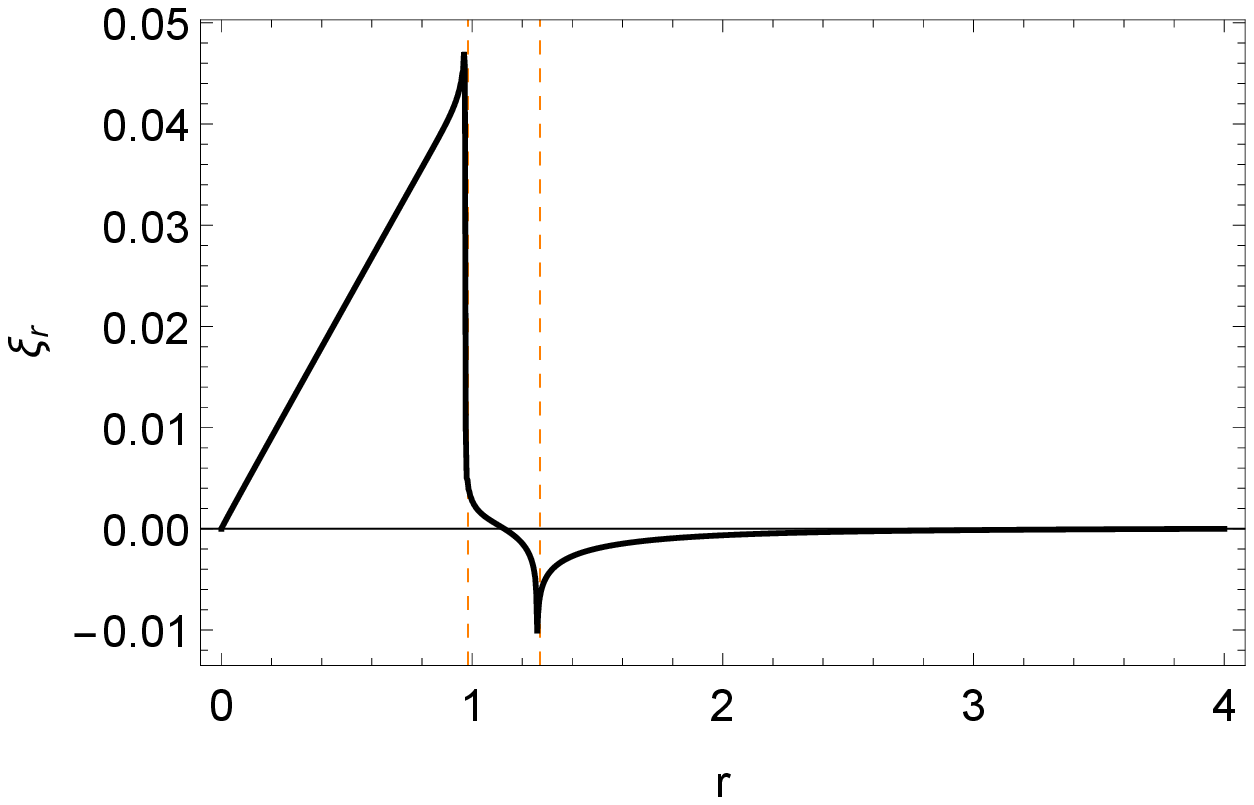}}         
 \hspace{+.2in}
          \caption{The radial displacement eigenfunctions for the CS-modes associated with the hyperbolic tangent vorticity profile (defined in \eqref{eq:ZTanh}); $\textrm{E}=0.1, m=2$  for (a), (c), (e); $\textrm{E}=1, m=2$ for (b), (d), (f). The wavespeeds are given by, $\omega_r=$ (a) $0.3$, (c) $0.8$  and (e) $1.2$ for $\textrm{E}=0.1$ and by $\omega_r=$ (b) $-0.7$, (d) $0.8$  and (f) $2$ for $\textrm{E}=1$. The analytical locations for the shear-wave singularities are marked as red (dashed) lines.}
\label{Fig:CSeigenfunctionstanh}
\end{figure}

\begin{figure}
     \centering
     \subfloat[]{
          \includegraphics[height=1.75in]{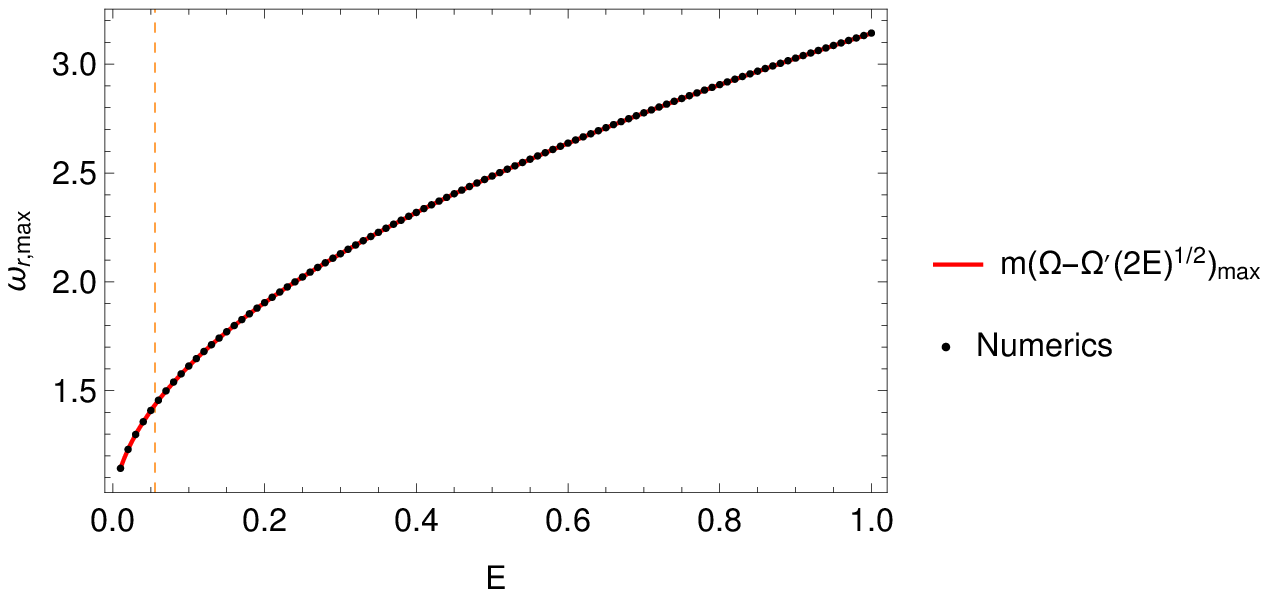}}
          \hspace{-.3in}
     \subfloat[]{
          \includegraphics[height=1.75in]{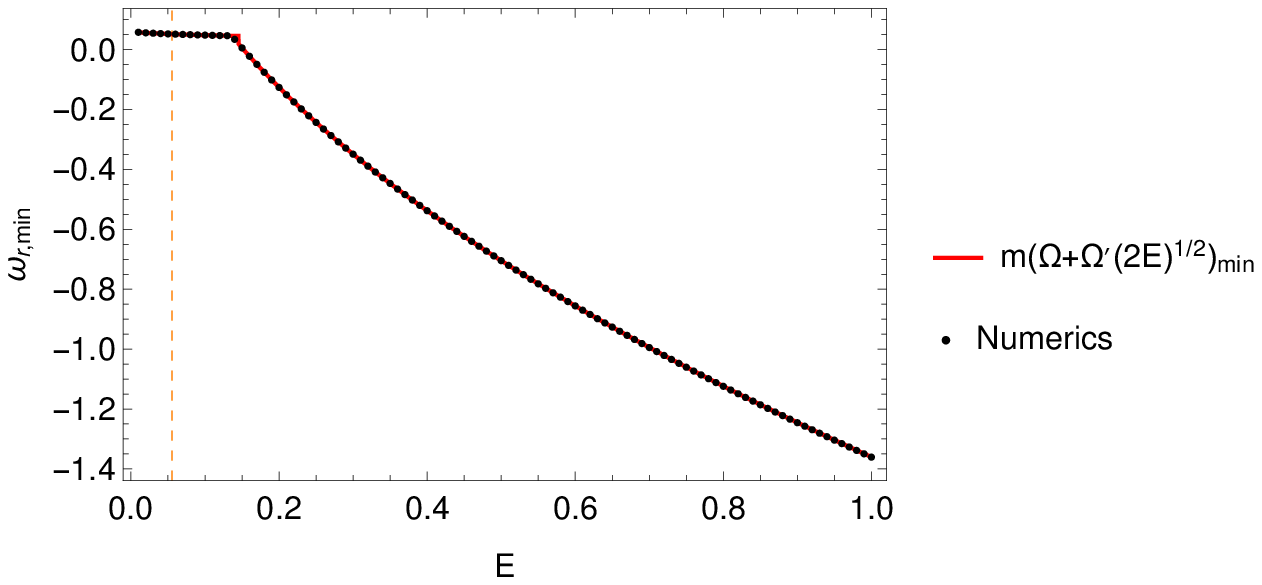}} 
          \caption{Variation of the (a) upper and (b) lower bounds of the SSWS and FSWS spectra, respectively, with $\textrm{E}$, for the hyperbolic tangent vorticity profile (defined in \eqref{eq:ZTanh}) with $d=0.025$.}\label{Fig:SSWS_FSWS}
\end{figure}

The point that needs emphasis, with regard to the alternate interpretation above, is its validity for arbitrary $\textrm{E}$. Although closed form expressions for the eigenfunctions, belonging to the two continuous spectra, can no longer be obtained when $\textrm{E}$ is not small, the pair of travelling wave singularities still exist and satisfy $\omega = m[\Omega(r) \pm 2 \Omega'(r)(2\textrm{E})^{\frac{1}{2}}]$. The elastic boundary layer solution given in (\ref{disp:EBL}) must now be interpreted as a Frobenius expansion in the vicinity of the relevant singular point. The FSWS spectrum corresponds to the interval $[0,m(1-2(2\textrm{E})^{\frac{1}{2}})]$ for $\textrm{E} < 1/8$, and to $[m(1-2(2\textrm{E})^{\frac{1}{2}}),0]$ for $\textrm{E} > 1/8$. The SSWS spectrum continues to be given by $[0,m(1+2(2\textrm{E})^{\frac{1}{2}})]$ for finite $\textrm{E}$, although there arises a degeneracy for $\textrm{E} > 1/18$ due to shear waves at a pair of radial locations propagating with the same frequency. Figures \ref{Fig:spectratanh} and \ref{Fig:CSeigenfunctionstanh} show the spectra, and a few representative eigenfunctions, for the hyperbolic tangent vorticity profile (defined in \eqref{eq:ZTanh}), determined numerically using a spectral method (cf. next section). Figure \ref{Fig:SSWS_FSWS} shows a comparison between the numerical and analytical estimates for the upper (and lower) bounds of the SSWS (and FSWS) intervals with $\textrm{E}$ for this vorticity profile. \\

\begin{figure}
     \centering
     \subfloat[]{
          \includegraphics[height=1.2in]{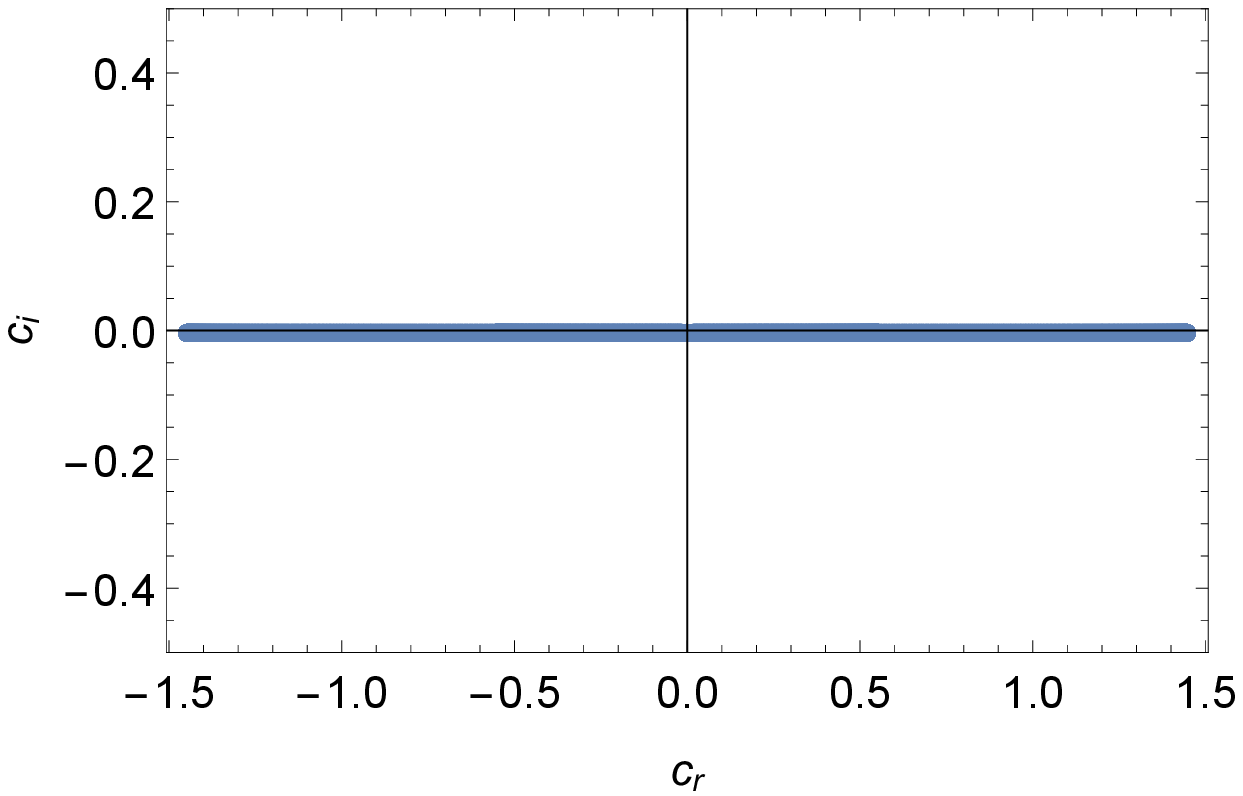}}
          \subfloat[]{
          \includegraphics[height=1.2in]{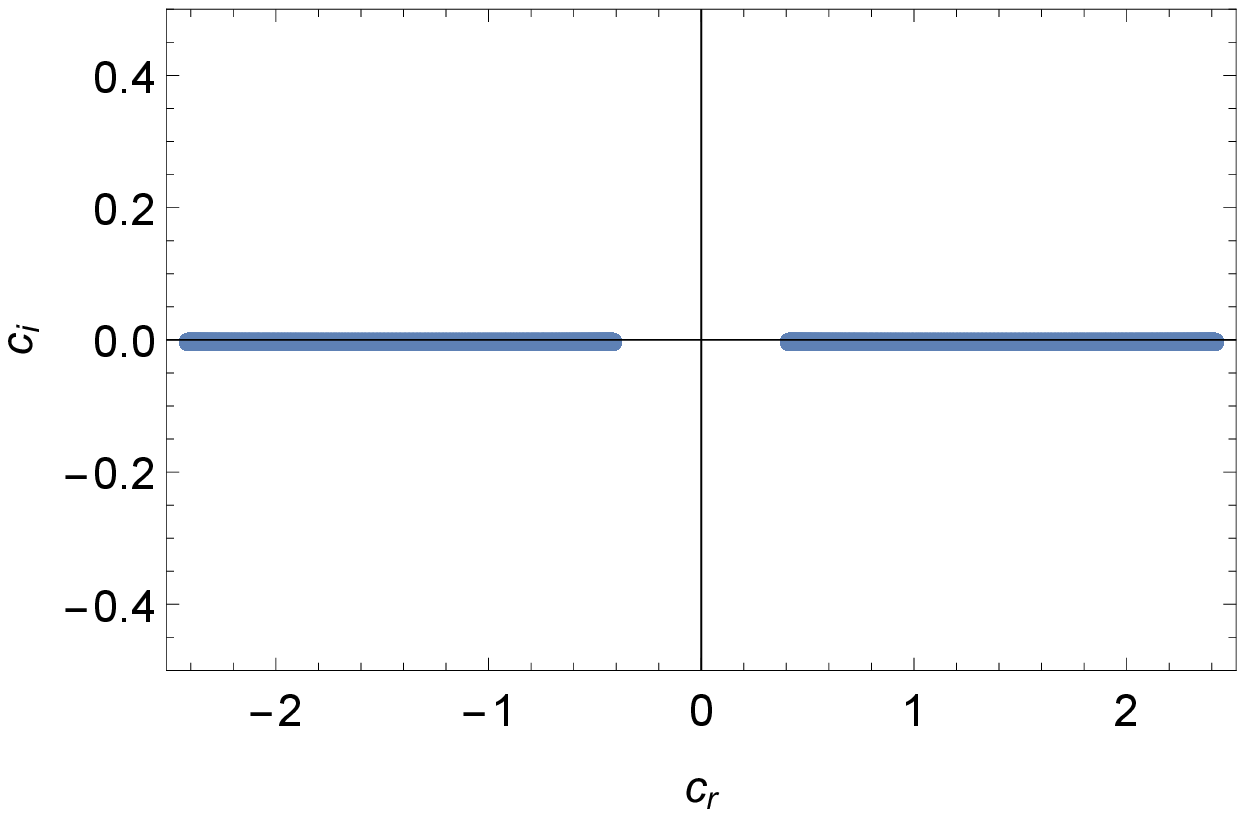}}
          
    \caption{The spectrum for plane Couette flow for (a) $\textrm{E}=0.1$ and (b) $\textrm{E}=1$; $k=2$ and $N=1500$.}\label{Fig:spectracouette}
\end{figure}

\begin{figure}
     \centering
     \subfloat[]{
          \includegraphics[height=1.2in]{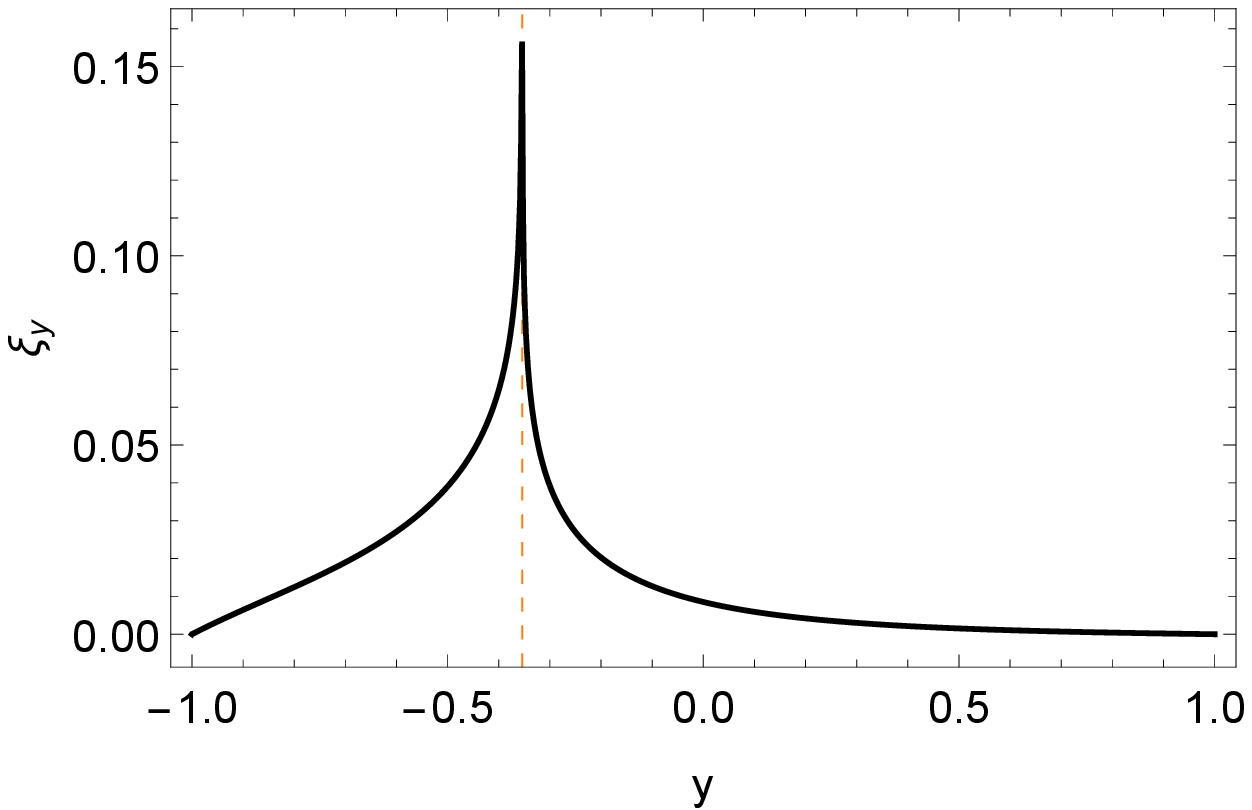}}
           \hspace{+.2in}
           \subfloat[]{
          \includegraphics[height=1.2in]{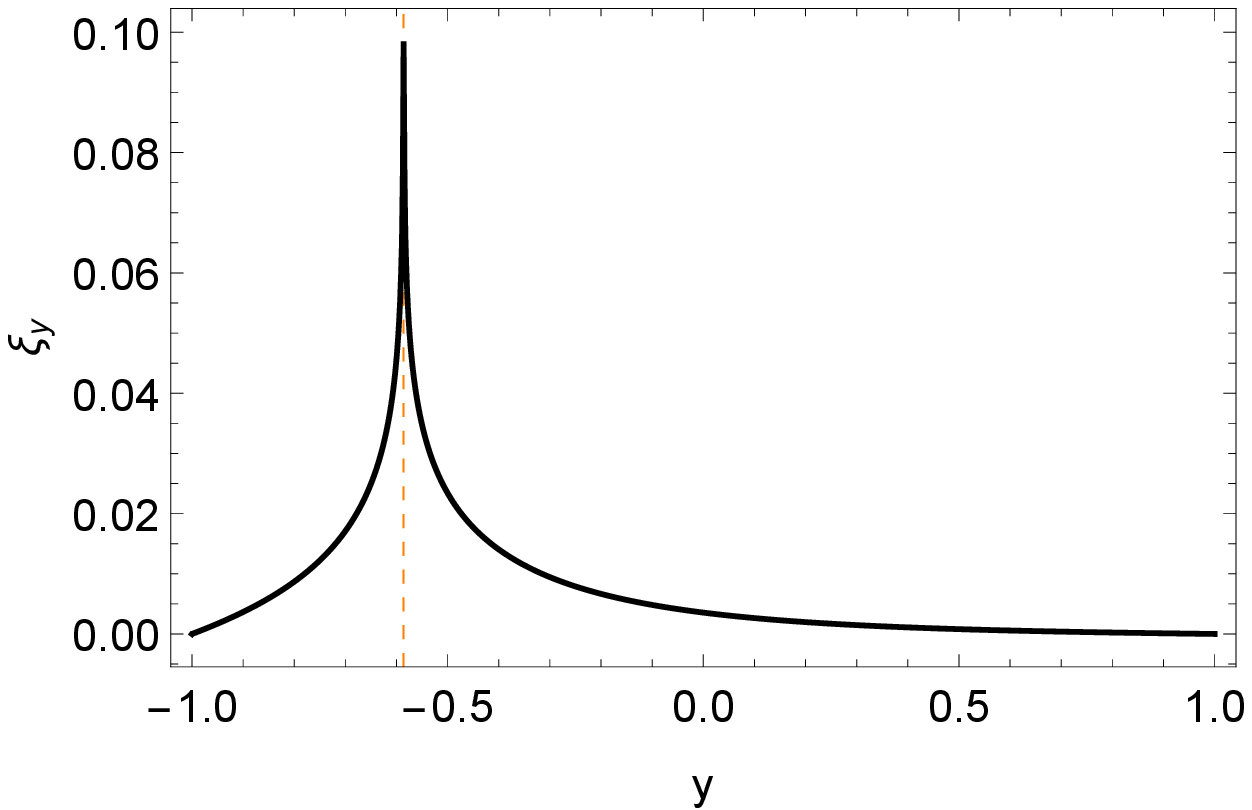}}
       \hspace{+.2in}
      \subfloat[]{
          \includegraphics[height=1.2in]{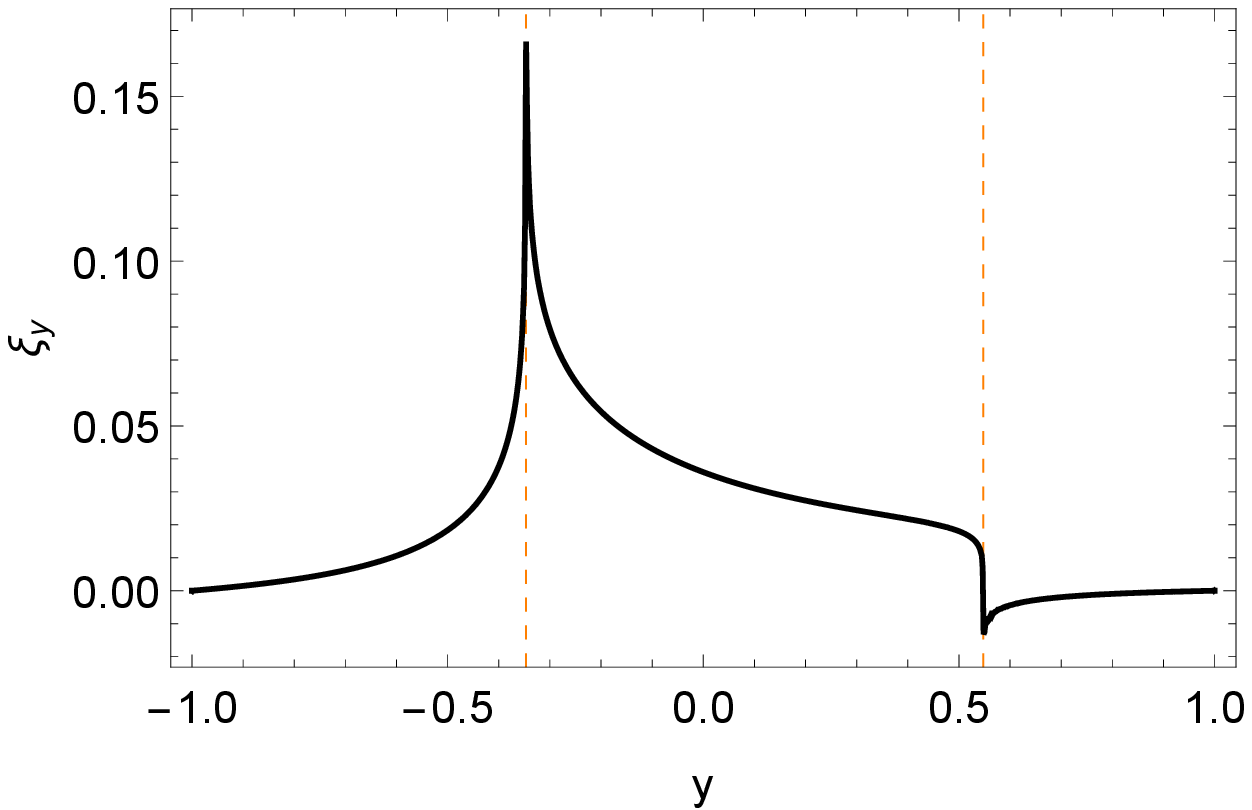}}  
           \hspace{+.2in} 
     \subfloat[]{
          \includegraphics[height=1.2in]{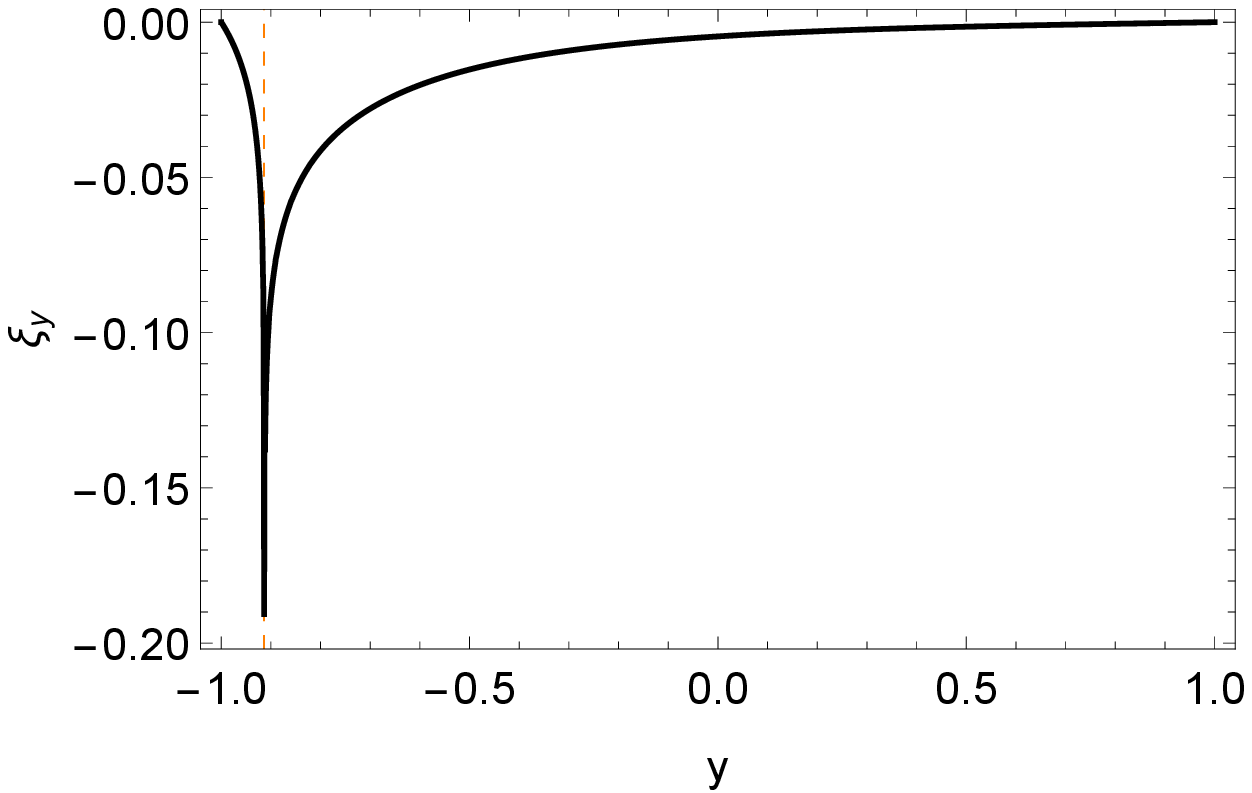}}
           \hspace{+.2in}
     \subfloat[]{
          \includegraphics[height=1.2in]{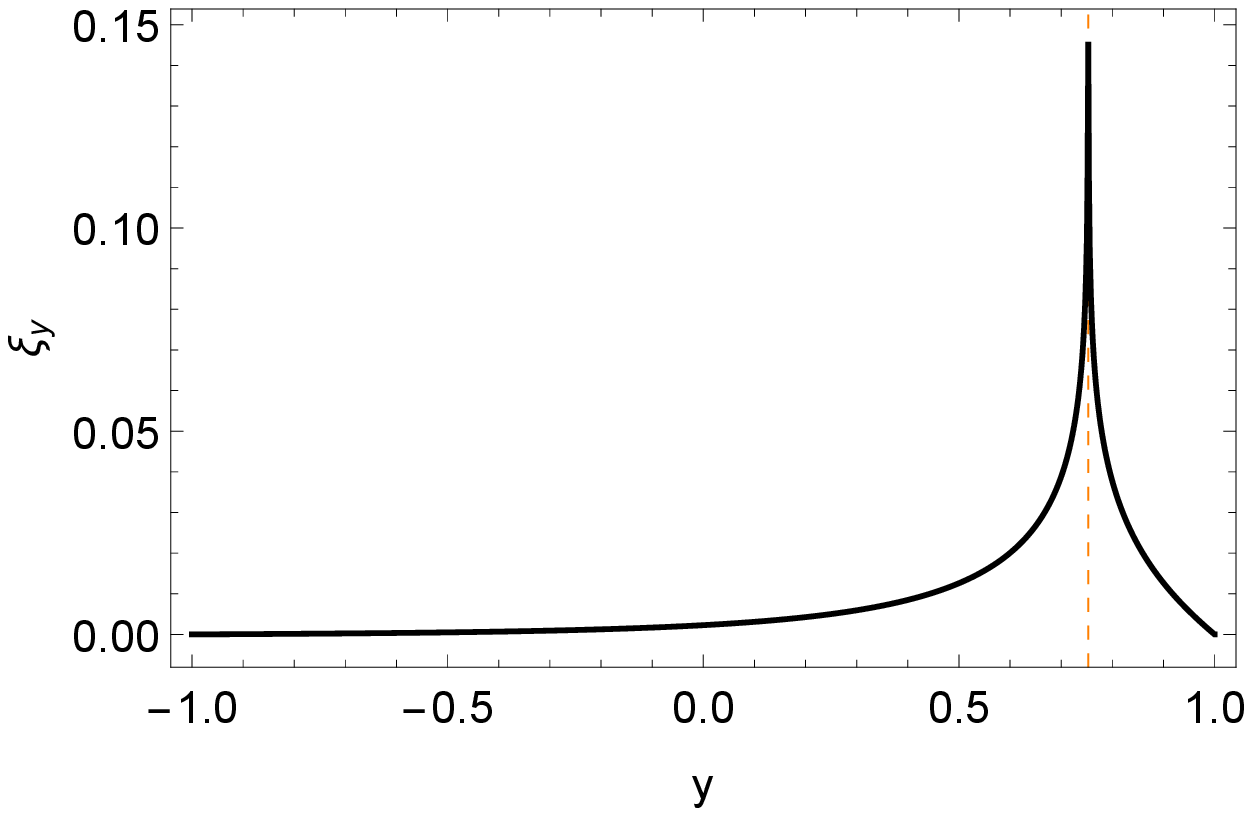}}
     \hspace{+.2in}
      \subfloat[]{
          \includegraphics[height=1.2in]{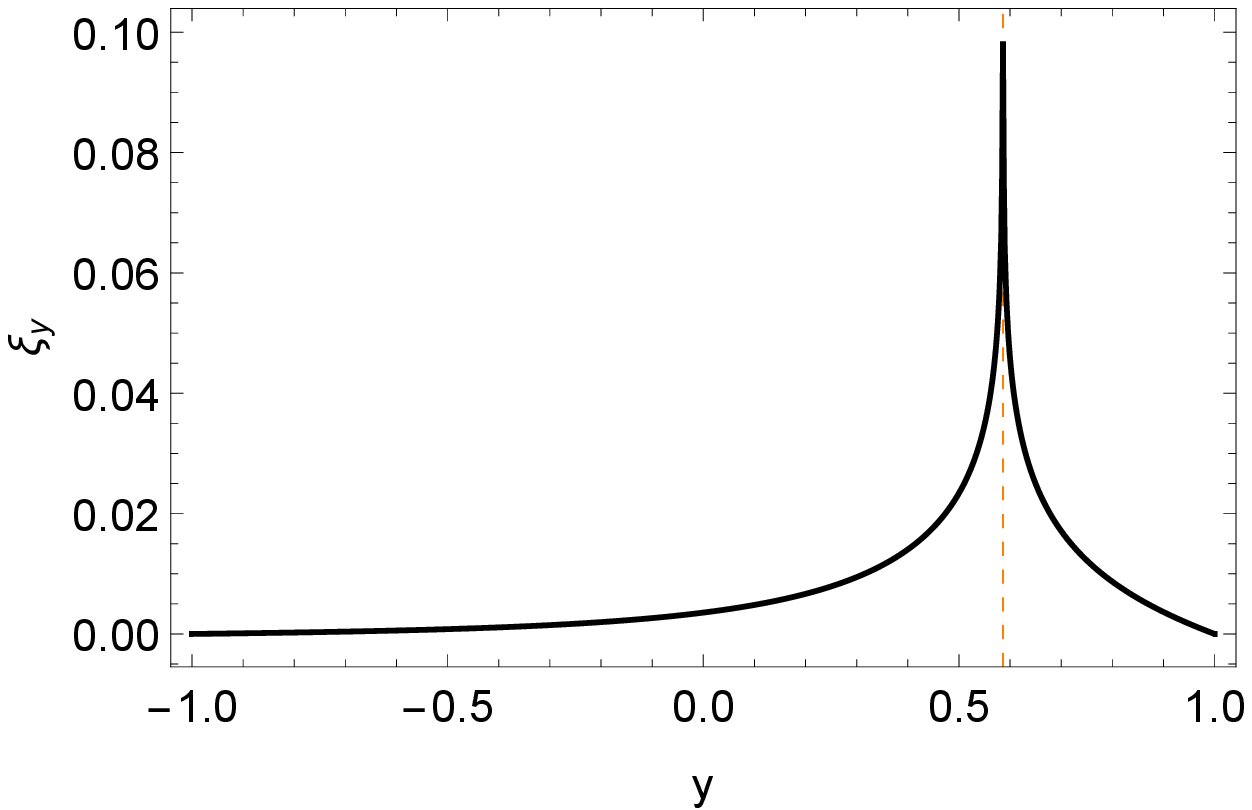}}        
          \hspace{+.2in}
          \caption{The wall-normal displacement eigenfunctions for plane Couette flow for the CS-modes; $\textrm{E}=0.1, k=2$  for (a), (c), (e); $\textrm{E}=1, k=2$ for (b), (d), (f). The wavespeeds are given by $c_r=$ (a) $-0.8$, (c) $0.1$  and (e) $1.2$ for $\textrm{E}=0.1$ and by $c_r=$ (b) $-2$, (d) $0.5$ and (f) $2$ for $\textrm{E}=1$. The analytical locations for the singularities are marked as red (dashed) lines.}\label{Fig:CSeigenfunctionscouette}
\end{figure}

The close analogy of the Rankine vortex with plane Couette flow\,(\cite{GS2014}) allows one to identify the travelling-wave spectra in the latter case too. For the (dimensionless)\,base-state profile $U(y) = y$ in the domain $(0,1)$, the FSWS and SSWS span the intervals $[-(2\textrm{E})^{\frac{1}{2}},1-(2\textrm{E})^{\frac{1}{2}}]$ and $[(2\textrm{E})^{\frac{1}{2}},1+(2\textrm{E})^{\frac{1}{2}}]$, both of which violate the semi-circle theorem. Unlike the Rankine vortex, the linearity of the Couette profile implies that the aforementioned spectral intervals remain valid for arbitrary $\textrm{E}$, and become disjoint for $\textrm{E} > 1/8$. An analysis similar to that detailed above, but simpler, can be carried out for plane Couette flow to obtain expressions for the singular eigenfunctions corresponding to the fast and slow shear wave spectra in the limit $\textrm{E} \ll 1$. The outer solutions, valid when $|y-y_c| \gg \textrm{E}^{\frac{1}{2}}$ with $y_c$ being the critical level, are the well-known Case eigenfunctions with the normal velocity given by $\hat{u}_y(y;y_c) = \sinh k(1-y_c) \sinh ky$ for $0< y < y_c$ and $\hat{u}_y(y;y_c) = \sinh ky_c \sinh k(1-y)$ for $y < y_c < 1$ \,(\cite{CASE60}). Within the elastic boundary layer, the SSWS eigenfunctions are given by:
\begin{align}
\tilde{u}_{y-}(\eta) =&\hat{B}_2 \mbox{Pf}.\eta\!\ln \! \frac{\eta\!-\!1}{\eta \!+\!1}\! +\! (2\textrm{E})^{\frac{1}{2}}\eta \hat{B}_{1-} \,\hspace*{0.05in} \eta < -1 , \label{SSWS:1couette}\\
\tilde{u}_{y+}(\eta) =&\hat{B}_2 \mbox{Pf}.\eta\!\ln \lvert\!\frac{\eta\!-\!1}{\eta\!+\!1}\rvert \!+\! (2\textrm{E})^{\frac{1}{2}}\eta \hat{B}_{1+} \,\hspace*{0.05in} \eta > - 1,
\label{SSWS:2couette}
\end{align}
and the FSWS eigenfunctions are given by:
\begin{align}
\tilde{u}_{y-}(\eta) =&\hat{B}_2\mbox{Pf}.\eta\!\ln \!\lvert \frac{\eta\!-\!1}{\eta\!+\!1} \rvert\!\!+\! (2\textrm{E})^{\frac{1}{2}}\eta \hat{B}_{1-} \,\hspace*{0.05in} \eta < 1, \label{FSWS:1couette}\\
\tilde{u}_{y+}(\eta) =&\hat{B}_2 \mbox{Pf}.\eta\!\ln \frac{\eta-1}{\eta+1} \!+\! (2\textrm{E})^{\frac{1}{2}}\eta \hat{B}_{1+} \,\hspace*{0.05in} \eta > 1,
\label{FSWS:2couette}
\end{align}
Here, $\eta = (y-y_c)/(2\textrm{E})^{\frac{1}{2}}$ is the boundary layer variable, and the constants appearing in (\ref{SSWS:1couette})-(\ref{FSWS:2couette}) are $\hat{B}_2 = -\frac{1}{2}\hat{u}_y(y_c;y_c)$ with $\hat{u}_y(y_c;y_c) = \sinh ky_c \sinh k(1-y_c)$ being the normal velocity at the critical level, and $\hat{B}_{1+} =-k \cosh k(1-y_c) \sinh ky_c$ and $\hat{B}_{1-}= k\cosh ky_c \sinh k(1-y_c)$. Since the original inviscid spectrum is purely continuous, the exceptional case of $\hat{B}_{1+} = \hat{B}_{1-}$, corresponding to the Kelvin mode for the Rankine vortex above, does not arise. Figures \ref{Fig:spectracouette} and \ref{Fig:CSeigenfunctionscouette} show the spectra, and a few representative eigenfunctions, for plane Couette flow, again determined using a spectral method (cf. next section). \\

The CS-modes arising from the multiple continuous spectra above, together with a possibly finite number of discrete modes, must form a complete basis for the independent fields required to completely characterize an initial state in the limit $\Rey, \Deb \rightarrow \infty$. For the Rankine vortex, as governed by the elastic Rayleigh equation, these may be taken as the radial velocity field and the two components of the polymeric force field, $\nabla \bcdot {\boldsymbol a}$; since the radial component of the normal stress, $a_{rr}$, does not enter in the elastic Rayleigh limit, one may equivalently consider the radial velocity field and the stress components $a_{r\theta}$ and $a_{\theta\theta}$. One therefore needs (at least)\,three continuous spectra in order to represent an arbitrary initial condition. The analysis detailed above, in choosing a continuous solution across $r=r_c\,(\eta = 0)$, does not account for the third spectrum needed. This is the Doppler spectrum corresponding to $\omega = \Omega(r_c)$. Now, the Frobenius exponents for $r = r_c$, for the radial displacement field, are $0$ and $1$, and there is no singularity at $r = r_c\,(\eta=0)$, as is also evident from the elastic boundary layer solutions above\,(see (\ref{disp:EBL}) above). However, in a manner similar to inviscid plane Couette flow\,(\cite{CASE60}), one nevertheless requires CS-modes with a singularity at $r = r_c$ to generate a complete basis. The Doppler spectrum eigenfunctions, for finite $\textrm{E}$, with $De \rightarrow \infty$, may be generated by different choices of one of the two regular solutions on either side of $r = r_c$\,(the one corresponding to a Frobenius exponent of $0$). On including the effects of relaxation, $r = r_c$ becomes a singular point as already indicated in section \ref{sec:elas_form}. The functional form of the CS-modes has been obtained earlier\,(\cite{GRAHAM98}). For large but finite $\Deb$, these modes, arising from the solution of a fourth-order ODE, will be valid in an inner layer of $O(\Deb^{-1})$\,(with $\Deb^{-1} \ll \ste$), and will transition to the elastic Rayleigh Doppler-spectrum eigenfunctions on scales much larger than $O(\Deb^{-1})$.

\section{Shear Wave Resonance Instability of a Vortex} \label{discrete:instability}

In this section, we show that a vortex in a visco-elastic fluid is susceptible to a two-dimensional instability. Towards this end, we first numerically examine the spectra of Rankine-like smooth vorticity profiles as a function of $\textrm{E}$ (section \ref{finiteE:numerics}). The characterization of the CS-spectra of the elastic Rayleigh equation above serves as a valuable aid in interpreting the results of the spectral code\,(one of three approaches used here to numerically characterize the finite-$\textrm{E}$ modal instability). For small $\textrm{E}$, the unstable eigenfunction is a regularized version of the CS-modes analyzed in section 3, with the singularities at the travelling wave locations cut-off due to a finite growth rate. Further, knowledge of the finite-E CS-spectrum also helps in isolating isolating the discrete mode from the numerical ballooned manifestation of the true CS-spectra. For the specific case of a Rankine vortex, the elastic instability is then analyzed via a matched asymptotic expansions approach valid for small $\textrm{E}$ (section \ref{smallE:analysis}).

\subsection{Numerical calculation of the unstable mode} \label{finiteE:numerics}
The Rankine vortex has a compact vorticity profile with a step discontinuity at the core radius. A smooth vorticity profile,  convenient for use in the numerical calculations below, is given by:
\begin{eqnarray}
 Z(r) = \frac{Z_0}{2}\left\{1-\tanh\left[\frac{r-a}{d}\right]\right\}. \label{eq:ZTanh}
\end{eqnarray}
Here, $d$ is the length scale over which there is a smooth transition from the core vortical region to the exterior irrotational one, with $d \rightarrow$ 0 denoting the limit of a Rankine vortex. 

\subsubsection{Details of the numerical method}
 \begin{figure}
     \centering
          \includegraphics[height=2.5in]{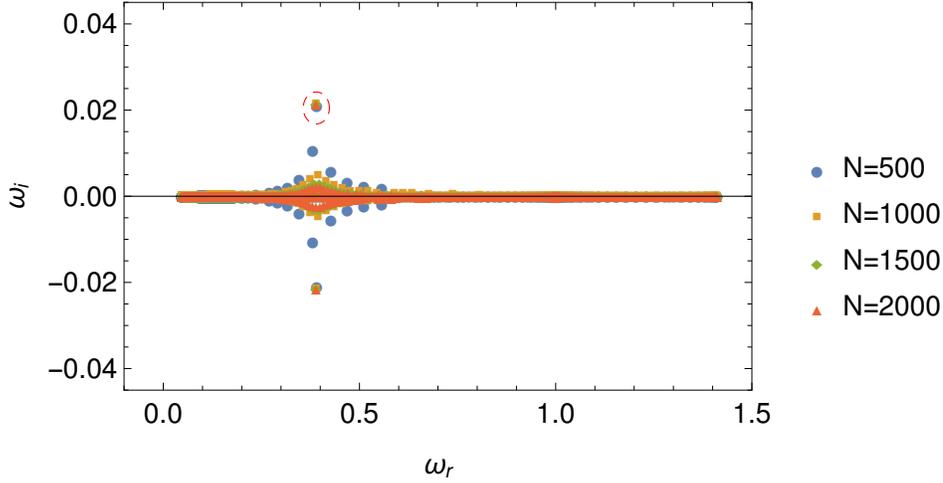}
          \caption{Collapse of the ballooned spectrum with increasing number of collocation points ($N$) for the vorticity profile defined in (\ref{eq:ZTanh}) with $d=0.025, a=1$ for $\textrm{E}=0.1$; the domain is $r \in (0,r_\infty)$ with $r_\infty = 4a$. The converged unstable mode is encircled.}\label{Fig:balloon}
\end{figure}

Stability investigations in the limit for large $\Rey$ and $\Deb$ suffer from numerical issues, largely due to the presence of the continuous spectra in the elastic Rayleigh equation \citep{MILLER05}. For the problem under consideration, two additional effects make the investigation of the unstable mode a difficult one for small E. First, the unstable mode asymptotes to a singular neutral mode (a part of the continuous spectrum) in the limit $\textrm{E} \rightarrow 0$. This is unlike the classical inviscid problem where the unstable mode approaches a regular neutral mode, the so-called $S$-wave, close to the threshold\,(\cite{DRAZINREID81}; \cite{DRAZINHOW66}). The second reason is the emergence of a transcendentally small length scale for small E (identified as part of the analysis in the next section). The inability to resolve this scale leads to a breakdown of the numerics below E $\sim 0.02$ regardless of the approach. \\

For the smooth vorticity profile in (\ref{eq:ZTanh}), two separate formulations of the eigenvalue problem in (\ref{eq:xir}) are studied. In the first formulation, the linear eigenvalue problem given by (\ref{eq_ur})-(\ref{eq_att}) (in the limit $\Rey,\Deb\rightarrow\infty$) is solved using Chebyshev collocation. In the second formulation, a solution of the nonlinear eigenvalue problem (\ref{eq:xir}) for $\xi$, is obtained using a compound matrix method wherein the original non-linear eigenvalue problem is written as a higher dimensional linear one, which is then solved using standard Chebyshev collocation \citep{BRIDMOR84, roy2014thesis}. Both formulations yield consistent results, and here we show results from the second formulation \citep{roy2014thesis}. The spectral method obtains the entire eigenspectrum including the singular continuous spectra. Since the continuous spectrum modes aren't $C^{\infty}$, the neutral continuous spectra manifest as balloons in the numerical spectrum and for a modest number of collocation points, the ballooned continuous spectrum ends up engulfing the unstable mode \citep{shankar2019, roy2014thesis}. As shown in figure \ref{Fig:balloon}, it is only for a sufficiently large number of collocation points ($N$) that the CS-balloon shrinks sufficiently for the converged discrete mode to be identified (see figure \ref{Fig:balloon}). Recall that the spectral method, using the second formulation above, was also used to characterize the continuous spectrum in section \ref{sec:CSrayleigh}. \\

To study the Rankine vortex, a regular shooting method is applied, using the inbuilt $bvp4c$ command in MATLAB. Owing to the difficulty arising from the extreme sensitivity of the eigenvalue found to the initial guess, we use a `carpet-bombing' technique to obtain a reasonably accurate initial guess of the eigenvalue \citep{MILLER05}. Herein, one of the boundary conditions is allowed to be an unknown, its difference from the true boundary condition being termed the `error'. This error is then minimized on a complex-$\omega$ grid to arrive at the initial guess (see \cite{roy2014thesis} for details). We solve (\ref{eq:xir}) only in the region $r>a$, with appropriate boundary conditions at the core-(irrotational) exterior interface (see \eqref{eq:bc_elas_1}-\eqref{eq:bc_elas_2} in section \ref{smallE:analysis}). \\

\subsubsection{Numerical results for the inertio-elastic instability}
\begin{figure}
     \centering
     \subfloat[]{
          \includegraphics[height=2.1in]{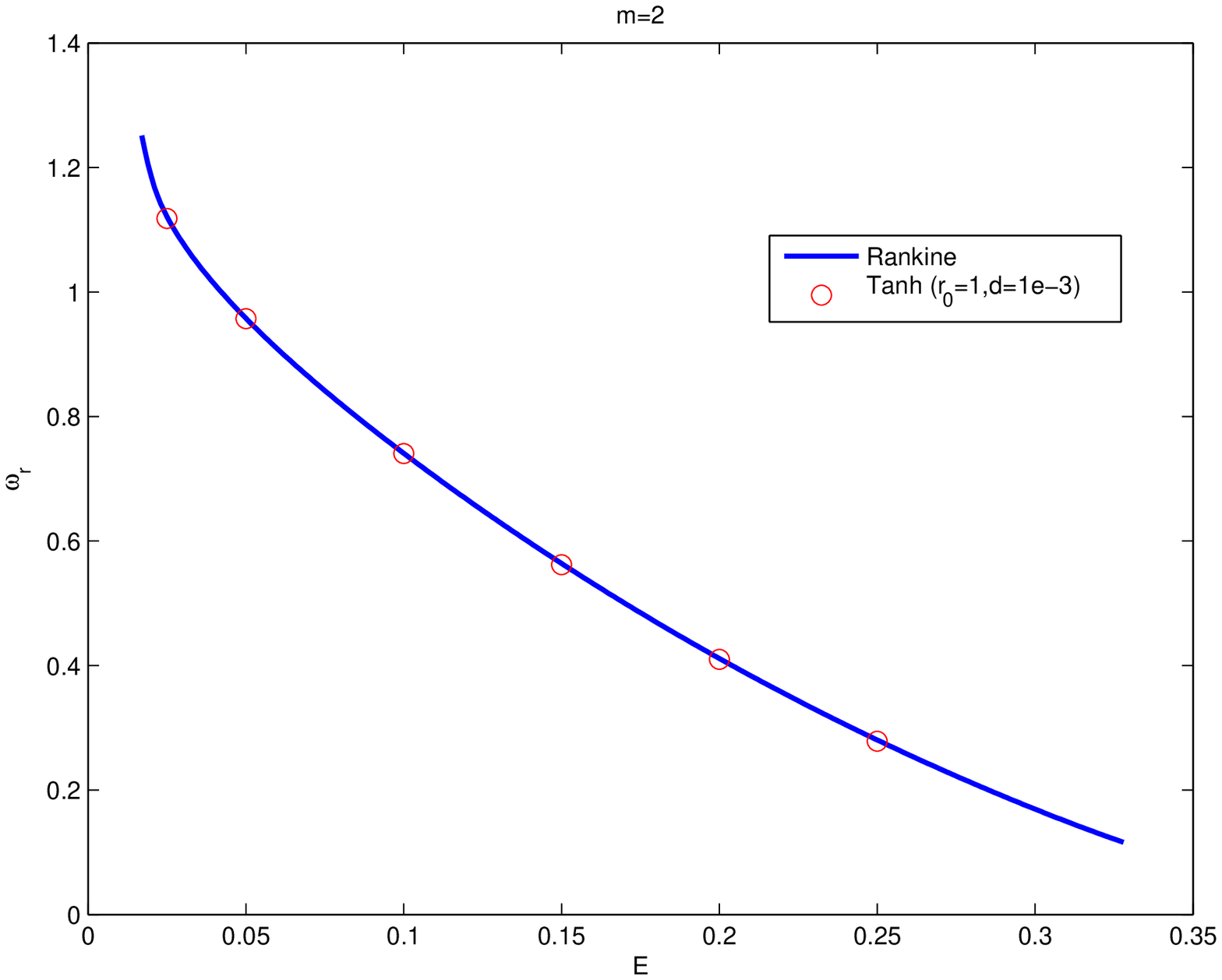}}
          \hspace{-.15in}
     \subfloat[]{
          \includegraphics[height=2.1in]{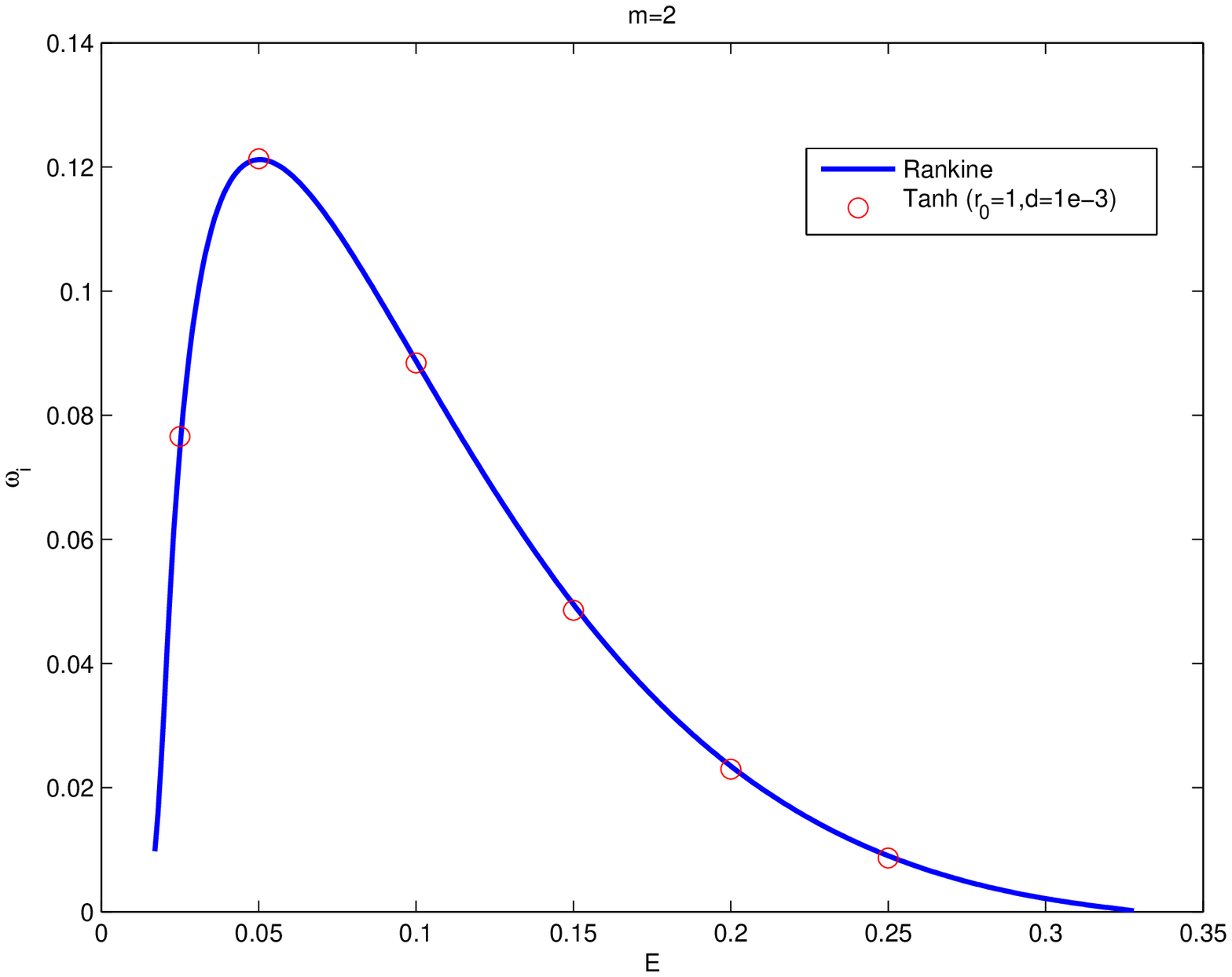}}        
          \caption{Wave speed and growth rate comparisons between a Rankine vortex (shooting method) and the smooth vortex in (\ref{eq:ZTanh}) with $a=1$ and $d=10^{-3}$ (spectral method), for $m=2$.}\label{Fig:Rankine_tanh}
\end{figure}

\begin{figure}
     \centering
     \subfloat[]{
          \includegraphics[height=2.1in]{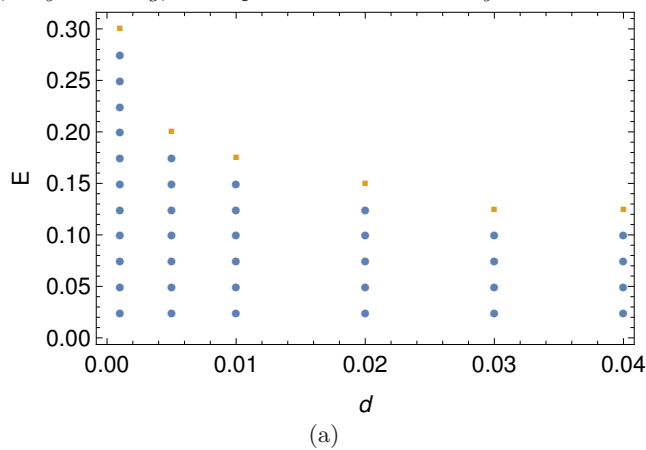}}      
          \caption{The region of instability for the smooth vortex (\ref{eq:ZTanh}) for varying $E$ and the smoothness parameter $d$, with $a=1$ and $m=2$; the domain is $r \in (0,r_\infty)$ with $r_\infty = 4a$. The circles denote the parameters for which a converged unstable mode was obtained, and the squares the parameters for which a converged unstable mode could not be found for $N$ upto $3000$.}\label{Fig:tanh_region}
\end{figure}

Figures \ref{Fig:Rankine_tanh}a and b show both results from the spectral code (red symbols), for the smooth vortex defined by (\ref{eq:ZTanh}) for $d=10^{-3}$, and that obtained from the shooting method for the Rankine profile. The close comparison of the growth rate  and the wave speed shows that an inertio-elastic instability exists for both a Rankine vortex and Rankine-like smooth vorticity profiles. For a smooth profile, one expects the transition width ($d$), the region over which the flow transitions from a rigid-body rotation to an irrotational straining one, to determine the existence of the instability. Expectedly, figure \ref{Fig:tanh_region} shows that instability exists for a narrower range of E with increasing $d$. \\ 

\begin{figure}
     \centering
     \subfloat[]{
          \includegraphics[height=2in]{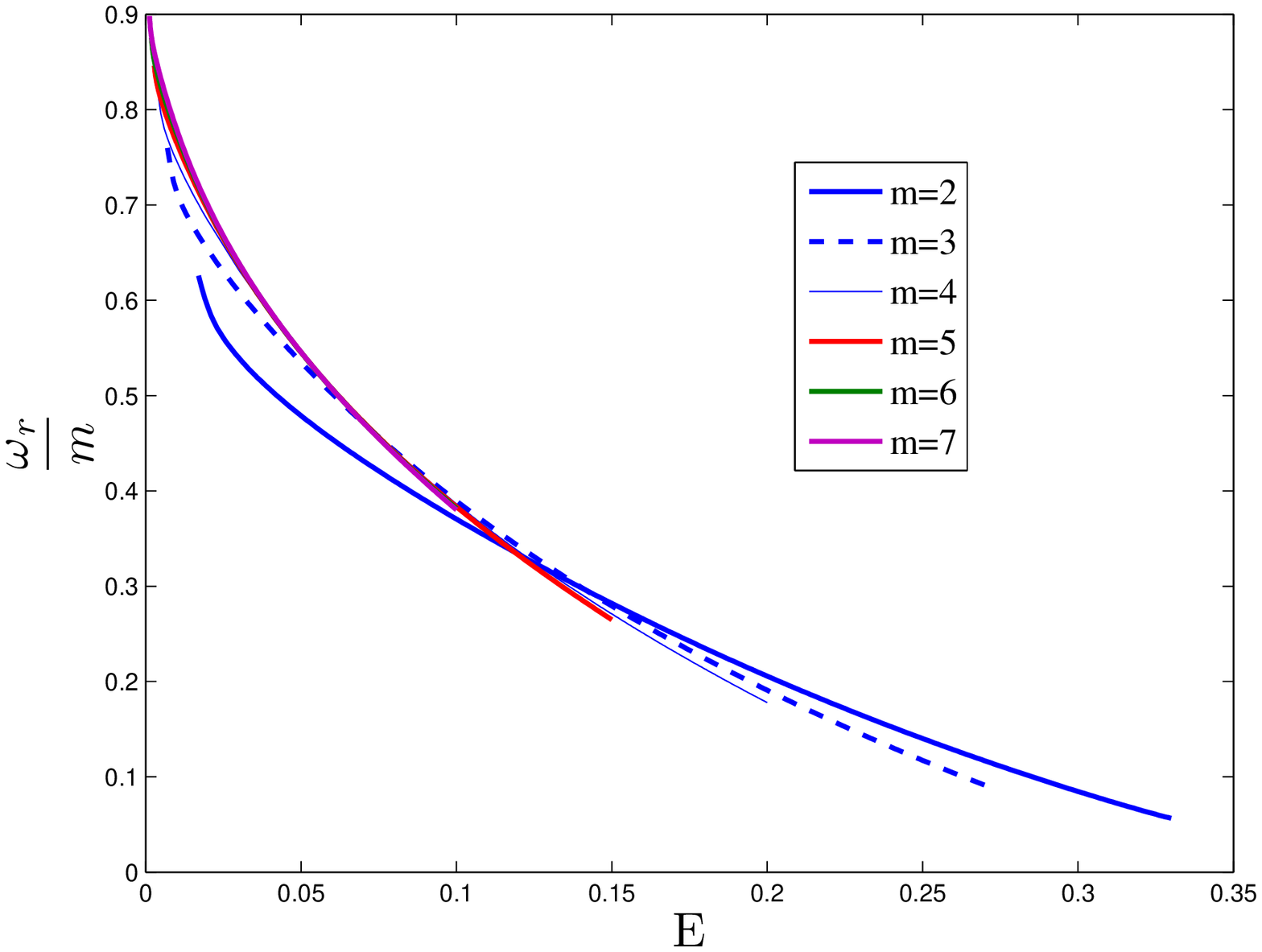}}
          \hspace{-.1in}
     \subfloat[]{
          \includegraphics[height=2in]{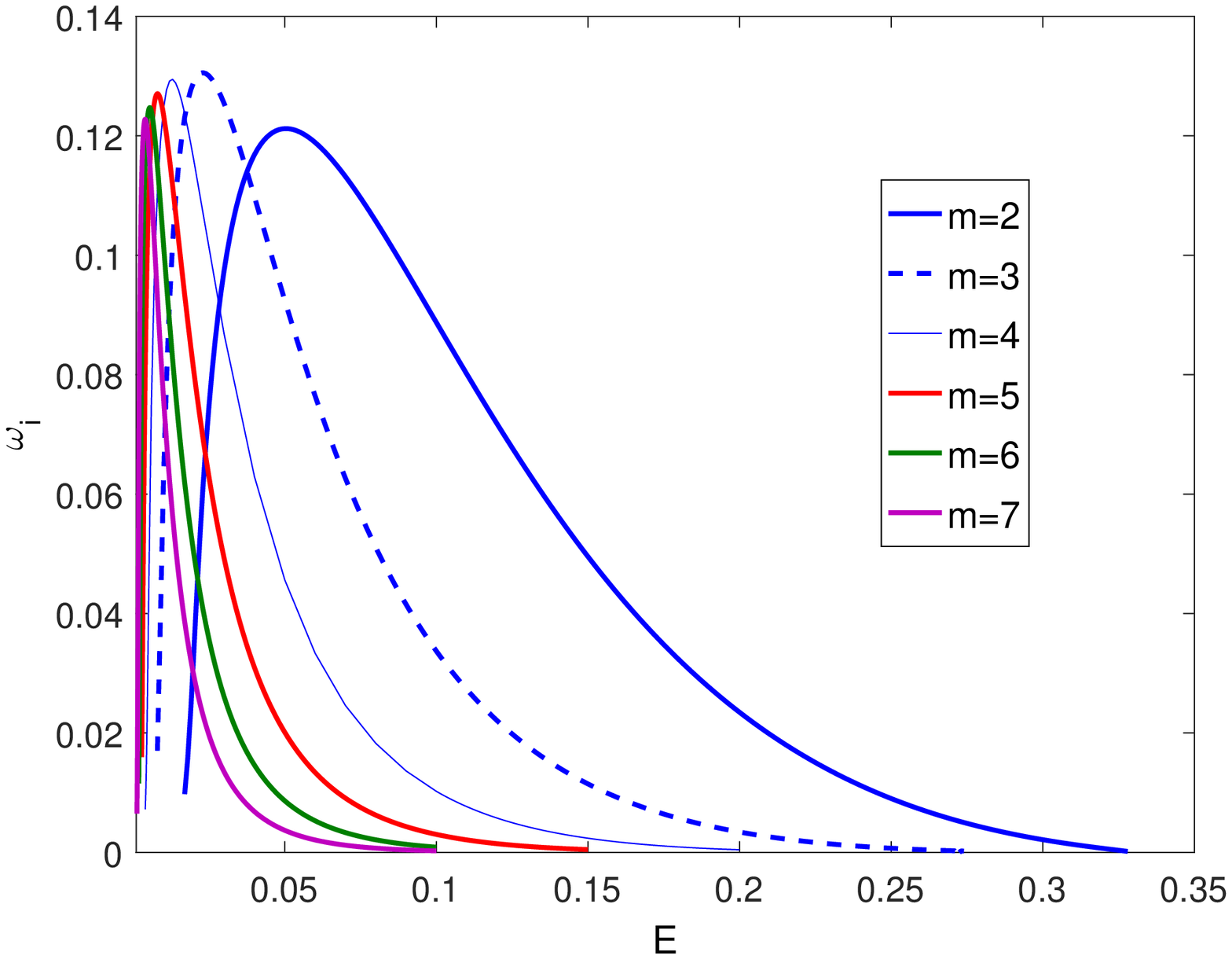}}\\
     \subfloat[]{
          \includegraphics[height=2in]{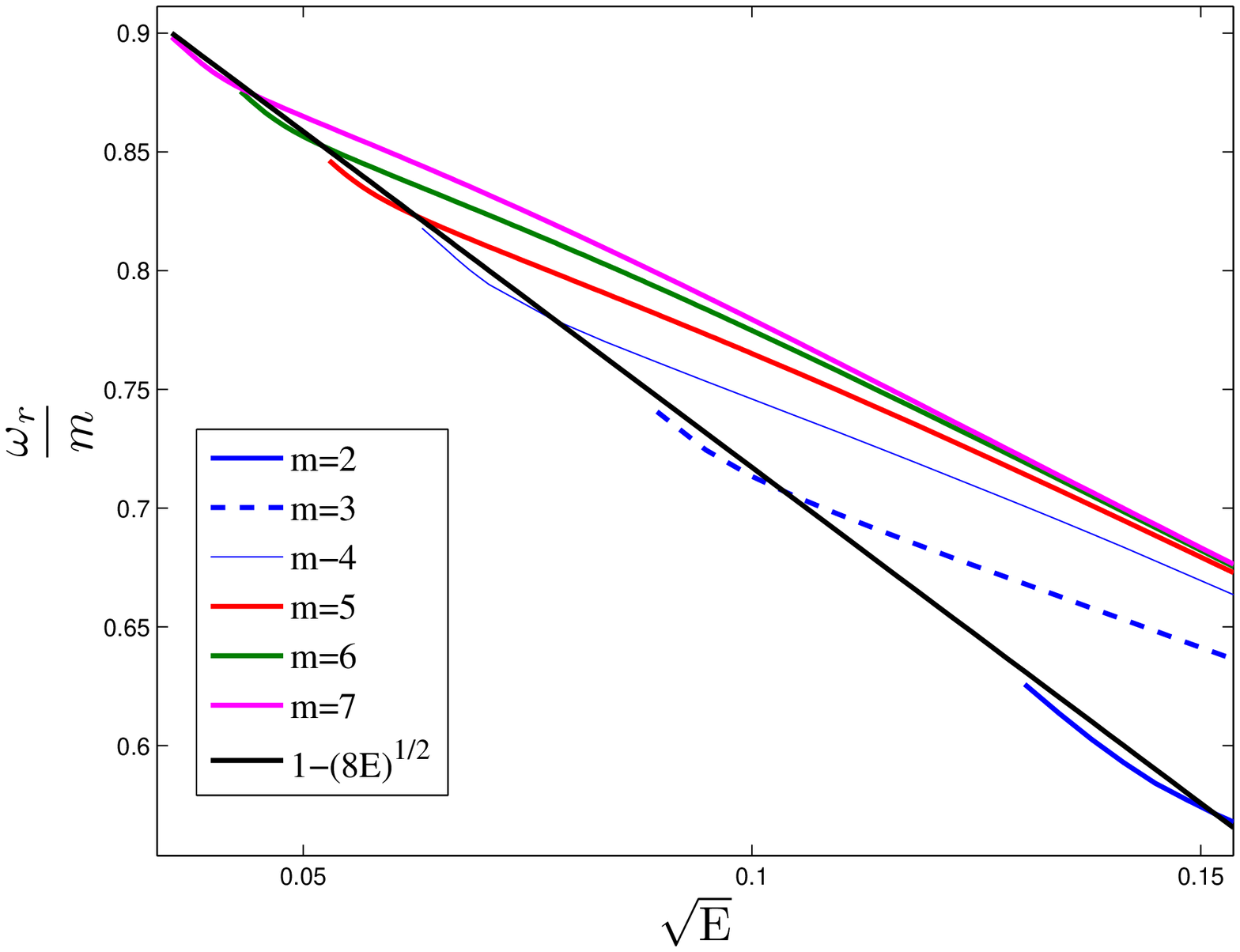}}
          \hspace{-.1in}
     \subfloat[]{
          \includegraphics[height=2in]{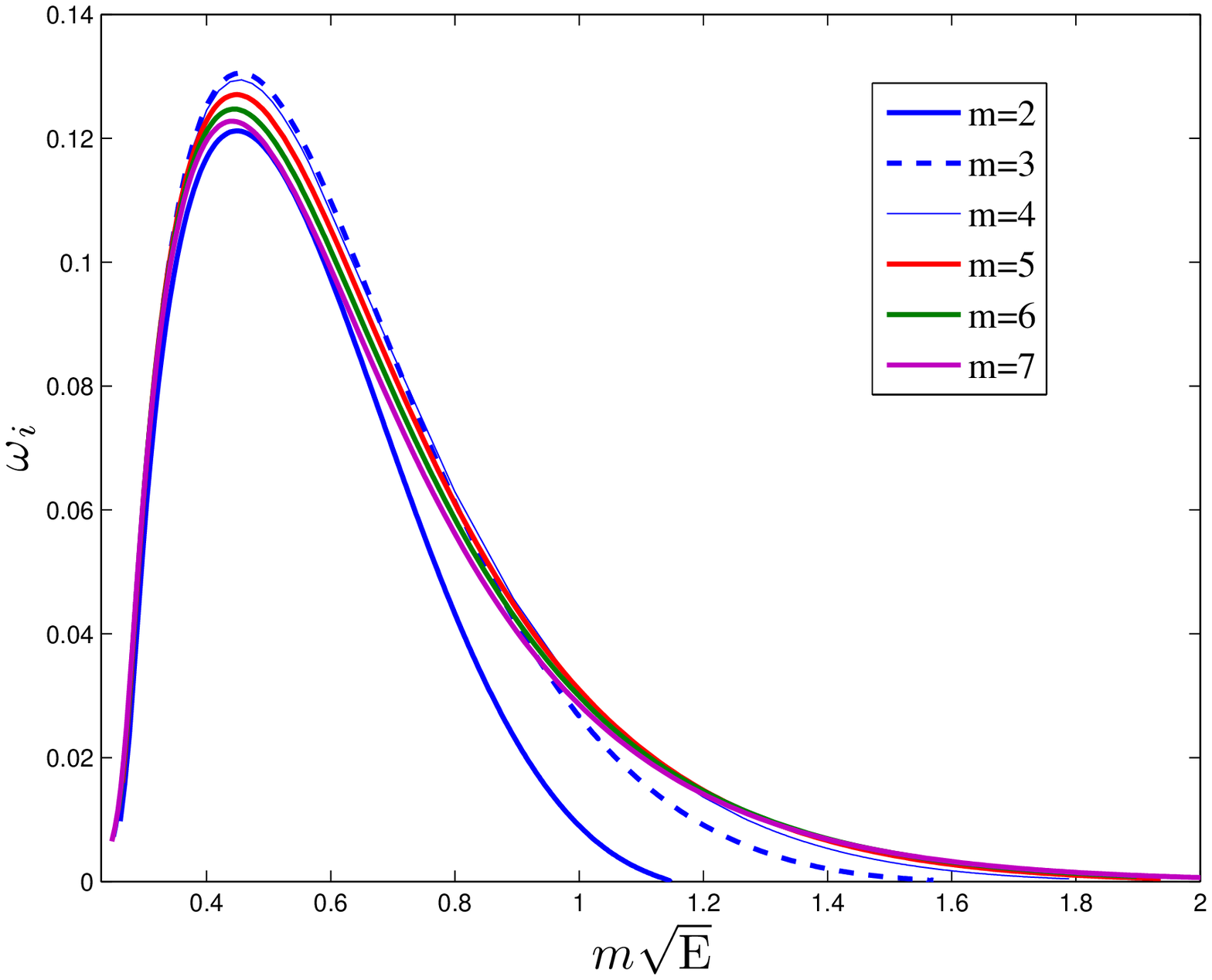}}        
          \caption{Dependence of the wave speed (a) and growth rate (b) on E for various values of $m$. In (c) the phase-speed is compared with the leading order asymptotic solution, $\omega_r=m(1-\sqrt{8\textrm{E}})$ (section \ref{smallE:analysis}), while (d) shows the collapse of the $\omega_i$ curves when plotted as a function of $m\ste$.}\label{Fig:Rankine_growth}
\end{figure}

From here on, we focus on the Rankine vortex, and thus all the results presented are computed using the shooting method. Figures \ref{Fig:Rankine_growth}a and b show the dependence of the wave speed and growth rate of the unstable mode on $\textrm{E}$ for different values of $m$. Figure \ref{Fig:Rankine_growth}c shows the convergence of the modal frequencies for different $m$ to a common asymptote, given by $1- \sqrt{8\textrm{E}}$, in the limit E $\rightarrow 0$. Figure \ref{Fig:Rankine_growth}d highlights the dependence of the growth rate on the re-scaled azimuthal wave-number $m\ste$. Note that the collapse occurs for $m\ste \sim$ O$(1)$ and $m$ sufficiently large, and accordingly, the curves for the lowest $m$'s $(m = 2$ and $3$) deviate from the scaled form. Figure \ref{Fig:Rankine_growth}d also shows a rather precipitous drop in growth rate for small $\textrm{E}$ leading to an eventual breakdown of the numerics below E $\sim 0.02$. As argued in the next section, the steep drop and the associated breakdown arise from a transcendental scaling of the eigenvalue for small $\textrm{E}$. \\
\begin{figure}
     \centering
     \subfloat[]{
          \includegraphics[height=2in]{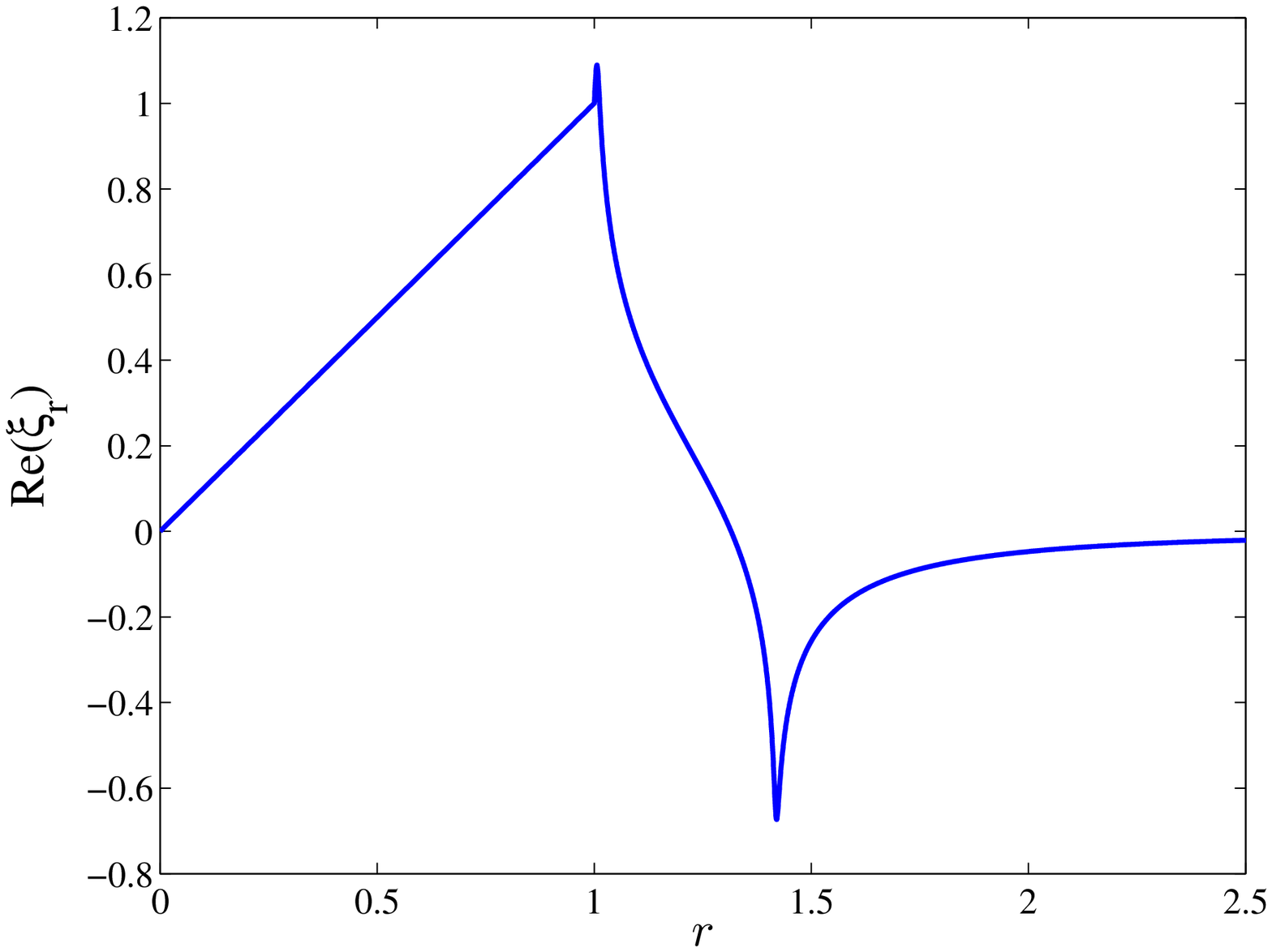}}
          \hspace{-.15in}
     \subfloat[]{
          \includegraphics[height=2in]{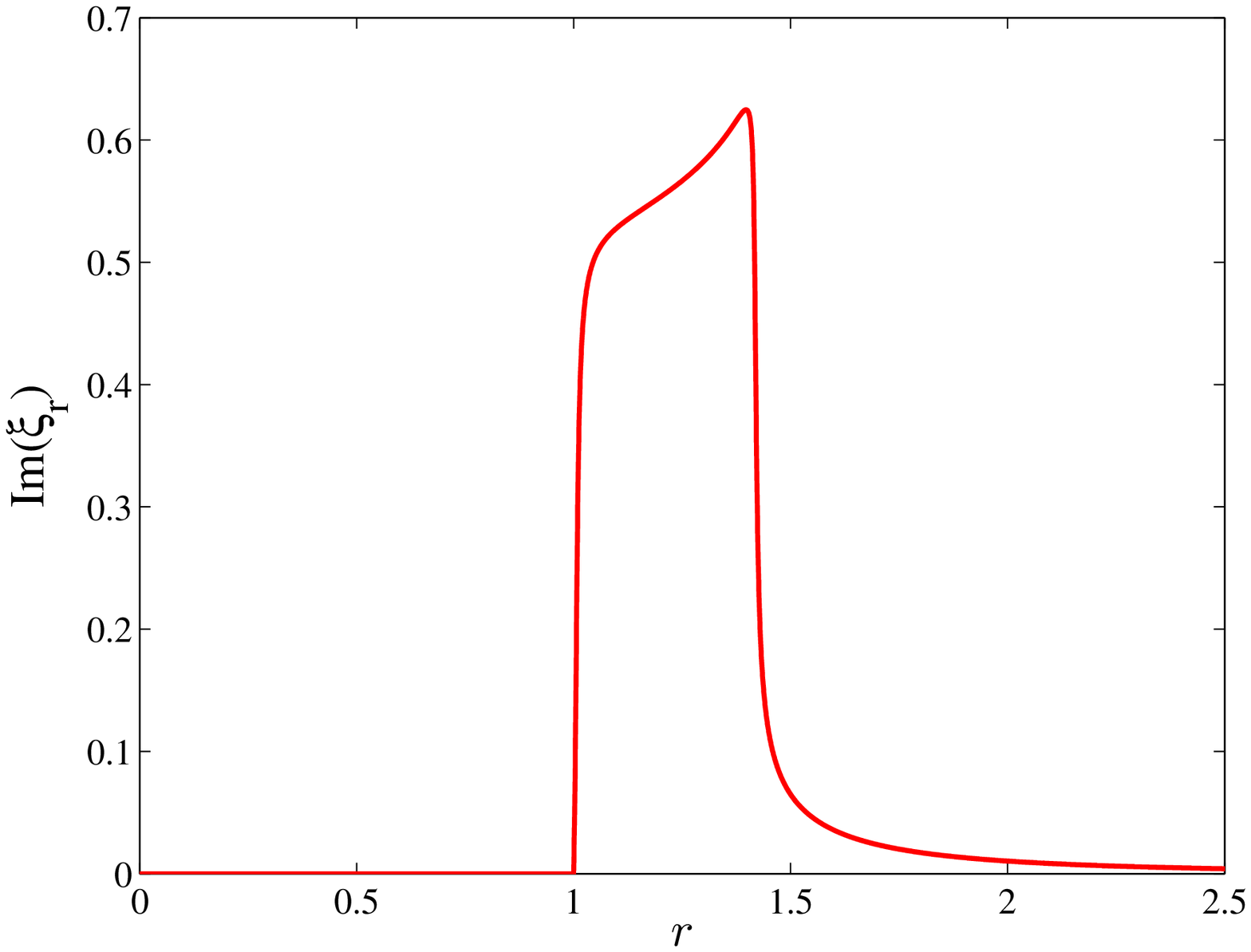}}\\
     \subfloat[]{
          \includegraphics[height=2in]{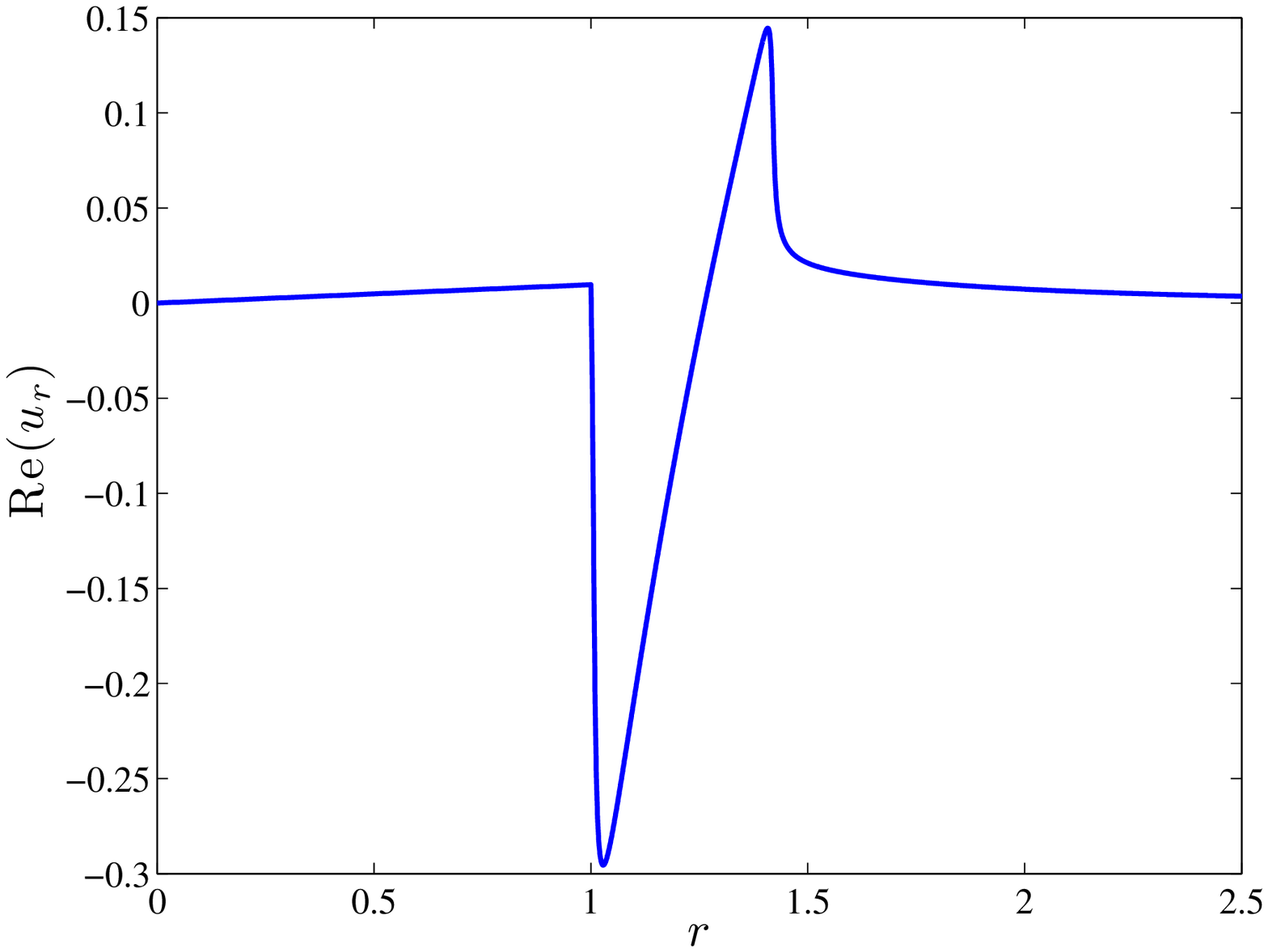}}
          \hspace{-.1in}
     \subfloat[]{
          \includegraphics[height=2in]{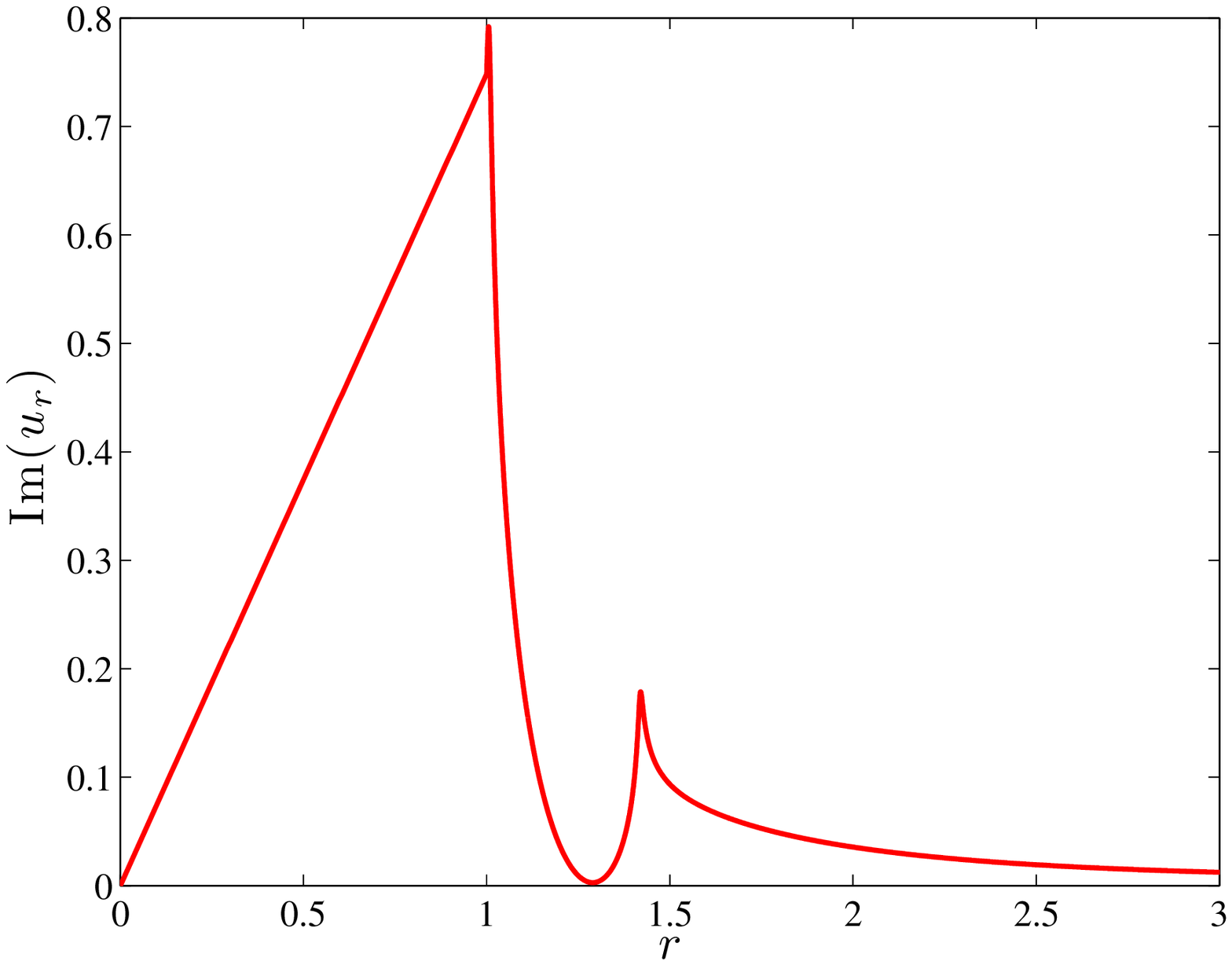}}        
          \caption{Real and imaginary parts of the radial displacement ((a) and (b)) and velocity eigenfunctions ((c) and (d)) for the unstable mode for $E=0.017$, $m$=2.}\label{Fig:Rankine_eigfunc}
\end{figure}

\begin{figure}
     \centering
          \includegraphics[height=2.5in]{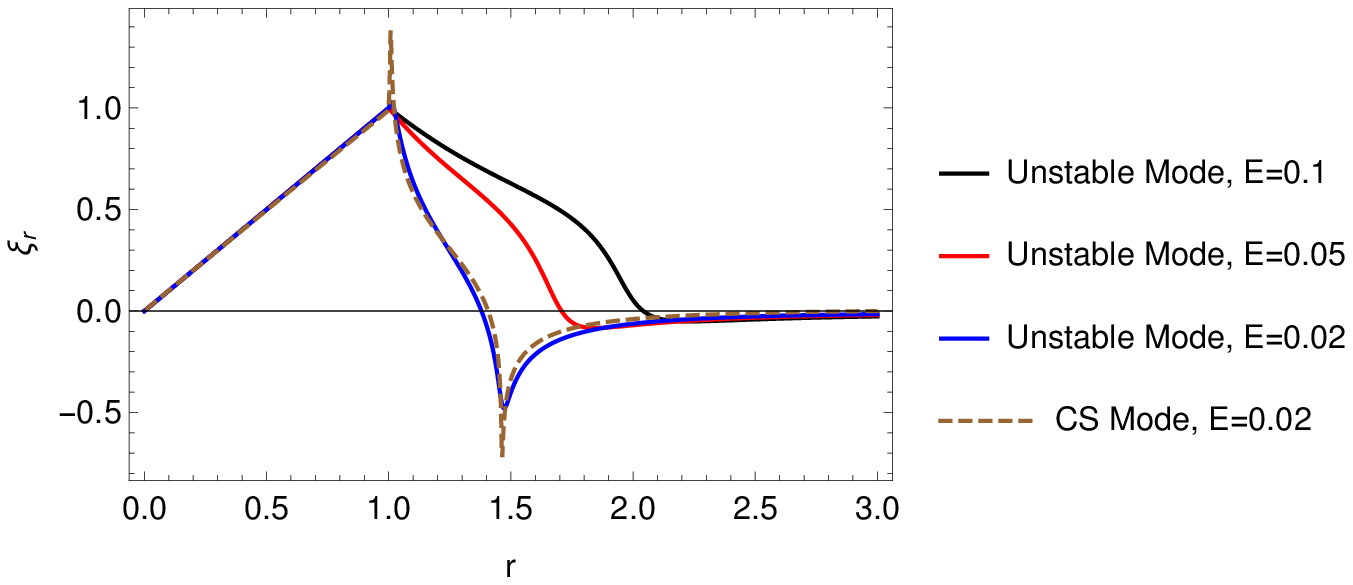}
          \caption{Real part of the radial displacement eigenfunction for the unstable mode for varying E; the continuous spectrum eigenfunction for E$=0.02$, and with the same wave-speed, is shown for comparison.}\label{Fig:eig_comp}
\end{figure}

The eigenfunctions for both radial displacement and velocity fields are shown in figure \ref{Fig:Rankine_eigfunc}. The twin peaks corresponding to the shear wave locations are clearly visible, and are reminiscent of the singular peaks for the CS-modes seen in section \ref{sec:CSrayleigh}. Figure \ref{Fig:eig_comp} quantifies this resemblance by plotting the real parts of the radial displacement eigenfunctions, for a sequence of E's approaching zero, along with a continuous spectrum eigenfunction for sufficiently small E. This resemblance, along with the conformance of the wave speed of the unstable mode to the elastic shear wave scaling ($\omega_r \sim 1- \sqrt{8\textrm{E}}$, see figure \ref{Fig:Rankine_growth}c), points to the inertio-elastic instability resulting from a resonant interaction between the elastic shear waves, which for small E propagate at nearly the same speed near the vortex core \citep{RALLHIN95, MILLER05}. The shear wave resonance argument also explains why the instability continues to exist for smooth Rankine-like profiles as is seen from figure \ref{Fig:tanh_region}.  \\

\subsection{A matched asymptotic analysis for the inertio-elastic instability of a Rankine vortex: $E\ll 1$} \label{smallE:analysis}
In this subsection, we analyze the elastic Rayleigh equation, (\ref{eq:xir}), for the Rankine profile. This profile exhibits a complete separation of the rotational and straining regions. The polymer molecules in the vortex core are unaffected by solid-body rotation, and retain their equilibrium distribution of conformations. Thus, the polymer stress is homogeneous and isotropic and the solution for $r < 1$ is identical to the inviscid case, with $u_r \propto r^{m-1}$. In order to obtain a normalized radial displacement, the constant of proportionality is chosen to be $\Sigma_0 = \omega-m$; as a result, $\xi = r^{m-1}$ and $\displaystyle\frac{d\xi}{dr}\displaystyle\Bigl|_{r=1-}\Bigr.=m-1$. \\

The radial displacement in the core, derived above, may now be used as part of a boundary condition for that in the exterior, governed by (\ref{eq:xir}) written in expanded form as:
\begin{eqnarray}
&&D\left[r^3\left\{(\omega-m\Omega)^2-2m^2\textrm{E}\Omega'^2\right\}D\xi\right]=r(m^2-1)\left\{(\omega-m\Omega)^2-2m^2\textrm{E}\Omega'^2\right\}\xi  \label{eq:xir_elas}. 
\end{eqnarray}\\
Noting that the radial displacement is continuous, that is, $\xi\Bigl|_{r=1+}\Bigr.=\xi\Bigl|_{r=1-}\Bigr.=1$ and integrating (\ref{eq:xir}) across $r=1$, one obtains:
\begin{eqnarray}
&&\left[r^3P\frac{d\xi}{dr}\right]_{r=1-}^{r=1+}=0\\
\Rightarrow&&\left\{\Sigma_0^2-8\textrm{E}(m\Omega_0)^2\right\}\frac{d\xi}{dr}\Bigl|_{r=1+}\Bigr.-\Sigma_0^2\frac{d\xi}{dr}\Bigl|_{r=1-}\Bigr.=0\\
\Rightarrow &&\frac{d\xi}{dr}\Bigl|_{r=1+}\Bigr.=\frac{(m-1)\,(\omega-m)^2}{(\omega-m)^2-8m^2\textrm{E}},
\end{eqnarray}
where the expression for the radial displacement gradient in the core is used in the last step. The above condition is combined with the usual requirement of decay in the far-field. To summarize the problem statement, the radial displacement in the irrotational exterior satisfies \eqref{eq:xir_elas}, and is subject to the following boundary conditions:
\begin{eqnarray}
\xi\Bigl|_{r=1}\Bigr.&=&1, \label{eq:bc_elas_1}\\
\frac{d\xi}{dr}\Bigl|_{r=1+}\Bigr.&=&\frac{(m-1)\,(\omega-m)^2}{(\omega-m)^2-8m^2\textrm{E}},\label{eq:bc_elas_2}\\
\xi&\rightarrow&0, \hspace{.2in}\textrm{as}\,\,r\rightarrow\infty.  \label{eq:bc_elas_3}
\end{eqnarray}

The collapse of the growth rate curves for various $m$, in figure \ref{Fig:Rankine_growth}d, suggests $m\ste$ as the parameter relevant to the asymptotics of the unstable mode for small $\textrm{E}$. This can also be seen in (\ref{eq:xir_elas}), where the term $(\omega-m\Omega)^2-2m^2\textrm{E}\Omega'^2$  involves the interplay of inertia and elasticity, with a balance implying $(\omega-m\Omega)\sim \pm m\sqrt{2\textrm{E}}\Omega'$. For a near-neutral mode, this balance occurs near the vortex core edge, $r\approx 1$, implying $(\omega-m)\sim \pm m\sqrt{8\textrm{E}}$ since $\Omega' =2$ at $r=1$. Thus, the elastic stresses for this mode are localized about the critical radius, $r_c\sim1+$O$(\ste)$. This suggests the introduction of a re-scaled boundary layer coordinate  $x = (r-1)/\ste$ with $\omega=m(1-a_1\ste)$, where $a_1$ is the unknown eigenvalue. In the limit $\textrm{E} \rightarrow 0$ with $m\ste$ fixed, this leads to the following simplified equation,
\begin{eqnarray}
\frac{d}{dx}\left[\left\{(a_1-2x)^2-8\right\}\frac{d\xi}{dx}\right]=m^2 \textrm{E} \left\{(a_1-2x)^2-8\right\}\xi, \label{OBL_eq1}
\end{eqnarray}
subject to the boundary conditions, 
 \begin{eqnarray}
\xi\Bigl|_{x=0}\Bigr.&=&1\\
\frac{d\xi}{dx}\Bigl|_{x=0}\Bigr.&=&\ste\frac{(m-1)\,(\omega-m)^2}{(\omega-m)^2-8m^2\textrm{E}} =m\ste\frac{\,a_1^2}{a_1^2-8}  \\
\xi&\rightarrow&0, \hspace{.2in}\textrm{as}\,\,x\rightarrow\infty .
\end{eqnarray}
The above equation is of the confluent Heun form, although recognition of this fact is not helpful from the perspective of obtaining closed form analytic solutions for $m\ste\sim $O$(1)$ \citep{RALLHIN95, RENARDY08}. \\

Thus, to make analytical progress we consider the alternate limit $\textrm{E}\ll 1$ for $m\sim$ O$(1)$. This also helps clarify the absence of a threshold $\textrm{E}$ for the instability - an aspect that, as already seen, isn't resolved by the numerics (as is evident in figure \ref{Fig:Rankine_growth}d). However, in this limit, apart from the obvious length scale of O($\ste$), a transcendentally small (in $\textrm{E}$) length scale emerges from consideration of the boundary conditions. To see this, consider the boundary layer equation \eqref{OBL_eq1} which takes the form,
\begin{eqnarray}
\frac{d}{dx}\left[\left\{(a_1-2x)^2-8\right\}\frac{d\xi}{dx}\right]=0 \label{OBL_eq0}.
\end{eqnarray}
The RHS of (\ref{OBL_eq1}) has been discarded, being asymptotically small for $m\sim O(1)$. (\ref{OBL_eq0}) has solutions of the form,
 \begin{eqnarray}
 \xi=c_1+c_2\log\left[\frac{2x-a_1-\sqrt{8}}{2x-a_1+\sqrt{8}}\right],
 \label{eq:OBL_eq0_sol}
 \end{eqnarray}  
 with the boundary conditions,
 \begin{eqnarray}
\xi\Bigl|_{x=0}\Bigr.&=&1 , \label{eq:bc_elast_1}\\
\frac{d\xi}{dx}\Bigl|_{x=0}\Bigr.&=&\ste\frac{(m-1)\,(\omega-m)^2}{(\omega-m)^2-8m^2\textrm{E}}=\ste\frac{(m-1)\,a_1^2}{a_1^2-8} , \label{eq:bc_elast_2}\\
\xi&\rightarrow&0, \hspace{.2in}\textrm{as}\,\,x\rightarrow\infty . \label{eq:bc_elast_3}
\end{eqnarray}
The far-field decay required by (\ref{eq:bc_elast_3}) implies that $c_1=0$ in \eqref{eq:OBL_eq0_sol}. Applying the gradient boundary condition, (\ref{eq:bc_elast_2}), we have $c_2=(m-1)a_1^2\ste/(8\sqrt{2}))$. Next, considering (\ref{eq:bc_elast_1}), one obtains 
 \begin{eqnarray}
  \xi\Bigl|_{x=0}=\frac{(m-1)a_1^2\ste}{8\sqrt{2}}\log\left[\frac{a_1+\sqrt{8}}{a_1-\sqrt{8}}\right]=1\label{eq:dirchlet_1}
 \end{eqnarray}
 Since $a_1\sim O(1)$ the above relation can only be satisfied if,
 \begin{eqnarray}
 a_1=\sqrt{8}+2\sqrt{8}\,e^{-\frac{1}{m-1}\sqrt{\frac{2}{\textrm{E}}}}a_2,
 \end{eqnarray}
where $a_2$ is an O$(1)$ constant. Although we have added a seemingly exponentially small quantity to the expected O$(1)$ estimate, this addition is crucial in the normal displacement boundary condition. With the transcendentally small addition, the normal displacement gradient, given by (\ref{eq:bc_elast_2}), is found to be exponentially large. This obviously contradicts the algebraic scaling assumed in (\ref{eq:bc_elast_2}), and highlights the subtle nature of the small $\textrm{E}$ limit. Physically, the locations $x=a_1+\sqrt{8}$ and $x=a_1-\sqrt{8}$ correspond, in re-scaled form, to the fore- and aft-moving shear wave singularities. Since $a_1-\sqrt{8}\sim e^{-\frac{1}{m-1}\sqrt{\frac{2}{\textrm{E}}}}$, the implication is that the aft-travelling shear wave is separated from the edge of the core only by a transcendentally small amount. Note that the above contradiction is not an artifact of the order in which we choose to satisfy the boundary conditions above. The underlying transcendental scaling also explains the precipitous drop in the growth rate (similar to that in figure \ref{Fig:Rankine_growth}d) of the elastic instability in a submerged jet observed by \cite{RALLHIN95} - see figure 7 therein. \\

\begin{figure}
     \centering
          \includegraphics[height=3.5in]{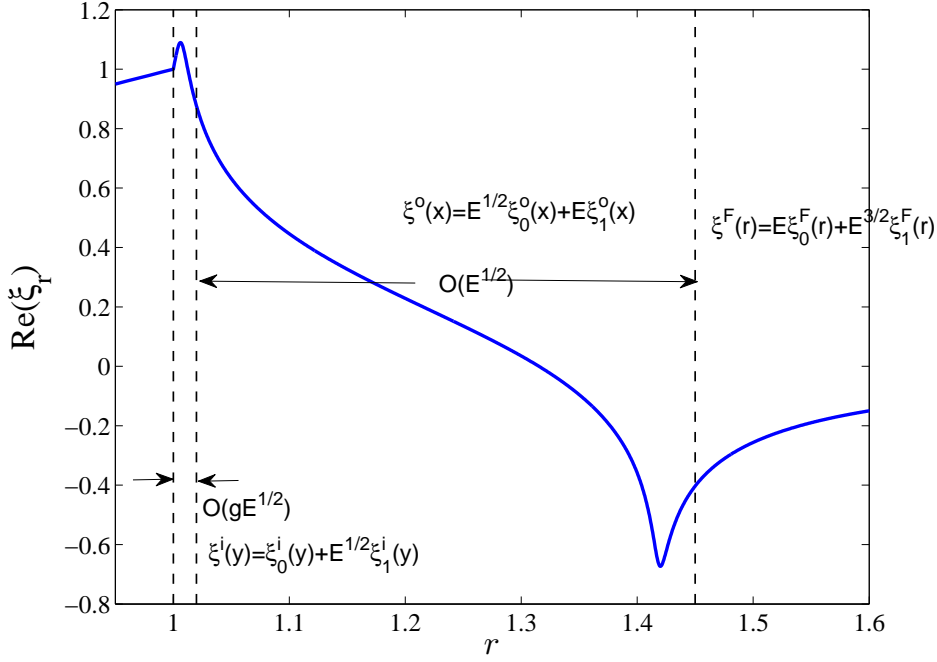}
          \caption{Numerically evaluated radial displacement eigenfunction illustrating the nested boundary layer structure discussed in the text: $r < 1$ - the vortex core, $r >1$ with $r - 1 \sim$ O$(g\sqrt{\textrm{E}})$ - the inner boundary layer, $r >1$ with $r-1\sim$ O$(\ste)$ - the outer boundary layer and $r >1$ with $r-1\sim$ O$(1)$ - the outer region.}\label{Fig:xir_BL}
\end{figure}

Thus, we see a crucial difference between the eigenfunction structure in the two limits analyzed above. For $\textrm{E} \rightarrow 0$ with $m\ste$ fixed, both the forward and backward travelling shear-waves lie within a boundary layer with a thickness of $\ste$, next to the vortex core, this being the only length scale of relevance. However, in the limit  $\textrm{E}\ll 1$ for $m\sim$ O$(1)$, the above arguments show that while the foreward-travelling shear wave is still localized in an O$(\ste)$ boundary layer, the backward-travelling wave is only separated from the edge of the core by a transcendentally small distance of O$ \left( e^{-\frac{1}{m-1}\sqrt{\frac{2}{\textrm{E}}}} \right)$. Thus, the perturbation analysis in this latter case needs to recognize two different length scales, one of them exponentially smaller than the other. The asymptotic framework must accordingly include an 
additional inner boundary layer with $r-1\sim$ O$(g\ste)$ satisfying the boundary conditions at $r=1$ ($g$ will turn out to be transcendentally small). The outer O$(\ste)$ boundary layer is therefore no longer constrained to satisfy the boundary conditions at $r=1$, resolving the contradiction in the naive approach above. Instead, it matches onto both the inner boundary layer and the outer regions in the appropriate limits. Figure \ref{Fig:xir_BL} shows the different asymptotic regions in a numerically evaluated radial displacement eigenfunction. We thus have the following double expansion, for the eigenvalue, in the limit $g \ll 1, \textrm{E} \ll 1$:
 \begin{eqnarray}
 \frac{\omega}{m}=1-\ste\left[\sqrt{8}+g \left\{c_0+c_1\sqrt{\textrm{E}}+c_2\textrm{E} + c_3\textrm{E}^{3/2}+\hdots\right\} +O(g^2) \right]\label{eq:full_omg_pred}.
\end{eqnarray}
Note that, in neglecting the contributions at O$(g^2)$ and higher, we anticipate the transcendental smallness of $g$ in \eqref{eq:full_omg_pred}, owing to which terms of O$(g)$ are, in principle, smaller than any algebraic order in E; that $g\sim e^{-\frac{1}{m-1}\sqrt{\frac{2}{\textrm{E}}}}$ emerges from the detailed analysis given below, and the expansion in \eqref{eq:full_omg_pred} conforms to the exponential asymptotics formalism \citep{boldyrev2009}. Note that the O$(g)$ contribution is crucial despite its transcendental smallness, since it contributes to the leading order growth rate. It is shown below that the growth rate is O$(\textrm{E}^2e^{-1/\ste})$ for small E, and therefore, also transcendentally small. Importantly, however, this establishes the absence of a threshold for the elastic instability in contrast to what is suggested by the numerical results above (figure \ref{Fig:Rankine_growth}d). 

The matched asymptotic expansions approach used here may be validated by consideration of an exactly soluble sub-problem - that governed solely by the LHS of (\ref{eq:xir_elas})- termed the LHS problem herein. The agreement of the asymptotic approach with the small-E expansion of the exact solution
serves as a useful validation; the analysis of the LHS problem is detailed in Appendix \ref{sec:app}. 

\subsubsection{Outer region- $r-1\sim O(1)$}
\label{sec:farsec}
To begin with, we study the solution in the region where $r-1\sim$ O$(1)$. Expanding $P$ in the elastic Rayleigh equation (\ref{eq:xir}) for small E, one obtains, 
\begin{equation}
\frac{P}{m^2}=\mathcal{S}_0+\ste \mathcal{S}_1+\textrm{E} \mathcal{S}_2 + \textrm{O}(g\ste),\\
\label{eq:P_exp}
\end{equation}
where $\mathcal{S}_0 = \left(1-\frac{1}{r^2}\right)^2$, $\mathcal{S}_1 = -2\sqrt{8}\left(1-\frac{1}{r^2}\right)$ and $\mathcal{S}_2 = 8\left(1-\frac{1}{r^6}\right)$. The expansion of $P$ has terms which scale algebraically with $\ste$ as well as terms that are transcendentally small (of O$(g)$). At leading order, we only need to consider the finite number (three) of terms in \eqref{eq:P_exp} that scale algebraically. The unknown eigenvalue constants (the $c_i$s) in \eqref{eq:full_omg_pred} are transcendentally small and hence do not enter the expansion for $P$ and the outer region analysis at leading order. Denoting the displacement by $\xi^{F}(r)$, the above points to the following expansion,
\begin{eqnarray}
\xi^F(r)=\textrm{E} \,\xi^F_0(r)+\textrm{E}^{3/2}\,\xi^F_1(r) + \textrm{O}(\textrm{E}^{2})+ \textrm{O}(g\ste),
\label{eq:xif}
\end{eqnarray}
where we again neglect the transcendentally smaller contributions. While the O(E) scaling of the leading order term is anticipated from the matching considerations below (see section \ref{sec:rhs_elas_match}), it may nevertheless be anticipated based on the discussion above. The aft-shear-wave is located in an exponentially small neighborhood of the edge of the core, and the (logarithmic) displacement due to this wave  decreases by O$(\ste)$ in the outer boundary layer which is exponentially distant when measured in units of the inner boundary layer thickness of O$(g \ste)$. Further, the near-cancellation between the O$(\ste)$ displacements associated with the fore- and aft-shear waves for $r - 1\sim$ O$(1)$ leads to an asymptotically smaller $\xi^F$ of O(E) in this outer region (see figure \ref{Fig:xir_BL}), as assumed in the expansion above. Plugging \eqref{eq:xif} in (\ref{eq:xir_elas}), one obtains the following equations at successive orders whose solutions are written alongside.

At O${(\textrm{E})}$,
\begin{equation}
\displaystyle\frac{d}{dr}\left[r^3S_0\displaystyle\frac{d\xi_0^F}{dr}\right] - S_0\xi_0^F r (m^2-1)=0,
\end{equation}
with $\xi_0^F\rightarrow 0$ for $r\rightarrow\infty$, which gives $\xi_0^F(r)=\displaystyle\frac{\mathcal{B}_0}{r^{m-1}(r^2-1)}$. Similarly, at O${(\textrm{E}^{3/2})}$,
\begin{equation}
\displaystyle\frac{d}{dr}\left[r^3 S_0\displaystyle\frac{d\xi_1^F}{dr}\right] - S_0\xi_1^F r (m^2-1) = -\displaystyle\frac{d}{dr}\left[r^3 S_1\displaystyle\frac{d\xi_0^F}{dr}\right] + S_1\xi_0^F   r (m^2-1),
\end{equation}
with $\xi_1^F\rightarrow 0$ for $r\rightarrow\infty$, which gives $\xi_1^F(r)=\displaystyle\frac{2\sqrt{2}\mathcal{B}_0}{r^{m-1}(r^2-1)^2}+\displaystyle\frac{\mathcal{B}_1}{r^{m-1}(r^2-1)}$.\\

The $\mathcal{B}_{i}$'s in the above expressions are integration constants which will be determined from matching considerations (see section \ref{sec:rhs_elas_match}). Using the expressions for $\xi_0^F$ and $\xi_1^F$ above, the solution in the outer region, to O$(\textrm{E}^{3/2})$, may be written as:
\begin{eqnarray}
\xi^F(r)&=&\textrm{E}\displaystyle\frac{\mathcal{B}_0}{r^{m-1}(r^2-1)}+\textrm{E}^{3/2}\left\{\displaystyle\frac{2\sqrt{2}\mathcal{B}_0}{r^{m-1}(r^2-1)^2}+\displaystyle\frac{\mathcal{B}_1}{r^{m-1}(r^2-1)}\right\}+\textrm{O}(\textrm{E}^2). \label{eq:_xi}
\end{eqnarray}
In the outer region solution ($\xi^F$), there are no signatures of the travelling shear-wave singularities. Note that we do not consider the O($\textrm{E}^2$) and higher order contributions to $\xi^F$ since they are not required to determine the growth rate at leading order, an insight that is obtained from the solution of the LHS problem in Appendix \ref{sec:app}.

\subsubsection{Outer boundary layer - $r-1\sim O(\ste)$}
\label{sec:outersec}
Having found the solution in the outer region, we now consider the outer boundary layer using the boundary layer coordinate $x = (r-1)/\ste $ already introduced at the beginning of this section. Denoting the boundary layer displacement as $\xi(r)=\xi^{o}(x)$, one obtains the rescaled elastic Rayleigh equation as:
\begin{eqnarray}
\frac{d}{dx}\left[\mathcal{Q}\frac{d\xi^o}{dx}\right]=\frac{\textrm{E}\mathcal{Q}(m^2-1)}{(1+\sqrt{\textrm{E}}x)^2}\xi^o, \label{eq:xir_OBL}
\end{eqnarray}
where $\mathcal{Q}=\displaystyle\frac{(1+\sqrt{\textrm{E}}x)^3P}{m^2\textrm{E}}$, and is further expanded as,
\begin{equation}
\mathcal{Q}=\mathcal{Q}_0+\ste \mathcal{Q}_1+\textrm{E} \mathcal{Q}_2 +\textrm{E}^{3/2} \mathcal{Q}_3+  \textrm{O}(\textrm{E}^{2})+ \textrm{O}(g\ste),
\end{equation}
with $\mathcal{Q}_0=4x(x-2\sqrt{2})$, $\mathcal{Q}_1=-12\sqrt{2}x(x-2\sqrt{2})$, $\mathcal{Q}_2=x^2(x^2-4\sqrt{2}x-24)$ and $\mathcal{Q}_3=x^3(88-x^2)$. Similar to the expansion of $P$ discussed earlier in section \ref{sec:farsec}, the transcendentally small terms can be neglected in the expansion of $Q$. 
The displacement $(\xi^o)$ may thus be expanded as,
\begin{eqnarray}
\xi^o(x)=\ste \,\xi^o_0(x)+\textrm{E}\,\xi^o_1(x)+\textrm{E}^{3/2}\,\xi^o_2(x) +\textrm{E}^{2}\,\xi^o_3(x) + \textrm{O}(\textrm{E}^{5/2})+ \textrm{O}(g\ste) .
\end{eqnarray}
where where the leading $O(\ste)$ scaling is consistent with the aforementioned physical arguments. Substituting the above expansion in (4.26), one obtains the following equations at successive orders whose solutions are written alongside. \\
At O${(\textrm{E}^{1/2})}$ we obtain,
\begin{equation}
\displaystyle\frac{d}{dx}\left[\mathcal{Q}_0\displaystyle\frac{d\xi_0^o}{dx}\right]=0
\end{equation}
\begin{equation}
\Rightarrow\,\,\xi_0^o(x)=\mathcal{G}_{10}+\mathcal{G}_{11}\log\left(\displaystyle\frac{x-2\sqrt{2}}{x}\right).
\label{eq:outer0}
\end{equation}
In $\xi_0^o(x)$ above, and in the solutions at higher orders below, the forward and backward travelling wave singularities correspond to $x = 2 \sqrt{2}$ and $x = 0$, respectively; the latter location is the edge of the core, since transcendentally small terms are now neglected. \\
At O${(\textrm{E})}$ we obtain,
\begin{equation}
\displaystyle\frac{d}{dx}\left[\mathcal{Q}_0\displaystyle\frac{d\xi_1^o}{dx}\right]=-\displaystyle\frac{d}{dx}\left[\mathcal{Q}_1\displaystyle\frac{d\xi_0^o}{dx}\right],
\end{equation}
\begin{equation}
\Rightarrow\,\,\xi_1^o(x)=\mathcal{G}_{21}+\mathcal{G}_{11}\log\left(\displaystyle\frac{x-2\sqrt{2}}{x}\right).
\label{eq:outer1}
\end{equation}
Note that $\xi_1^o$ is again the homogeneous solution since $\mathcal{Q}_1$ has the same $x$-dependence as $\mathcal{Q}_0$.
At O${(\textrm{E}^{3/2})}$ we obtain,
\begin{equation}
\displaystyle\frac{d}{dx}\left[\mathcal{Q}_0\displaystyle\frac{d\xi_2^o}{dx}\right]=-\displaystyle\frac{d}{dx}\left[\mathcal{Q}_1\displaystyle\frac{d\xi_1^o}{dx}\right]-\displaystyle\frac{d}{dx}\left[\mathcal{Q}_2\displaystyle\frac{d\xi_0^o}{dx}\right]+(m^2-1) \mathcal{Q}_0 \xi_0^o,
\end{equation}
\begin{align}
\Rightarrow\,\,\xi_2^o(x)&=\mathcal{G}_{22}+\mathcal{G}_{12}\log\left(\displaystyle\frac{x-2\sqrt{2}}{x}\right)-\displaystyle\frac{\mathcal{G}_{10}}{\sqrt{2}}\left(\displaystyle\frac{32}{x-2\sqrt{2}}+x-2\sqrt{2}\right) - \mathcal{G}_{10} (m^2-1) \int \displaystyle\frac{x^2-3\sqrt{2}x}{3(x-2\sqrt{2})} \log \left( x\right) dx \nonumber\\
 &+\mathcal{G}_{10} (m^2-1) \int \left(\displaystyle\frac{x^2-3\sqrt{2}x}{3(x-2\sqrt{2})}+\displaystyle\frac{8\sqrt{2}}{3x(x-2\sqrt{2})} \right) \log \left(x-2\sqrt{2} \right) dx \nonumber\\
&- \mathcal{G}_{10} (m^2-1) \left(\displaystyle\frac{\sqrt{2}x}{3} - \displaystyle\frac{4}{3} \log \left( \displaystyle\frac{x-2\sqrt{2}}{x} \right) \right).
\label{eq:outer2}
\end{align}
At O${(\textrm{E}^{2})}$ we obtain,
\begin{equation}
\displaystyle\frac{d}{dx}\left[\mathcal{Q}_0\displaystyle\frac{d\xi_3^o}{dx}\right]=-\displaystyle\frac{d}{dx}\left[\mathcal{Q}_1\displaystyle\frac{d\xi_2^o}{dx}\right]-\displaystyle\frac{d}{dx}\left[\mathcal{Q}_2\displaystyle\frac{d\xi_1^o}{dx}\right]-\displaystyle\frac{d}{dx}\left[\mathcal{Q}_3\displaystyle\frac{d\xi_0^o}{dx}\right]+(m^2-1) \mathcal{Q}_1 \xi_0^o - 2 x (m^2-1) \mathcal{Q}_0 \xi_0^o+(m^2-1) \mathcal{Q}_0 \xi_1^o,
\end{equation}
\begin{align}
\Rightarrow\,\,\xi_3^o(x)&=\mathcal{G}_{23}+\mathcal{G}_{13}\log\left(\displaystyle\frac{x-2\sqrt{2}}{x}\right) \nonumber\\
&+3\sqrt{2}\mathcal{G}_{12}\log\left(\displaystyle\frac{x-2\sqrt{2}}{x}\right)-3\mathcal{G}_{10}\left(x+\displaystyle\frac{32}{x-2\sqrt{2}}\right)-\displaystyle\frac{\mathcal{G}_{11}}{\sqrt{2}}\left(\displaystyle\frac{32}{x-2\sqrt{2}}+x-2\sqrt{2}\right)\nonumber\\
&+\displaystyle\frac{\mathcal{G}_{10}}{\sqrt{2}}\left(\displaystyle\frac{(x-2\sqrt{2})^2}{2}+6\sqrt{2}\left(x-2\sqrt{2}\right)-64\log\left(x-2\sqrt{2}\right)+\displaystyle\frac{160\sqrt{2}}{x-2\sqrt{2}}\right) \nonumber\\
&- (\mathcal{G}_{11}-3\mathcal{G}_{10})  (m^2-1) \int \displaystyle\frac{x^2-3\sqrt{2}x}{3(x-2\sqrt{2})} \log \left( x\right) dx \nonumber\\
 &+(\mathcal{G}_{11}-3\mathcal{G}_{10}) (m^2-1) \int \left(\displaystyle\frac{x^2-3\sqrt{2}x}{3(x-2\sqrt{2})}+\displaystyle\frac{8\sqrt{2}}{3x(x-2\sqrt{2})} \right) \log \left(x-2\sqrt{2} \right) dx \nonumber\\
 &- (\mathcal{G}_{11}-3\mathcal{G}_{10}) (m^2-1) \left( \displaystyle\frac{\sqrt{2}x}{3} - \displaystyle\frac{4}{3} \log \left( \displaystyle\frac{x-2\sqrt{2}}{x} \right) \right) \nonumber\\
 &+\displaystyle\frac{1}{3}(m^2-1)\mathcal{G}_{21} \left(\displaystyle\frac{x^2}{2}- \sqrt{2} x -4 \log \left( x- 2 \sqrt{2} \right) \right)\nonumber\\
 &+\displaystyle\frac{\sqrt{2}}{3}(m^2-1)\mathcal{G}_{10} \left( \displaystyle\frac{x^2}{2} + \sqrt{2} x -4 \log \left( x- 2 \sqrt{2} \right) \right) \nonumber\\
 &+\mathcal{G}_{10}  (m^2-1) \int \displaystyle\frac{3 x^3-8\sqrt{2}x^2}{6(x-2\sqrt{2})} \log \left( x\right) dx \nonumber\\
 &-\mathcal{G}_{10}  (m^2-1) \int \left( \displaystyle\frac{3 x^3-8\sqrt{2}x^2}{6(x-2\sqrt{2})} +\displaystyle\frac{32}{3 x(x-2\sqrt{2})} \right) \log \left( x-2\sqrt{2}\right) dx. 
 \label{eq:outer3}
\end{align}
The constants $\mathcal{G}_{10},\mathcal{G}_{11},\hdots$ in \eqref{eq:outer0}-\eqref{eq:outer3} are determined via matching in section \ref{sec:rhs_elas_match}. In deriving the limiting forms of the above solutions for $x \rightarrow 0$, required for matching, one has to account for the multivaluedness of the logarithm in the displacement fields ($\xi_0^o(x), \xi_1^o(x), \hdots$) in (\ref{eq:outer0}) - (\ref{eq:outer3}). Recall that $x = 2 \sqrt{2}$ marks the location (singularity) of the neutrally stable forward travelling shear-wave. The neutral stability arises because the (imaginary) growth rate appears at a higher (and transcendentally small) order in the perturbation expansion. This leads to an ambiguity in the phase jump associated with the logarithm, that is well known in inviscid hydrodynamic stability \citep{DRAZINREID81}. The resolution involves displacing the aforementioned shear-wave singularity off the real axis to $x = 2 \sqrt{2} - i \epsilon$, with $\epsilon > 0$ representative of the small but finite growth rate. As a result, one has:
\begin{eqnarray}
\log(x-2\sqrt{2} + i \epsilon)&=&\log|x-2\sqrt{2}|\hspace{.5in}x>2\sqrt{2},\nonumber\\
&=&\log|x-2\sqrt{2}|+i \pi \hspace{.2in}x<2\sqrt{2}.
\end{eqnarray}
The above relation may now be used in obtaining the limiting forms of the outer boundary layer solutions.

\subsubsection{Inner boundary layer - $r-1 \sim$ O$(g\ste)$}
\label{sec:innersec}
Finally, we introduce an inner boundary layer in an exponentially small neighborhood of the core, corresponding to O$(1)$ values of the boundary layer coordinate $y = (r-1)/g \ste $ with $g, \textrm{E} \ll 1$. Denoting the inner boundary layer displacement as $\xi(r)=\xi^{i}(y)$, we have from (\ref{eq:xir_elas})-(\ref{eq:bc_elas_2}): 
\begin{eqnarray}
\frac{d}{dy}\left[\mathcal{R}\frac{d\xi^i}{dy}\right]=\frac{g^2\textrm{E}\mathcal{R}(m^2-1)}{(1+g\ste y)^2}\xi^i, \label{eq:xir_IBL}
\end{eqnarray}
with $\mathcal{R}=(1+g\ste y)^3P/(m^2g\textrm{E})$, which is further expanded as,
\begin{equation}
\mathcal{R}=\mathcal{R}_0+\ste \mathcal{R}_1+ \textrm{O}(\textrm{E}) ,
\end{equation}
where $\mathcal{R}_0=4\sqrt{2}(c_0-2y)$ and $\mathcal{R}_1=4\sqrt{2}(c_1+6\sqrt{2}y)$. Note that this is the first instance where the unknown eigenvalue constants (the $c_i$s) enter the expansion. In anticipation of the transcendental smallness of $g$, we assume $g\textrm{E}^{-\alpha}\rightarrow 0$ as $\textrm{E}\rightarrow 0$, $\forall\, \alpha>0$, which allows the neglect of the RHS term in (\ref{eq:xir_elas}) at all orders. The boundary conditions at the core-exterior interface, (\ref{eq:bc_elas_1}) and (\ref{eq:bc_elas_2}), take the form:
\begin{eqnarray}
\xi^i(y=0)=1 \label{eq:IBL_BC1}\\
\frac{d\xi^i}{dy}\Bigl(y=0\Bigr)&=&(m-1)\ste\left\{\frac{\sqrt{2}}{c_0}-\frac{\sqrt{2}c_1\ste}{c_0^2} +\hdots\right\} . \label{eq:IBL_BC2}
\end{eqnarray}
Guided by the above expansions the boundary layer variable, $\xi^i(y)$, is expanded as follows:
\begin{eqnarray}
\xi^i(y)=\xi_0^i(y)+\ste\, \xi_{1}^i(y)+\textrm{E}\, \xi_2^i(y)+ \textrm{O}(\textrm{E}^{3/2})+ \textrm{O}(g\ste),
\label{eq:innerexp}
\end{eqnarray}
where we note that the radial displacement is now O$(1)$, as dictated by the boundary condition (\ref{eq:IBL_BC1}). Plugging the above expansion in (\ref{eq:xir_IBL}) and using the boundary conditions (\ref{eq:IBL_BC1}) and (\ref{eq:IBL_BC2}), we have the following equations (and boundary conditions) and solutions at successive orders.\\
At O${(1)}$ we obtain,
\begin{equation}
\displaystyle\frac{d}{dy}\left[\mathcal{R}_0\displaystyle\frac{d\xi_0^i}{dy}\right]=0,
\end{equation}
with $\xi_0^i(y=0)=1$ and $\displaystyle\frac{d\xi_0^i}{dy}(y=0)=0$, which gives
\begin{equation}
\xi_0^i(y)=1 .
\label{eq:inner0}
\end{equation}
At O${(\textrm{E}^{1/2})}$ we obtain, 
\begin{equation}
\displaystyle\frac{d}{dy}\left[\mathcal{R}_0\displaystyle\frac{d\xi_1^i}{dy}\right]=0,
\end{equation}
with $\xi_1^i(y=0)=0$ and $\displaystyle\frac{d\xi_1^i}{dy}(y=0)=\displaystyle\frac{\sqrt{2}(m-1)}{c_0}$, which gives,
\begin{equation}
\xi_1^i(y)=-\displaystyle\frac{(m-1)}{\sqrt{2}}\log\left(\frac{c_0-2y}{c_0}\right).
\end{equation}
From the expression for $\xi_1^i(y)$ and the solutions at higher orders below, we see that the singularity associated with the backward travelling shear-wave is now resolved, and corresponds to $y = c_0/2$ ($x = g c_0/2$), where $c_0$ still needs to be determined. \\
At O${(\textrm{E})}$ we obtain, 
\begin{equation}
\displaystyle\frac{d}{dy}\left[\mathcal{R}_0\displaystyle\frac{d\xi_2^i}{dy}\right]=-\displaystyle\frac{d}{dy}\left[\mathcal{R}_1\displaystyle\frac{d\xi_1^i}{dy}\right],
\end{equation}
with $\xi_2^i=0$ and $\displaystyle\frac{d\xi_2^i}{dy}=-\displaystyle\frac{\sqrt{2}(m-1)c_1}{c_0^2}$, which gives,
\begin{equation}
\xi_2^i(y)=-3(m-1)\log\left(\displaystyle\frac{c_0-2y}{c_0}\right)-\displaystyle\frac{(m-1)}{2}(\sqrt{2}c_1+6c_0)\left\{\displaystyle\frac{1}{c_0-2y}-\frac{1}{c_0}\right\}.
\label{eq:inner2}
\end{equation}
The expansion (\ref{eq:innerexp}) with the solutions given by (\ref{eq:inner0})-(\ref{eq:inner2}),  satisfies both the boundary conditions at the vortex core ($y=0$), and further, needs to be matched to the solution in the outer boundary layer, which will provide us with values of the constants $c_0,c_1,\hdots$ in the eigenvalue expansion. In the matching region ($y\gg1$), one again needs to account for the phase jump associated with the logarithm. This is done by noting that the singularity of the unstable mode, associated with the backward travelling shear-wave, lies off the real axis at $y = c_0/2 - i \epsilon^{\prime}$  $(\epsilon^{\prime} > 0)$, and one has therefore the relations: 
\begin{eqnarray}
\log(c_0-2y - 2 i \epsilon^{\prime})&=&\log|c_0-2y|\hspace{.5in}y<\frac{c_0}{2}\nonumber\\
&=&\log|c_0-2y| - i \pi \hspace{.2in}y>\frac{c_0}{2} .
\end{eqnarray}

\subsubsection{Matching}\label{sec:rhs_elas_match}
With the inner, outer boundary layer and far-field solutions in place, we proceed to derive the necessary constants via matching each of the solutions. First we expand the outer region solutions for small values of $r-1$, writing it in terms of the outer boundary layer coordinate, $x=(r-1)/\ste$, \\
\begin{eqnarray}
\textrm{E} \xi_0^F &\sim& \ste \frac{\mathcal{B}_0}{2x} \left(1-  \frac{\ste x}{2} (2m-1)  + O\left(x^2\right) \right) ,
\end{eqnarray}
\begin{eqnarray}
\textrm{E}^{3/2} \xi_1^F &\sim& \textrm{E} \frac{\mathcal{B}_1}{2x} \left(1-  \frac{\ste x}{2} (2m-1) O\left(x^2\right) \right) +\nonumber\\
&&\ste \frac{\sqrt{2}\mathcal{B}_0}{2x^2} \left(1- m \ste x + O\left(x^2\right) \right) .
\end{eqnarray} \\
Next, the outer boundary layer solution is expanded both for large and small $x$ for matching with the far-field and inner boundary layer, respectively. For $x \gg 1$, \\
\begin{eqnarray}
\xi_0^o &\sim& \mathcal{G}_{20}+\mathcal{G}_{10} \left( \frac{-2\sqrt{2}}{x} + O\left(\frac{1}{x^2}\right) \right) ,
\end{eqnarray}
\begin{eqnarray}
\xi_1^o &\sim& \mathcal{G}_{21} + \mathcal{G}_{11} \left( \frac{-2\sqrt{2}}{x} + O\left(\frac{1}{x^2}\right) \right) ,
\end{eqnarray}
and
\begin{eqnarray}
\xi_2^o &\sim& \mathcal{G}_{22} +2\mathcal{G}_{10} + \mathcal{G}_{12} \left( \frac{-2\sqrt{2}}{x} + O\left(\frac{1}{x^2}\right) \right) -\mathcal{G}_{10} \frac{x}{\sqrt{2}}  \nonumber\\
&&-\mathcal{G}_{10} \left( \frac{16\sqrt{2}}{x} + O\left(\frac{1}{x^2}\right) \right) +\mathcal{G}_{10}(m^2-1) \left( \frac{-\sqrt{2}x}{3} + \frac{8\sqrt{2}}{3x} \right)  \nonumber\\
&& +\frac{4}{3} \mathcal{G}_{10}(m^2-1) \left(1+ \mathcal{Z}_1 + \mathcal{Z}_2  + \log(2\sqrt{2})(\log(2\sqrt{2})- \log(i \gamma)) \right) ,
\label{eq:xi_2_o}
\end{eqnarray}
where $\mathcal{Z}_1$ and $\mathcal{Z}_2$ are real constants defined by the integrals,
\begin{eqnarray}
\mathcal{Z}_1 = \int_{2\sqrt{2}}^{\infty} \frac{\log(x-2\sqrt{2})-\log(x)}{x} dx ,
\end{eqnarray}
\begin{eqnarray}
\mathcal{Z}_2 = \int_{2\sqrt{2}}^{\infty} \frac{2\sqrt{2}\log x-2\sqrt{2}\log 2\sqrt{2}}{x(x-2\sqrt{2})} dx .
\end{eqnarray}
The integrals in the exact expression, \eqref{eq:outer2}, for $\xi_2^o$ appear to have a singularity when taking the limit $x \gg 1$. To resolve this, we again recognize that the travelling wave singularity must be displaced into the complex plane for an unstable mode and the constant $\gamma$ indicates this small but finite displacement. Note that the last term in \eqref{eq:xi_2_o} when combined with the constant $\mathcal{G}_{22}$ (the first term), shows that $\xi_2^o$ is independent of the arbitrary constant $\gamma$, which is only used as an intermediate step.

For $gc_0 \ll x\ll 1$,
\begin{eqnarray}
\xi_0^o &\sim& \mathcal{G}_{10} \left(\log 2\sqrt{2} - \log x + \phi \right) ,
\end{eqnarray}
\begin{eqnarray}
\xi_1^o &\sim& \mathcal{G}_{21}+ \mathcal{G}_{11} \left(\log 2\sqrt{2} - \log x + \phi \right) ,
\end{eqnarray}
\begin{eqnarray}
\xi_2^o &\sim& \mathcal{G}_{22} + 10\mathcal{G}_{10}+ \mathcal{G}_{12} \left(\log 2\sqrt{2} - \log x + \phi \right)   \nonumber\\
&&- \frac{4}{3} \mathcal{G}_{10} (m^2-1) \left(\log 2\sqrt{2} - \log x + \phi \right)   \nonumber\\
&&+ \mathcal{G}_{10}(m^2-1) \left(\frac{2}{3} + \frac{4}{3} (\mathcal{Z}_3-\mathcal{Z}_4)  - \frac{4}{3} \log(x)(\phi+\log(2\sqrt{2})) \right) \nonumber\\
&&+ \mathcal{G}_{10}(m^2-1) \left(\frac{8}{3} \log 2\sqrt{2}(\Phi_1+\log 2\sqrt{2}) -\frac{4}{3} \log(2\sqrt{2}) \log(i\gamma) \right) ,
\end{eqnarray}
where $\mathcal{Z}_3$ and $\mathcal{Z}_4$ are real constants defined by the integrals
\begin{eqnarray}
\mathcal{Z}_3= \int_{0}^{2\sqrt{2}} \frac{\log(2\sqrt{2}-x)-\log 2\sqrt{2}}{x} dx ,
\end{eqnarray}
and
\begin{eqnarray}
\mathcal{Z}_4 = \int_{0}^{2\sqrt{2}} \frac{\log x-\log 2\sqrt{2}}{x-2\sqrt{2}} dx .
\end{eqnarray} \\
The inner boundary layer solution is is written in terms of the outer boundary layer coordinate, using $y = x/g$, and then expanded for $g \ll 1$ (corresponding to $y \gg 1$) as, \\
\begin{eqnarray}
\xi_0^i &\sim& 1 , \nonumber
\end{eqnarray}
\begin{eqnarray}
\xi_1^i &\sim& -\frac{m-1}{\sqrt{2}}\left(\log x +\log\left(\frac{2}{c_0}\right)-\log g + \phi^{\prime}\right) ,
\end{eqnarray}
\begin{eqnarray}
\xi_2^i &\sim& -3\left(m-1\right)\left(\log x+\log\left(\frac{2}{c_0}\right)-\log g + \phi^{\prime}\right) +\frac{m-1}{\sqrt{2}} \left(\sqrt{2} \frac{c_1}{c_0} +6 \right) .
\end{eqnarray} \\
At O$\left(1\right)$, we note that the only term contributing is from the inner boundary layer. Thus to achieve a consistent balance a term from the eigenfunction at O$\left(\ste\right)$ must contribute. This gives, 
\begin{eqnarray}
&&g=e^{-\frac{1}{(m-1)}\sqrt{\frac{2}{\textrm{E}}}}.
\end{eqnarray}
This implies that the $\log g$ term contributes at a lower order as opposed to the order in which it appears in the inner-boundary layer expansion. A detailed explanation for this matching step is provided in Appendix \ref{sec:app}. \\

The growth rate only results at O$(\textrm{E}^2)$. By looking at the lower order matching results, it is evident that only the functional form of $\log \left( x- 2 \sqrt{2} \right) $, and not $\log \left( x- 2 \sqrt{2} \right) - \log x$, results in an imaginary part for the eigenvalue. Such a form only occurs at O$(\textrm{E}^2)$. This is also evident in the calculation that is carried out in Appendix \ref{sec:app}. Thus one may directly carry out the matching procedure for the imaginary term at O$(\textrm{E}^2)$ between the inner and outer boundary layer solutions. The eigenvalue constant, $c_3$, enters the inner boundary layer solution at O$(\textrm{E}^2)$ and writing $c_3 = c_{3r}+ i c_{3i}$, gives immediately an expression for the growth rate ($c_{3i}$):
\begin{eqnarray}
\displaystyle\frac{ (m-1) c_{3i}}{\sqrt{2} c_0}=-32\sqrt{2} \pi \mathcal{G}_{10} -\displaystyle\frac{4}{3} (m^2-1) \pi \mathcal{G}_{21}-\displaystyle\frac{8\sqrt{2}}{3} (m^2-1) \pi \mathcal{G}_{10}.
\label{eq:dispersion}
\end{eqnarray}
The RHS of \eqref{eq:dispersion} results from the limiting form of the outer boundary layer solution at O($\textrm{E}^2$) \eqref{eq:outer3}. As noted earlier, only terms of the functional form $\log \left( x- 2 \sqrt{2} \right) $ in \eqref{eq:outer3} contribute to the imaginary part and hence only these are considered when evaluating the limiting form. The matching procedure is detailed in \ref{sec:lhs_elas_match}, and to O$(\textrm{E})$, yields the following expressions for the various constants:
\begin{eqnarray}
&&c_0=4\sqrt{2}e^{\frac{6}{m-1}},\,c_1=16(m-2)e^{\frac{6}{m-1}}, \nonumber\\
&&\mathcal{G}_{20}=0,\,\,\mathcal{G}_{10}=\frac{m-1}{\sqrt{2}},\,\,\mathcal{G}_{21}=(2m-1)(m-1),\,\,\mathcal{G}_{11}=3(m-1),\nonumber\\
&&\mathcal{B}_0=-4(m-1),\,\mathcal{B}_1=4\sqrt{2}(m-1)(m-3). \nonumber
\end{eqnarray}
These are sufficient to find the growth rate at O$(\textrm{E}^2)$. Recall from section \ref{sec:innersec}, that the location of the backward travelling shear-wave is given by $y = c_0/2$ in terms of the inner boundary layer coordinate. Using the expression for $c_0$, we thus get the radial location of the backward travelling shear wave as $r = 1+ 2 \sqrt{2 \textrm{E}}  e^{-\frac{1}{(m-1)}\sqrt{\frac{2}{\textrm{E}}}+ \frac{6}{m-1}} $. The shear wave is indeed a transcendentally small distance away from the vortex core, as anticipated by the simplistic approach earlier.

\begin{figure}
     \centering
          \includegraphics[height=2.5in]{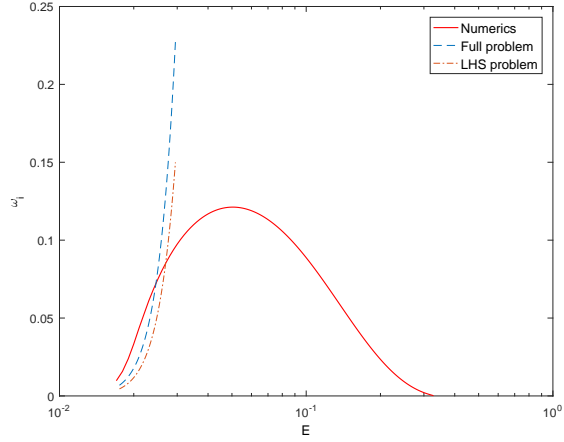}
          \caption{Comparison of numerically evaluated growth rate and the asymptotic expression obtained in (\ref{fingrowth}) and the growth rate of the LHS problem, derived in Appendix \ref{sec:app} in (\ref{eq:LHS_omg_App}) for $m$=2.}\label{Fig:Rankine_growth_comp}
\end{figure}

Substituting the above results in the growth rate expression yields,
\begin{eqnarray}
c_{3i}=-4\sqrt{2} \pi e^{\frac{6}{m-1}} \left(32\sqrt{2} +\displaystyle\frac{4}{3} (m^2-1) (2m-1)+\displaystyle\frac{8\sqrt{2}}{3} (m^2-1)\right) 
\label{fingrowth}
\end{eqnarray}
which corresponds to an unstable mode. Figure \ref{Fig:Rankine_growth_comp} shows a comparison between the asymptotic expression in \eqref{fingrowth} and the numerical results. Unfortunately, the transcendentally small growth rate for small E does not allow for a quantitative comparison. Figure \ref{Fig:Rankine_growth_comp} also shows a comparison with the growth rate obtained from the LHS problem in Appendix \ref{sec:app}.  

\section{Conclusions}\label{sec:conclu}
In this paper, we have described a novel inertio-elastic instability of a vortex column in a dilute polymer solution. The regime analyzed pertains to large Reynolds ($\Rey$) and Deborah ($\Deb$) numbers, involving a balance of inertia and elasticity at leading order, and the instability is governed by the elastic equivalent of the Rayleigh equation for swirling flows; momentum diffusion and relaxation of the disturbance polymeric stresses being neglected. The instability arises due to the resonance of elastic shear waves made possible by the background shear in the irrotational exterior of the vortex column. The dimensionless parameter that appears in the elastic Rayleigh equation, and that governs the growth rate of the unstable mode, is the elasticity number E $= \Deb/\Rey(1-\beta)$. \\

The existence of the instability is first demonstrated numerically for the Rankine vortex, for finite E, as a function of the azimuthal wavenumber $m$. A matched asymptotic expansions approach, valid for small E, helps extend the numerical results down to E $= 0$, and thereby shows the absence of an elasticity threshold for instability. That is, to say, the Rankine vortex is unstable for any finite E in the limit $\Deb, \Rey \rightarrow \infty$, although the growth rate is transcendentally small, scaling as O$(\textrm{E}^2e^{-1/\textrm{E}^{\frac{1}{2}}})$ for E $\rightarrow 0$. The numerical investigation also shows that the instability persists persists for smooth vorticity profiles - at least the `intense' Rankine-like profiles. This finding is in contrast to previous work in this regard for parallel shear flows. For instance, the inertio-elastic instability for the specific case of submerged jets was believed to be essentially dependent on the discontinuity in the first normal stress profile that arises from the assumed abrupt transition of the base-state jet profile to a quiescent ambient at either end (where the jet velocity equals zero; see \cite{MILLER05}). Thus, the instability was speculated to be absent for smoothed versions of such jet profiles. However, the present results suggest that a slightly smoothed version of the original parabolic jet will also be subject to an analogous inertio-elastic instability. We have verified the same by analyzing a smoothed version of the parabolic jet profile using a spectral calculation (not shown); an analytical approach is not possible in this case, owing to the insolubility of the inviscid (E$ = 0$) Rayleigh equation for the plane Poiseuille profile.\\

Here, we have only considered two dimensional disturbances and we anticipate novel features to emerge, both with regard to the continuous spectrum and in terms of a denumerable infinity of discrete modes, when three dimensional perturbations, with a finite axial wavenumber, are considered. Extending the analysis for three-dimensional perturbations, to the case of large but finite $\Deb$ and $\Rey$, would help precisely delineate the domain of existence of the inertio-elastic instability. Earlier computations (see \cite{OGIPOT08}) and experiments  \citep{groisman1996, boldyrev2009, muller2011, muller2013, bai2015} have only explored the  domain of existence for moderate $\Rey$ and $\Deb$, and for the Taylor-Couette geometry.\\

There is an exact analogy between the viscoelastic flows studied here, in the limit $De \rightarrow \infty$, and magnetohydrodynamic (MHD) flows in the limit of infinite magnetic Reynolds number $(Re_m)$ (\cite{OGI03, OGIPOT08}, \cite{mutabazi2019}); here Re$_m=\mu_s/(\rho\lambda_m)Re$ for a conducting fluid of density $\rho$, viscosity $\mu_s$ and magnetic diffusivity $\lambda_m$. Thus, the elastic shear waves examined here are the analog of Alfv\'en waves in the MHD context. An analog of the instability discussed in this paper exists in the astrophysical setting of accretion discs (\cite{OGIPOT08}, \cite{BALBUS91}). In particular, accretion discs with a rapid transition in the azimuthal magnetic field would be susceptible to a Alfv\'en-wave resonance instability, similar to the shear-wave resonance instability described herein. Further, the matched asymptotic expansions approach, involving multiple boundary layers, presented here may be extended to gain insight into the classical non-axisymmetric magnetorotational instability (\cite{ogilvie1996}, \cite{OGIPOT08}, \cite{velikhov2006}) as well as magnetohydrodynamic instabilities in parallel shear flows (\cite{umurhan2015, stern1963}). Although a detailed comparison must await an analysis of three-dimensional perturbations, there are features associated with earlier numerical spectral calculations in the former case that bear the hallmark of a transcendentally small length scale \citep{ogilvie1996}.\\

The present inertio-elastic instability also shares similarities with the widely studied stratorotational instability, which is also postulated as a mechanism for (outward) angular momentum transport in cold weakly ionized accretion disks \citep{dubrulle2005}. The latter arises from a resonance between Kelvin waves and/or inertia-gravity waves and also exhibits a transcendentally small scaling for small Rossby number \citep{yavneh2001, vanneste2007}. Nevertheless, there are important differences between the two cases. Most importantly, the stratorotational instability only occurs for three dimensional perturbations. Further, the locations of the Kelvin waves are fixed in the neighborhood of the two walls and hence the spacing between the resonating modes is fixed by the gap width; this, along with the exponential trapping of the individual waves explains the transcendental scaling above. On the other hand, for the inertio-elastic instability, the spacing is fixed by the region where inertial and elastic stresses are of comparable magnitude (the elastic boundary layer). Finally, the eigenmode of the usual stratorotational instability does not become singular in the limit of zero growth rate \citep{yavneh2001, vanneste2007}. The present scenario thus more closely resembles the recently studied case where the stratorotational instability results from a resonant interaction between a Kelvin/inertia-gravity wave with a baroclinic critical level \citep{balmforth2018}. \\

Finally, the inertio-elastic instability examined here may be important to the general dynamics of polymeric flows at large $\Rey$. Simplistically speaking, the vortex column analyzed in this paper may be likened to an eddy in the turbulent cascade scenario where the time scale is short enough for elastic stresses to become comparable to inertial stresses, while at the same time being much more important than viscosity. The interaction between vortices and polymers has been shown to play a crucial role in the buffer layer structure of wall-bounded turbulent flows of dilute polymer solutions \citep{roy2006mechanism, kim2007, MUNG08, TABDEGENNES86}. The instability studied in this paper can be a starting point to form a deeper mechanistic understanding of the same. \\

\appendix
\section{The LHS problem}
\label{sec:app}
As indicated in the main text, the approach used in section \ref{smallE:analysis}, may be validated by the exactly soluble LHS problem - that governed solely by the LHS of (\ref{eq:xir_elas}) for the Rankine profile. The governing equation for the LHS problem is:
\begin{eqnarray}
 D\left[r^3\left\{(\omega-m\Omega)^2-2m^2\textrm{E}\Omega'^2\right\}D\xi\right]=0 \label{eq:LHS}
\end{eqnarray} 
with $\Omega = 1/r^2$, $\Omega^{\prime} = -2/r^3$ and the boundary conditions:
\begin{eqnarray}
\xi\Bigl|_{r=1}\Bigr.&=&1, \label{eq:lhs_bc_elas_1}\\
\frac{d\xi}{dr}\Bigl|_{r=1+}\Bigr.&=&\frac{(m-1)\,(\omega-m)^2}{(\omega-m)^2-8m^2\textrm{E}},\label{eq:lhs_bc_elas_2}\\
\xi&\rightarrow&0, \hspace{.2in}\textrm{as}\,\,r\rightarrow\infty.  \label{eq:lhs_bc_elas_3}
\end{eqnarray} 
As will be seen below, the eigenvalue expression obtained from the LHS problem agrees with the complete problem to O$(g\ste)$. \\

The solution of (\ref{eq:LHS}) may be readily written as,
\begin{eqnarray}
\xi=\frac{\displaystyle\int_r^{\infty}\displaystyle\frac{dr'}{r'^3P(r')}}{\displaystyle\int_1^{\infty}\displaystyle\frac{dr'}{r'^3P(r')}},
\label{lhssol}
\end{eqnarray}
where $P(r)=(\omega-m\Omega(r))^2-2m^2\textrm{E}\Omega'(r)^2$, defined earlier following (\ref{eq:xir}). (\ref{lhssol}) satisfies the first and third boundary conditions above by construction. Applying the remaining boundary condition (\ref{eq:lhs_bc_elas_2}), leads to the following dispersion relation, valid for arbitrary $\textrm{E}$:
\begin{eqnarray}
\mathcal{D}(\omega,m;\textrm{E})\equiv1+(m-1)(\omega-m)^2\int_1^{\infty}\frac{dr'}{r'^3P(r')}=0 \label{eq:disp_LHS}
\end{eqnarray} 
  \begin{figure}
     \centering
     \subfloat[Contour $\mathcal{C}$ in complex $\omega$ plane]{
          \includegraphics[height=2.45in]{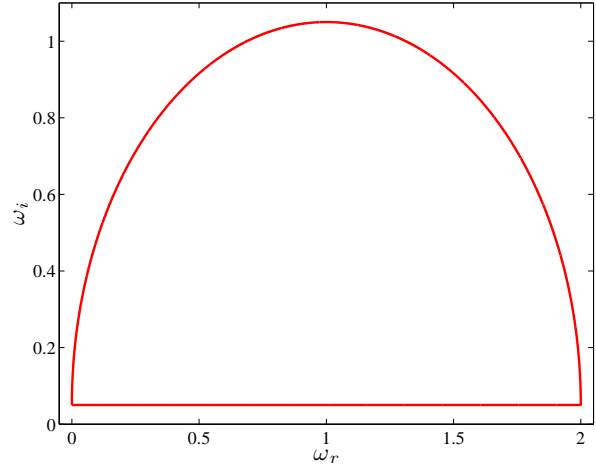}}
          \hspace{-.1in}
     \subfloat[Contour $\mathcal{C}'$ in complex $\mathcal{D}$ plane]{
          \includegraphics[height=2.45in]{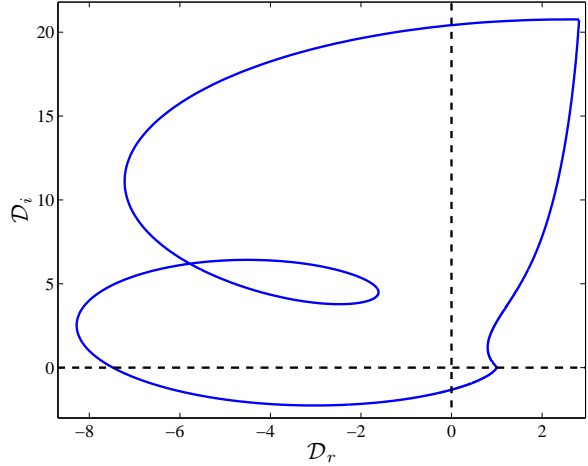}}\caption{Diagnosis of the unstable elastic mode via the Nyquist method: $m=2$, E=0.1}\label{Fig:nyquist}
\end{figure}
Before proceeding with a detailed analysis of the dispersion relation for small $\textrm{E}$, we use a Nyquist method (\cite{BALM98}) to establish the presence of an unstable mode. Consider a contour $\mathcal{C}$ in the complex $\omega$-plane (see figure \ref{Fig:nyquist}a) with an image $\mathcal{C}'$ (see figure \ref{Fig:nyquist}b) in the complex $\mathcal{D}$ plane via the mapping $\mathcal{D}(\omega)$. Based on the elastic version of Howard's semi-circle theorem for swirling flows, the contour $\mathcal{C}$ is chosen to be a suitably large semi-circle with its diameter an infinitesimal distance above the $\omega_r$-axis. This ensures that an unstable mode, if it exists, would be contained inside $\mathcal{C}$, while also avoiding the branch-cut on the real axis (associated with the elastic continuous spectra). The Nyquist criterion states that the number of times the image contour $\mathcal{C}'$ in the D plane loops the origin in the $\mathcal{D}$ plane equals  the number of zeros of $\mathcal{D}(\omega,m;\textrm{E})$ within $\mathcal{C}$. Thus, figure 17b confirms the existence of a single exponentially growing mode in the LHS problem. \\

\subsection{Small-E expansion of the exact solution}
Next, we analyze (\ref{eq:disp_LHS}) to obtain the scaling for the growth rate in the limit of small $\textrm{E}$. (\ref{eq:disp_LHS}), explicitly written out after evaluation of the integral in closed form, takes the form
 \begin{eqnarray}
\mathcal{D}(\omega,m;\textrm{E})&\equiv&1+\frac{(m-1)f^2}{(\eta_1-\eta_2)(\eta_2-\eta_3)(\eta_3-\eta_1)}\left\{\eta_1(\eta_2-\eta_3)\log(1-\eta_1)+\eta_2(\eta_3-\eta_1)\log(1-\eta_2)+\right.\nonumber\\
&&\left.\eta_3(\eta_1-\eta_2)\log(1-\eta_3)\right\}=0, \label{eq:disp2_LHS}
\end{eqnarray} 
 where $f=m/\omega-1$ and $\eta_{1,2,3}$ are the roots of the cubic: $\eta(\eta-1-f)^2-8(1+f)^2 \textrm{E}=0$. The exact expressions for $\eta_{1,2,3}$ are omitted for brevity. We make the \emph{a priori} assumption of $f\ll1$ but do not specify its smallness relative to E; implying that the unstable mode exists in the vicinity of the core. As already argued in section \ref{smallE:analysis}, this is a reasonable assumption since the balance between elastic and inertial terms in P occurs when $\omega-m\sim O(\ste)$. Expanding the exact expressions of $\eta_{1,2,3}$ for small values of E and $f$, one obtains,
 \begin{eqnarray}
 \eta_1&=&1+2 \sqrt{2\textrm{E} }-4 \textrm{E} +10 \sqrt{2} \textrm{E} ^{3/2}-64 \textrm{E} ^2+f \left(1+\sqrt{2\textrm{E} }-5 \sqrt{2} \textrm{E} ^{3/2}+64 \textrm{E} ^2+\hdots\right)\nonumber\\
 &&+f^2 \left(-\frac{\sqrt{\textrm{E} }}{2 \sqrt{2}}+\frac{15 \textrm{E} ^{3/2}}{2 \sqrt{2}}-64 \textrm{E} ^2+\hdots\right)+\hdots,\\
 \eta_2&=&8 \textrm{E} +128 \textrm{E} ^2-128 f \textrm{E} ^2+128 f^2 \textrm{E} ^2+\hdots,\\
 \eta_3&=&1-2\sqrt{2\textrm{E} }-4 \textrm{E} -10 \sqrt{2} \textrm{E} ^{3/2}-64 \textrm{E} ^2+f \left(1-\sqrt{2\textrm{E} }+5 \sqrt{2} \textrm{E} ^{3/2}+64 \textrm{E} ^2+\hdots\right)\nonumber\\
 &&+f^2 \left(\frac{\sqrt{\textrm{E} }}{2 \sqrt{2}}-\frac{15 \textrm{E} ^{3/2}}{2 \sqrt{2}}-64 \textrm{E} ^2+\hdots\right)+\hdots
 \end{eqnarray}
Substituting the above expansions in (\ref{eq:disp2_LHS}), again expanding for small E, yields the asymptotic expansion for $f$. The corresponding expression for $\omega$ is found to be,
 \begin{eqnarray}
 \frac{\omega}{m}\sim1-\ste\left[\sqrt{8}+e^{-\frac{1}{m-1}\left(\sqrt{\frac{2}{\textrm{E} }}-6\right)} \left\{2\sqrt{8}-16\sqrt{\textrm{E}}+4\sqrt{8}\textrm{E}-128\textrm{E} ^{3/2}\log(32\textrm{E})-64(3+4i\pi)\textrm{E}^{3/2}\right\} \right]\nonumber\\ \label{eq:LHS_omg_App}
\end{eqnarray}
It is evident that $f=m/\omega-1 \sim O(\ste)\ll 1$, ensuring self-consistency. The above expression highlights the transcendentally small nature of the growth rate; specifically $\omega_i =  256 \pi \textrm{E}^2 e^{-\frac{1}{m-1}\left(\sqrt{\frac{2}{\textrm{E} }}-6\right)}$ in the limit $E\ll 1$. Note that in contrast to the growth rate, the correction to the leading order wave speed has an algebraic scaling, $1- \omega_r \sim O(\ste)$, consistent with the initial scaling arguments in section \ref{smallE:analysis}.

\subsection{Matched Asymptotics Expansions approach}
We now solve the LHS problem using using a matched asymptotics expansions approach, and show that the answer obtained to $ O(\textrm{E}^2e^{-\frac{1}{m-1}\sqrt{\frac{2}{\textrm{E} }}})$ matches exactly with that obtained from the exact solution (\ref{eq:LHS_omg_App} above). This serves to validate use of the technique for the full problem for which an exact solution is not available. We assume the following double expansion for the eigenvalue for $E \ll 1$ and $g \ll 1$:
 \begin{eqnarray}
 \frac{\omega}{m}=1-\ste\left[\sqrt{8}+g \left\{c_0+c_1\sqrt{\textrm{E}}+c_2\textrm{E}+\hdots\right\} +O(g^2) \right]\label{eq:omg_pred}
\end{eqnarray}
As will be seen, the transcendental scaling for $g$ naturally emerges in this approach, implying that \eqref{eq:omg_pred} conforms to the exponential asymptotics ansatz; the transcendentally small terms of O$(g)$ in this expansion are important since they determine the growth rate at leading order. The boundary layer structure is the same as that of the full problem, and has been discussed in detail in the main text in section \ref{smallE:analysis}. 

\subsubsection{Outer region- $r-1\sim $O$(1)$}
Proceeding similar to section \ref{smallE:analysis}, one expands $P$ for small E as, 
\begin{equation}
\frac{P}{m^2}=\mathcal{S}_0+\ste \mathcal{S}_1+\textrm{E} \mathcal{S}_2 + \textrm{O}(g\ste),
\end{equation}
where $\mathcal{S}_0= \left(1-\frac{1}{r^2}\right)^2$, $\mathcal{S}_1=-2\sqrt{8}\left(1-\frac{1}{r^2}\right)$ and $\mathcal{S}_2=8\left(1-\frac{1}{r^6}\right)$. This implies an expansion for the radial displacement of the form,
\begin{eqnarray}
\xi^F(r)=\textrm{E} \,\xi^F_0(r)+\textrm{E}^{3/2}\,\xi^F_1(r)+\textrm{E}^{2}\,\xi^F_2(r)+\textrm{O}(\textrm{E}^{5/2}) + \textrm{O}(g\ste).
\end{eqnarray}
From section \ref{sec:farsec}, we see that the RHS of (\ref{eq:xir_elas}) enters at all orders in the outer region. Thus, the outer region solutions of the full problem and LHS problem differ at all orders. For the LHS problem, at O$(\textrm{E})$,
\begin{equation}
\displaystyle\frac{d}{dr}\left[r^3S_0\displaystyle\frac{d\xi_0^F}{dr}\right]=0
\end{equation}
with $\xi_0^F\rightarrow 0$ for $r\rightarrow\infty$, which gives $\xi_0^F(r)=\displaystyle\frac{\mathcal{B}_0}{(r^2-1)}$. 

At O$(\textrm{E}^{3/2})$,
\begin{equation}
\displaystyle\frac{d}{dr}\left[r^3 S_0\displaystyle\frac{d\xi_1^F}{dr}\right]  = -\displaystyle\frac{d}{dr}\left[r^3 S_1\displaystyle\frac{d\xi_0^F}{dr}\right],
\end{equation}
with $\xi_1^F\rightarrow 0$ for $r\rightarrow\infty$, which gives $\xi_1^F(r)=\displaystyle\frac{2\sqrt{2}\mathcal{B}_0}{(r^2-1)^2}+\displaystyle\frac{4\sqrt{2}\mathcal{B}_0 + \mathcal{B}_1}{(r^2-1)}$.

At O$(\textrm{E}^{2})$,
\begin{equation}
\displaystyle\frac{d}{dr}\left[r^3 S_0\displaystyle\frac{d\xi_2^F}{dr}\right]  = -\displaystyle\frac{d}{dr}\left[r^3 S_1\displaystyle\frac{d\xi_1^F}{dr}\right]- \displaystyle\frac{d}{dr}\left[r^3 S_2\displaystyle\frac{d\xi_0^F}{dr}\right],
\end{equation}
with $\xi_2^F\rightarrow 0$ for $r\rightarrow\infty$, which gives, $\xi_2^F(r)=\displaystyle\frac{32\mathcal{B}_0}{3(r^2-1)^3}+\displaystyle\frac{20\mathcal{B}_0 + 2\sqrt{2}\mathcal{B}_1}{(r^2-1)^2}+\displaystyle\frac{32\mathcal{B}_0 + 4\sqrt{2}\mathcal{B}_1 + \mathcal{B}_2}{(r^2-1)}+8\mathcal{B}_0\log{\displaystyle\frac{r^2-1}{r^2}}$. The $\mathcal{B}_i$'s in the above expressions are integration constants, which are determined by matching to the outer boundary-layer solution.

\subsubsection{Outer boundary layer - $r-1\sim$ $ \ste$}
In terms of the rescaled coordinate $x = (r-1)/\ste$, the LHS problem in the outer boundary layer takes the form:
\begin{eqnarray}
\frac{d}{dx}\left[\mathcal{Q}\frac{d\xi^o}{dx}\right]=0, \label{eq:xir_OBLA}
\end{eqnarray}
where $\mathcal{Q}$, for small E, is expanded as,
\begin{equation}
\mathcal{Q}=\mathcal{Q}_0+\ste \mathcal{Q}_1+\textrm{E} \mathcal{Q}_2 +\textrm{E}^{3/2} \mathcal{Q}_3 + \textrm{O}(\textrm{E}^{2})+ \textrm{O}(g\ste),
\end{equation}
with $\mathcal{Q}_0=4x(x-2\sqrt{2})$, $\mathcal{Q}_1=-12\sqrt{2}x(x-2\sqrt{2})$, $\mathcal{Q}_2=x^2(x^2-4\sqrt{2}x-24)$ and $\mathcal{Q}_3=x^3(88-x^2)$.
For the displacement, we assume the expansion,
\begin{align}
\xi^o(x)&= \ste \,\xi^o_0(x)+\textrm{E}\,\xi^o_1(x)+\textrm{E}^{3/2}\,\xi^o_2(x) + \nonumber \\ 
&  \textrm{E}^{2} \log (32 \textrm{E}) \,\xi^o_{31}(x) +\textrm{E}^{2}\,\xi^o_3(x)+\textrm{O}(\textrm{E}^{5/2}) + \textrm{O}(g\ste) .
\label{eq:app_outer_bl}
\end{align}
The $\log (32 \textrm{E}) $ term in \eqref{eq:app_outer_bl} is necessitated by the $\log$ term in the outer region solution at $\textrm{E}^{2}$ viz. $\xi_2^F$.

At O${(\textrm{E}^{1/2})}$,
\begin{equation}
\displaystyle\frac{d}{dx}\left[\mathcal{Q}_0\displaystyle\frac{d\xi_0^o}{dx}\right]=0
\end{equation}
\begin{equation}
\Rightarrow\,\,\xi_0^o(x)=\mathcal{G}_{10}+\mathcal{G}_{11}\log\left(\displaystyle\frac{x-2\sqrt{2}}{x}\right).
\label{eq:outer_app_sol_1}
\end{equation}
At O${(\textrm{E})}$,
\begin{equation}
\displaystyle\frac{d}{dx}\left[\mathcal{Q}_0\displaystyle\frac{d\xi_1^o}{dx}\right]=-\displaystyle\frac{d}{dx}\left[\mathcal{Q}_1\displaystyle\frac{d\xi_0^o}{dx}\right],
\end{equation}
\begin{equation}
\Rightarrow\,\,\xi_1^o(x)=\mathcal{G}_{21}+\mathcal{G}_{11}\log\left(\displaystyle\frac{x-2\sqrt{2}}{x}\right).
\end{equation}
Note that at both O$(\textrm{E}^{1/2})$ and O$(\textrm{E})$ one obtains the same governing equation since $\mathcal{Q}_0$ and $\mathcal{Q}_1$ have an identical dependence on $x$ (to within a multiplicative constant). 
At O$(\textrm{E}^{3/2})$,
\begin{equation}
\displaystyle\frac{d}{dx}\left[\mathcal{Q}_0\displaystyle\frac{d\xi_2^o}{dx}\right]=-\displaystyle\frac{d}{dx}\left[\mathcal{Q}_1\displaystyle\frac{d\xi_1^o}{dx}\right]-\displaystyle\frac{d}{dx}\left[\mathcal{Q}_2\displaystyle\frac{d\xi_0^o}{dx}\right],
\end{equation}
\begin{equation}
\Rightarrow\,\,\xi_2^o(x)=\mathcal{G}_{22}+\mathcal{G}_{12}\log\left(\displaystyle\frac{x-2\sqrt{2}}{x}\right)-\displaystyle\frac{\mathcal{G}_{10}}{\sqrt{2}}\left(\displaystyle\frac{32}{x-2\sqrt{2}}+x-2\sqrt{2}\right).
\end{equation}
At  O$(\textrm{E}^{2} \log (32 \textrm{E}) )$,
\begin{equation}
\displaystyle\frac{d}{dx}\left[\mathcal{Q}_0\displaystyle\frac{d\xi_{31}^o}{dx}\right]=0,
\end{equation}
\begin{equation}
\Rightarrow\,\,\xi_{31}^o(x)=\tilde{\mathcal{G}}_{23}+\tilde{\mathcal{G}}_{13}\log\left(\displaystyle\frac{x-2\sqrt{2}}{x}\right).
\end{equation}
At  O$(\textrm{E}^{2})$,
\begin{equation}
\displaystyle\frac{d}{dx}\left[\mathcal{Q}_0\displaystyle\frac{d\xi_3^o}{dx}\right]=-\displaystyle\frac{d}{dx}\left[\mathcal{Q}_1\displaystyle\frac{d\xi_2^o}{dx}+\mathcal{Q}_2\displaystyle\frac{d\xi_1^o}{dx}+\mathcal{Q}_3\displaystyle\frac{d\xi_0^o}{dx}\right]
\end{equation}
\begin{align}
\Rightarrow\,\,\xi_3^o(x)&=\mathcal{G}_{23}+\mathcal{G}_{13}\log\left(\displaystyle\frac{x-2\sqrt{2}}{x}\right)+3\sqrt{2}\mathcal{G}_{12}\log\left(\displaystyle\frac{x-2\sqrt{2}}{x}\right) \nonumber\\
& -3\mathcal{G}_{10}\left(x+\displaystyle\frac{32}{x-2\sqrt{2}}\right) -\displaystyle\frac{\mathcal{G}_{11}}{\sqrt{2}}\left(\displaystyle\frac{32}{x-2\sqrt{2}}+x-2\sqrt{2}\right) \nonumber\\
& +\displaystyle\frac{\mathcal{G}_{10}}{\sqrt{2}}\left(\displaystyle\frac{(x-2\sqrt{2})^2}{2}+6\sqrt{2}\left(x-2\sqrt{2}\right)-64\log\left(x-2\sqrt{2}\right)+\displaystyle\frac{160\sqrt{2}}{x-2\sqrt{2}}\right).
\end{align}
The $\mathcal{G}_{ij}$'s in the above expressions are integration constants, which are determined by matching to the far-field and inner boundary-layer solutions.

\subsubsection{Inner boundary layer - $r-1 \sim$ O$(g\ste)$}
Introducing the inner boundary layer coordinate, $y = (r-1)/g \ste $ with $g, \textrm{E} \ll 1$ and the inner boundary layer displacement as $\xi(r)=\xi^{i}(y)$, we have from (\ref{eq:xir_elas})-(\ref{eq:bc_elas_2}): 
\begin{eqnarray}
\frac{d}{dy}\left[\mathcal{R}\frac{d\xi^i}{dy}\right]=\frac{g^2\textrm{E}\mathcal{R}(m^2-1)}{(1+g\ste y)^2}\xi^i, \label{eq:xir_IBL_app}
\end{eqnarray}
with $\mathcal{R}=(1+g\ste y)^3P/(m^2g\textrm{E})$, which is further expanded as,
\begin{equation}
\mathcal{R}=\mathcal{R}_0+\ste \mathcal{R}_1+ \textrm{E} \mathcal{R}_2 + \textrm{E}^{3/2} \log (32  \textrm{E} ) \mathcal{R}_{30}  +  \textrm{E}^{3/2} \mathcal{R}_{31} + \textrm{O}(\textrm{E}^{2}) ,
\label{eq:innerexp_r_app}
\end{equation}
where $\mathcal{R}_0=4\sqrt{2}(c_0-2y)$, $\mathcal{R}_1=4\sqrt{2}(c_1+6\sqrt{2}y)$, $\mathcal{R}_2=4\sqrt{2}c_2$, $\mathcal{R}_{30}=4\sqrt{2}c_{30}$ and $\mathcal{R}_{31}=4\sqrt{2}c_{31}$, where the 32 is retained in the log term in \eqref{eq:innerexp_r_app} for convenience. The boundary conditions are given by,
\begin{eqnarray}
\xi^i(y=0)&=&1 \label{eq:IBL_BC1_app}\\
\frac{d\xi^i}{dy}\Bigl(y=0\Bigr)&=&(m-1)\ste\left\{\frac{\sqrt{2}}{c_0}-\frac{\sqrt{2}c_1\ste}{c_0^2}+\frac{\sqrt{2} \left(c_1^2-c_0 c_2\right)\textrm{E}}{c_0^3}-\frac{\sqrt{2}  \left(c_1^3-2 c_0 c_1 c_2+c_0^2 c_{31}\right)\textrm{E} ^{3/2} }{c_0^4}-\right.\nonumber\\
&&\left.\frac{\sqrt{2} c_{30} \textrm{E} ^{3/2} \log(32 \textrm{E})}{c_0^2} +\hdots\right\}\label{eq:IBL_BC2_app}
\end{eqnarray}
$\xi^i(y)$ can thus be expanded as,
\begin{align}
\xi^i(y)&= \xi_0^i(y)+\ste\, \xi_{1}^i(y)+\textrm{E}\, \xi_2^i(y) +\textrm{E}^{3/2} \, \xi_3^i(y) \nonumber\\
& +  \textrm{E}^{2} \log (32  \textrm{E} ) \, \xi_{40}^i(y) + \textrm{E}^{2} \, \xi_{41}^i(y)  + \textrm{O}(\textrm{E}^{5/2})+ \textrm{O}(g\ste) ,
\label{eq:innerexp_app}
\end{align}
where the leading order radial displacement is now O$(1)$. Importantly, as already argued in the main text, the RHS in \eqref{eq:xir_IBL_app} is transcendentally small (on account of $g$), and may therefore be neglected to all algebraic orders considered below. Thus, at O(1), one obtains, \\
Finally, at O${(1)}$ we obtain,
\begin{equation}
\displaystyle\frac{d}{dy}\left[\mathcal{R}_0\displaystyle\frac{d\xi_0^i}{dy}\right]=0,
\end{equation}
with $\xi_0^i(y=0)=1$ and $\displaystyle\frac{d\xi_0^i}{dy}(y=0)=0$, which gives
\begin{equation}
\xi_0^i(y)=1 .
\end{equation}
At O${(\textrm{E}^{1/2})}$ we obtain, 
\begin{equation}
\displaystyle\frac{d}{dy}\left[\mathcal{R}_0\displaystyle\frac{d\xi_1^i}{dy}\right]=0,
\end{equation}
with $\xi_1^i(y=0)=0$ and $\displaystyle\frac{d\xi_1^i}{dy}(y=0)=\displaystyle\frac{\sqrt{2}(m-1)}{c_0}$, which gives,
\begin{equation}
\xi_1^i(y)=-\displaystyle\frac{(m-1)}{\sqrt{2}}\log\left(\frac{c_0-2y}{c_0}\right).
\label{eq:inner_app_sol_1}
\end{equation}
At O${(\textrm{E})}$ we obtain, 
\begin{equation}
\displaystyle\frac{d}{dy}\left[\mathcal{R}_0\displaystyle\frac{d\xi_2^i}{dy}\right]=-\displaystyle\frac{d}{dy}\left[\mathcal{R}_1\displaystyle\frac{d\xi_1^i}{dy}\right],
\end{equation}
with $\xi_2^i=0$ and $\displaystyle\frac{d\xi_2^i}{dy}=-\displaystyle\frac{\sqrt{2}(m-1)c_1}{c_0^2}$, which gives,
\begin{equation}
\xi_2^i(y)=-3(m-1)\log\left(\displaystyle\frac{c_0-2y}{c_0}\right)-\displaystyle\frac{(m-1)}{2}(\sqrt{2}c_1+6c_0)\left\{\displaystyle\frac{1}{c_0-2y}-\frac{1}{c_0}\right\}.
\end{equation}
At O${(\textrm{E}^{3/2})}$ we obtain, 
\begin{equation}
\displaystyle\frac{d}{dy}\left[\mathcal{R}_0\displaystyle\frac{d\xi_3^i}{dy}\right]=-\displaystyle\frac{d}{dy}\left[\mathcal{R}_1\displaystyle\frac{d\xi_2^i}{dy}\right]-\displaystyle\frac{d}{dy}\left[\mathcal{R}_2\displaystyle\frac{d\xi_1^i}{dy}\right],
\end{equation}
with $\xi_3^i=0$ and $\displaystyle\frac{d\xi_3^i}{dy}=\displaystyle\frac{\sqrt{2}(m-1)(c_1-c_0c_2)}{c_0^3}$, which gives,
\begin{align}
\xi_3^i(y)&=-9\sqrt{2}(m-1)\log\left(\displaystyle\frac{c_0-2y}{c_0}\right)-\displaystyle\frac{(m-1)}{\sqrt{2}}(c_2+6\sqrt{2}c_1+36c_0)\left\{\displaystyle\frac{1}{c_0-2y}-\frac{1}{c_0}\right\}\nonumber\\ 
& +\displaystyle\frac{(m-1)}{2\sqrt{2}}(c_1+3\sqrt(2)c_0)^2\left\{\displaystyle\frac{1}{(c_0-2y)^2}-\frac{1}{c_0^2}\right\}.
\end{align}
At O${(\textrm{E}^{2}\log(32\textrm{E}))}$ we obtain,
\begin{equation}
\displaystyle\frac{d}{dy}\left[\mathcal{R}_0\displaystyle\frac{d\xi_{40}^i}{dy}\right]=-\displaystyle\frac{d}{dy}\left[\mathcal{R}_{30}\displaystyle\frac{d\xi_1^i}{dy}\right],
\end{equation}
with $\xi_{40}^i=0$ and $\displaystyle\frac{d\xi_{40}^i}{dy}=-\displaystyle\frac{\sqrt{2}(m-1)c_{30}}{c_0^2}$, which gives,
\begin{equation}
\xi_{40}^i(y)=-\displaystyle\frac{(m-1)}{\sqrt{2}}c_{30}\left\{\displaystyle\frac{1}{c_0-2y}-\frac{1}{c_0}\right\}.
\end{equation}
At O${(\textrm{E}^{2})}$ we obtain,
\begin{equation}
\displaystyle\frac{d}{dy}\left[\mathcal{R}_0\displaystyle\frac{d\xi_{41}^i}{dy}\right]=-\displaystyle\frac{d}{dy}\left[\mathcal{R}_1\displaystyle\frac{d\xi_3^i}{dy}+\mathcal{R}_2\displaystyle\frac{d\xi_2^i}{dy}+\mathcal{R}_{31}\displaystyle\frac{d\xi_1^i}{dy}\right],
\end{equation}
with $\xi_{41}^i=0$ and $\displaystyle\frac{d\xi_{41}^i}{dy}=-\displaystyle\frac{\sqrt{2}(m-1)(c_1^3-2c_0c_1c_2+c_0^2c_3)}{c_0^4}$, which gives,
\begin{align}
\xi_{41}^i(y)&=-54(m-1)\log\left(\displaystyle\frac{c_0-2y}{c_0}\right)-\displaystyle\frac{(m-1)}{\sqrt{2}}\{c_3+6\sqrt{2}c_2+54(c_1+3\sqrt{2}c_0)\}\left\{\displaystyle\frac{1}{c_0-2y}-\frac{1}{c_0}\right\} \nonumber\\
& +\displaystyle\frac{(m-1)}{2\sqrt{2}}\{2c_2(c_1+3\sqrt{2}c_0)+9\sqrt{2}(c_1+3\sqrt(2)c_0)^2\}\left\{\displaystyle\frac{1}{(c_0-2y)^2}-\frac{1}{c_0^2}\right\} \nonumber\\
& -\displaystyle\frac{(m-1)}{3\sqrt{2}}(c_1+3\sqrt(2)c_0)^3\left\{\displaystyle\frac{1}{(c_0-2y)^3}-\frac{1}{c_0^3}\right\}.
\end{align}

\subsubsection{Matching}\label{sec:lhs_elas_match}
With the inner and outer boundary layer solutions, and the solution in the outer region, in place, we proceed to derive the necessary constants via matching.
To begin with, we expand the far-field solution for small values of $r-1$ by writing it in terms of the outer boundary layer coordinate as $r=1+ \ste x$, which leads to the following limiting forms:\\
\begin{eqnarray}
\textrm{E} \xi_0^F &\sim& \ste \frac{\mathcal{B}_0}{2x} \left(1-  \frac{\ste x}{2}  + O\left(x^2\right) \right) ,
\label{eq:far_limit_app_1}
\end{eqnarray}
\begin{eqnarray}
\textrm{E}^{3/2} \xi_1^F &\sim& \textrm{E} \frac{\mathcal{B}_1+4\sqrt{2}\mathcal{B}_0}{2x} \left(1-  \frac{\ste x}{2} O\left(x^2\right) \right) +\nonumber\\
&&\ste \frac{\sqrt{2}\mathcal{B}_0}{2x^2} \left(1-  \ste x  + O\left(x^2\right) \right) ,
\label{eq:far_limit_app_2}
\end{eqnarray}
\begin{eqnarray}
\textrm{E}^{2} \xi_2^F &\sim& \textrm{E}^{3/2} \frac{\mathcal{B}_2+4\sqrt{2}\mathcal{B}_1+32\mathcal{B}_0}{2x} \left(1-  \frac{\ste x}{2}  + O\left(x^2\right) \right) +\nonumber\\
&& \textrm{E} \frac{2\sqrt{2}\mathcal{B}_1+20\mathcal{B}_0}{4x^2} \left(1-  \ste x  + O\left(x^2\right) \right) +\nonumber\\
&& \ste \frac{4 \mathcal{B}_0}{3 x^3} \left(1- \ste \frac{3x}{2} + O\left(x^2)\right) \right)  +\nonumber\\
&& 8 \mathcal{B}_0  \textrm{E}^{2} \left(\log \ste+\log 2x +\log\left(1+\ste\frac{x}{2}\right)-2 \log\left(1+\ste x\right)\right) .
\end{eqnarray}
Next, the outer boundary layer solution needs to be expanded both for large and small $x$. The large and small $x$ limiting forms are needed for matching with the limiting forms of the outer region and inner boundary layer, respectively. For $x \gg 1$ we obtain the following limiting forms, \\
\begin{eqnarray}
\xi_0^o &\sim& \mathcal{G}_{20}+\mathcal{G}_{10} \left( \frac{-2\sqrt{2}}{x} + O\left(\frac{1}{x^2}\right) \right) ,
\label{eq:outer_limit_app_0}
\end{eqnarray}
\begin{eqnarray}
\xi_1^o &\sim& \mathcal{G}_{21} + \mathcal{G}_{11} \left( \frac{-2\sqrt{2}}{x} + O\left(\frac{1}{x^2}\right) \right) ,
\label{eq:outer_limit_app_0b}
\end{eqnarray}
\begin{eqnarray}
\xi_2^o &\sim& \mathcal{G}_{22} +2\mathcal{G}_{10} + \mathcal{G}_{12} \left( \frac{-2\sqrt{2}}{x} + O\left(\frac{1}{x^2}\right) \right) -\mathcal{G}_{10} \frac{x}{\sqrt{2}} \nonumber \\
&&-\mathcal{G}_{10} \left( \frac{16\sqrt{2}}{x} + O\left(\frac{1}{x^2}\right) \right) ,
\end{eqnarray}
\begin{eqnarray}
\xi_3^o &\sim& \left(-2\sqrt{2}\mathcal{G}_{13} -12\mathcal{G}_{12} -16 \sqrt{2}\mathcal{G}_{11} +64\mathcal{G}_{10} \right)\frac{1}{x} + \left(\mathcal{G}_{23} +2 \mathcal{G}_{11} - 10\sqrt{2}\mathcal{G}_{10} \right) \nonumber\\
&&+\left(-3\mathcal{G}_{10} - \mathcal{G}_{11} \frac{1}{\sqrt{2}} + 4\mathcal{G}_{10} \right) x -32 \sqrt{2} \mathcal{G}_{10} \log x + O\left(\frac{1}{x^2}\right) .
\end{eqnarray}
In the limit $g \ll x \ll 1$, one needs to account for the multi-valuedness of the logarithm in the $ \log(x - 2 \sqrt{2})$ term  (see (\ref{eq:outer_app_sol_1})). Here, $x = 2\sqrt{2}$ denotes the singularity of the forward travelling shear wave in the limit of neutral stability. For $x$ crossing $2\sqrt{2}$ along the real axis, there is an ambiguity in the sign of the phase jump associated with the logarithm, an aspect familiar from inviscid hydrodynamic stability \citep{DRAZINREID81}. The resolution lies in recognizing that the singularity associated with the unstable mode must lie in the complex plane, so that $x = 2\sqrt{2} - i \epsilon$ ($\epsilon >0$), even if the displacement from the real axis ($\epsilon$) is transcendentally small. This resolves the sign of the phase jump; the logarithmic term is now $ \log(x - 2 \sqrt{2} + i \epsilon)$, which leads to a phase jump of $i \pi$ in the limit $x \ll 1$. With this in place, one obtains the following small-$x$ forms for the outer boundary layer solutions: \\
\begin{eqnarray}
\xi_0^o &\sim& \mathcal{G}_{10} \left(\log 2\sqrt{2} - \log x + i \pi \right) ,
\label{eq:outer_limit_app_1}
\end{eqnarray}
\begin{eqnarray}
\xi_1^o &\sim& \mathcal{G}_{21}+ \mathcal{G}_{11} \left(\log 2\sqrt{2}  - \log x + i \pi \right) ,
\label{eq:outer_limit_app_1b}
\end{eqnarray}
\begin{eqnarray}
\xi_2^o &\sim& \mathcal{G}_{22} + 10\mathcal{G}_{10}+ \mathcal{G}_{12} \left(\log 2\sqrt{2} - \log x + i \pi \right),
\end{eqnarray}
\begin{eqnarray}
\xi_{31}^o &\sim& \tilde{\mathcal{G}}_{23} + \tilde{\mathcal{G}}_{13} \left(\log 2\sqrt{2} - \log x + i \pi \right) ,
\end{eqnarray}
\begin{eqnarray}
\xi_3^o &\sim& \mathcal{G}_{23} - 26\sqrt{2}\mathcal{G}_{10} +10\mathcal{G}_{11}+ \left( \mathcal{G}_{13} +3 \sqrt{2}\mathcal{G}_{12} \right) \left(\log 2\sqrt{2} - \log x + i \pi \right)  \nonumber\\ 
&& -32\sqrt{2} \mathcal{G}_{10} \left(\log 2\sqrt{2} + i \pi \right) .
\label{eq:outer_limit_app_2}
\end{eqnarray}
Finally, the inner boundary layer solution is expanded for large values of $y$ and towards this end, is written in terms of the outer boundary layer coordinate, $y=x/g$ with $x \sim $ O$(1)$ and $g \ll 1$. The multi-valuedness of the logarithmic is again accounted for by noting that the backward travelling shear-wave must lie at $y = c_0/2 - i \epsilon^{\prime}$ ($ \epsilon^{\prime} > 0$), which leads to a logarithmic term of the form $\log ( c_0-2y - 2 i \epsilon^{\prime} )$ in \eqref{eq:inner_app_sol_1} for instance. This in turn leads to a phase jump of $- i \pi$ across $y = c_{0}/2$. The large-y forms of the inner boundary layer solutions are given by: \\
\begin{eqnarray}
\xi_0^i &\sim& 1 ,
\label{eq:inner_limit_app_1}
\end{eqnarray}
\begin{eqnarray}
\xi_1^i &\sim& -\frac{m-1}{\sqrt{2}}\left(\log x +\log\left(\frac{2}{c_0}\right)-\log g - i \pi \right) ,
\label{eq:inner_limit_app_2}
\end{eqnarray}
\begin{eqnarray}
\xi_2^i &\sim& -3\left(m-1\right)\left(\log x +\log\left(\frac{2}{c_0}\right)-\log g - i \pi \right) +\frac{m-1}{\sqrt{2}} \left(\sqrt{2} \frac{c_1}{c_0} +6 \right) ,
\label{eq:inner_limit_app_2b}
\end{eqnarray}
\begin{eqnarray}
\xi_3^i &\sim& -9 \sqrt{2}\left(m-1\right)\left(\log x +\log\left(\frac{2}{c_0}\right)-\log g - i \pi\right) +\frac{m-1}{\sqrt{2}} \left(\frac{c_2}{c_0} +6 \sqrt{2}\frac{c_1}{c_0} +36 \right) \nonumber\\
&&-\frac{m-1}{2\sqrt{2}} \left(\frac{c_1}{c_0} +3\sqrt{2} \right)^2 ,
\end{eqnarray}
\begin{eqnarray}
\xi_{40}^i &\sim& \frac{m-1}{\sqrt{2}} \frac{c_{30}}{c_0} ,
\end{eqnarray}
\begin{align}
\xi_{41}^i(y)& \sim -54(m-1) \left(\log x +\log\left(\frac{2}{c_0}\right)-\log g - i \pi\right)  +\displaystyle\frac{(m-1)}{\sqrt{2}c_0}\{c_{31}+6\sqrt{2} c_2+54( c_1+3\sqrt{2}c_0)\}  \nonumber\\
& - \displaystyle\frac{(m-1)}{2\sqrt{2} c_0^2}\{2c_2(c_1+3\sqrt{2}c_0)+9\sqrt{2}(c_1+3\sqrt(2)c_0)^2\} -\displaystyle\frac{(m-1)}{3\sqrt{2}c_0^3}(c_1+3\sqrt(2)c_0)^3. 
\label{eq:inner_limit_app_3}
\end{align}

Having determined the appropriate limiting forms, we proceed to match the appropriate terms. In matching (\ref{eq:inner_limit_app_1}-\ref{eq:inner_limit_app_3}) to (\ref{eq:outer_limit_app_1}-\ref{eq:outer_limit_app_2}), an inconsistency appears in that the only term at O$(1)$ is that from the inner boundary layer ($\xi_0^i$ in (\ref{eq:inner_limit_app_1})), and there are no terms for this to match on to, in the outer boundary layer solutions. The resolution involves recognizing that one of the terms at O($\ste$), the one proportional to $ \log g$ in (\ref{eq:inner_limit_app_2}), jumps order to cancel the aforementioned O$(1)$ contribution. This implies,
\begin{eqnarray}
&&g=e^{-\frac{1}{(m-1)}\sqrt{\frac{2}{\textrm{E}}}},
\end{eqnarray}
and confirms the transcendental smallness of $g$ anticipated earlier. Owing to this transcendental smallness, the $\log g$ term contributes at an algebraic order lower than the nominal one at which it appears. Next, at O$(\ste)$, we first match the $\log x$ terms in the inner and outer boundary layer solutions  ((\ref{eq:inner_limit_app_2}) and (\ref{eq:outer_limit_app_1}), respectively), to determine the unknown constant ($\mathcal{G}_{10}$) in the outer boundary layer solution as, 
\begin{equation}
\mathcal{G}_{10}=\frac{m-1}{\sqrt{2}}.
\end{equation}
The constant terms in the outer boundary layer \eqref{eq:outer_limit_app_0} and the outer region \eqref{eq:far_limit_app_1} solutions are matched to determine the unknown constant ($\mathcal{G}_{20}$) in the outer boundary layer solution as,
\begin{equation}
\mathcal{G}_{20}=0.
\end{equation}
Then we match the constant term in the inner boundary layer (\ref{eq:inner_limit_app_2}) and the outer boundary layer solutions (\ref{eq:outer_limit_app_1}),  to determine the leading order term in the eigenvalue expansion ($c_0$) as,
\begin{equation}
c_0=4\sqrt{2}e^{\frac{6}{m-1}}.
\end{equation}
Finally, matching the coefficient of the $\frac{1}{x}$ term in the outer boundary layer  \eqref{eq:outer_limit_app_0} and the outer region \eqref{eq:far_limit_app_1} solutions is used to determine the unknown constant ($\mathcal{B}_0$) in the outer region solution as,
\begin{equation}
\mathcal{B}_{0}=-4(m-1).
\end{equation}
This completes the matching at O$(\ste)$. For all higher orders, the matching thus proceeds in the same sequence as above. For instance, at O(E), matching the $ \log x$ term from the inner (\ref{eq:inner_limit_app_2b}) and outer boundary layer solution (\ref{eq:outer_limit_app_1b}) leads to,
\begin{equation}
\mathcal{G}_{11}=3(m-1).
\end{equation}
Next, matching the constant terms in the outer boundary layer \eqref{eq:outer_limit_app_0b} and the outer region \eqref{eq:far_limit_app_2} solutions gives,
\begin{equation}
\mathcal{G}_{21}=(m-1).
\end{equation}
Matching the constant term in the inner boundary layer (\ref{eq:inner_limit_app_2b}) and the outer boundary layer solutions (\ref{eq:outer_limit_app_1b}) gives,
\begin{equation}
c_1=-16e^{\frac{6}{m-1}}.
\end{equation}
Finally, matching the $\frac{1}{x}$ term in the outer boundary layer \eqref{eq:outer_limit_app_0b} and the outer region \eqref{eq:far_limit_app_2} solutions gives,
\begin{equation}
\mathcal{B}_{1}=0.
\end{equation}
In a similar manner, one determines all unknown constants involved. The results are summarized below:
\begin{eqnarray}
&&g=e^{-\frac{1}{(m-1)}\sqrt{\frac{2}{\textrm{E}}}},\nonumber\\
&&c_0=4\sqrt{2}e^{\frac{6}{m-1}},\,c_1=-16e^{\frac{6}{m-1}},\,c_2=8\sqrt{2}e^{\frac{6}{m-1}},\,c_3=\left(-656\log\left(2\right)-192-256 i \pi\right) e^{\frac{6}{m-1}}\nonumber\\
&&\mathcal{G}_{20}=0,\,\,\mathcal{G}_{10}=\frac{m-1}{\sqrt{2}},\,\,\mathcal{G}_{21}=(m-1),\,\,\mathcal{G}_{11}=3(m-1),\nonumber\\
&&\mathcal{G}_{22}=\frac{3\sqrt{2}(m-1)}{2},\,\,\mathcal{G}_{12}=9\sqrt{2}(m-1),\nonumber\\
&&\mathcal{G}_{23}=\frac{81(m-1)}{3}-34\log\left(2\right)(m-1),\,\,\mathcal{G}_{13}=0,\nonumber\\
&&\mathcal{B}_0=-4(m-1),\,\mathcal{B}_1=0,\,\mathcal{B}_2=0.\nonumber\\
\end{eqnarray}
As a result, one has the following asymptotic expression for the eigenvalue, 
\begin{eqnarray}
\frac{\omega}{m}\sim1-\ste\left[\sqrt{8}+e^{-\frac{1}{m-1}\left(\sqrt{\frac{2}{\textrm{E} }}-6\right)} \left\{2\sqrt{8}-16\sqrt{\textrm{E}}+4\sqrt{8}\textrm{E}-128\textrm{E} ^{3/2}\log(32\textrm{E})-64(3+4i\pi)\textrm{E}^{3/2}\right\} \right] , \nonumber\\ 
\end{eqnarray}
which is the same as the expression as that obtained by direct expansion of the exact solution, for small E, in \eqref{eq:LHS_omg_App}.

\bibliographystyle{jfm}

\bibliography{refer}

\end{document}